\documentclass[3p,preprint]{elsarticle}

\usepackage{graphicx}
\usepackage{amssymb}
\usepackage{amsmath}
\usepackage{comment}
\usepackage{rotating}
\usepackage{epstopdf}
\usepackage[capitalise]{cleveref}

\usepackage{tikz}
\usetikzlibrary{shapes,arrows}

\graphicspath{{./img/}}

\newcommand{\II}{{\mathcal I}} %
\newcommand{\OO}{{\mathcal O}} %

\begin{document}
\begin{frontmatter}


\title{A Fast-Slow Model of Banded Vegetation Pattern Formation in Drylands}

\author[1]{Punit Gandhi} 
\ead{gandhipr@vcu.edu}
\address[1]{ Department of Mathematics and Applied Mathematics, Virginia Commonwealth University, Richmond, VA 23284, USA}

\author[2,3]{Sara Bonetti}
\ead{sara.bonetti@usys.ethz.ch}
\address[2]{Department of Environmental Systems Science, ETH Z\"urich, 8092 Z\"urich, Switzerland}
\address[3]{Bartlett School of Environment, Energy and Resources, University College London,WC1H 0NN London, UK}

\author[4]{Sarah Iams}
\ead{siams@seas.harvard.edu}
\address[4]{John A. Paulson School of Engineering and Applied Sciences, Harvard University, Cambridge, MA 02138, USA}

\author[5]{Amilcare Porporato}
\ead{aporpora@princeton.edu}
\address[5]{Department of Civil and Environmental Engineering and Princeton Environmental Studies, Princeton University, Princeton, NJ 08540, USA}
\author[6]{Mary Silber}
\ead{msilber@uchicago.edu}
\address[5]{Department of Statistics and Committee on Computational and Applied Mathematics, University of Chicago, Chicago, IL 60637, USA}



\begin{abstract}

From infiltration of water into the soil during rainstorms to seasonal plant growth and death, the ecohydrological processes that are thought to be relevant to the formation of banded vegetation patterns in drylands occur across multiple timescales.  We propose a new fast-slow switching model in order to capture these processes on appropriate timescales within a conceptual modeling framework based on reaction-advection-diffusion  equations.  The fast system captures hydrological processes that occur on minute to hour timescales during and shortly after major rainstorms, assuming a fixed  vegetation distribution. These include key feedbacks between vegetation biomass and downhill surface water transport, as well as between biomass and infiltration rate. The slow system acts between rain events, on a timescale of days to months, and evolves vegetation and soil moisture.  Modeling processes at the appropriate timescales allows parameter values to be set by the actual processes they capture. This reduces the number of parameters that are chosen expressly to fit pattern characteristics, or to artificially slow down fast processes by the orders of magnitude required to align their timescales with the biomass dynamics. We explore the fast-slow switching model through numerical simulation on a one-dimensional hillslope, and find agreement with certain observations about the pattern formation phenomenon, including band spacing and upslope colonization rates.  We also find that the predicted soil moisture dynamics are consistent with time series data that has been collected at a banded vegetation site. This fast-slow model framework introduces a tool for investigating the possible impact of changes to frequency and intensity of rain events in dryland ecosystems.
\vspace{2mm}\\

%
%
%
%


\end{abstract}




\end{frontmatter}

\section{Introduction}
\label{sec:introduction}

Dryland ecosystems are subject to infrequent and largely unpredictable rain inputs~\cite{noy1973desert}.  
These rain events act as pulses to the system that drive dynamics across multiple scales in time and space~\cite{Schwinning2004,collins2014multiscale}. 
When there is insufficient rain to support uniform vegetation coverage, the vegetation may become patchy, and, in some  
water-controlled regions these patches exhibit striking regularity in spatial arrangement~\cite{deblauwe2008global}. 
This is particularly the case for gently sloped terrain, where patterns often appear as bands of dense vegetation growth alternating with bare soil, each aligned transverse to the elevation grade. The vegetation bands can be tens of meters wide with spacing on the order of a hundred meters, and form a regular striped pattern that often occupies tens of square kilometers on the landscape~\cite{deblauwe2012determinants}. 

\begin{figure*}
    \centering
       \setlength{\unitlength}{\linewidth}
    \begin{picture}(1,0.53)
    \put(0.0,0.3){\includegraphics[width=\textwidth]{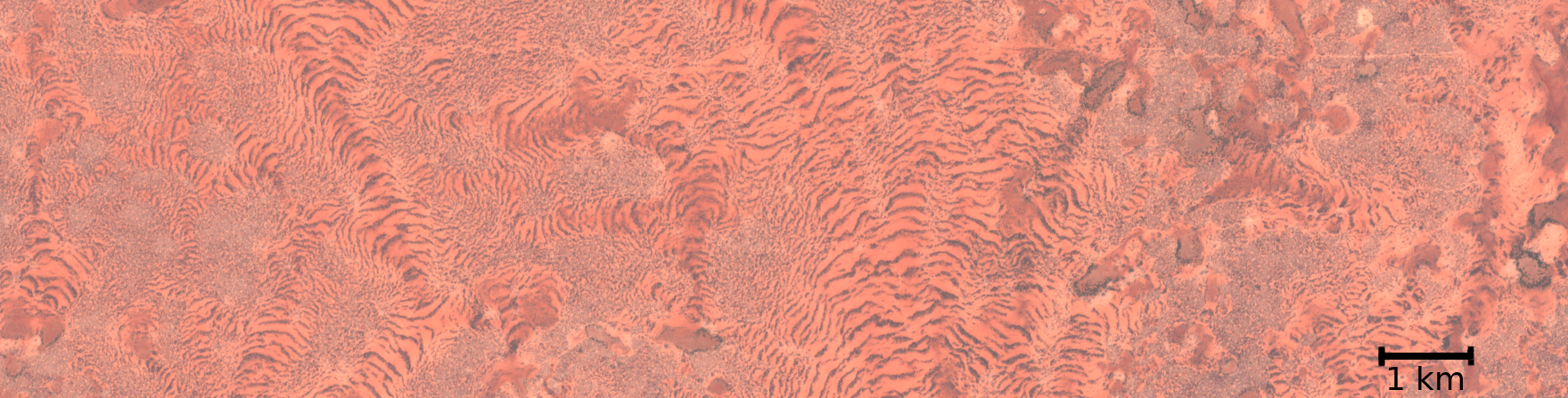}}
    \put(-0.06, -0.0){\includegraphics[width=1.1\textwidth]{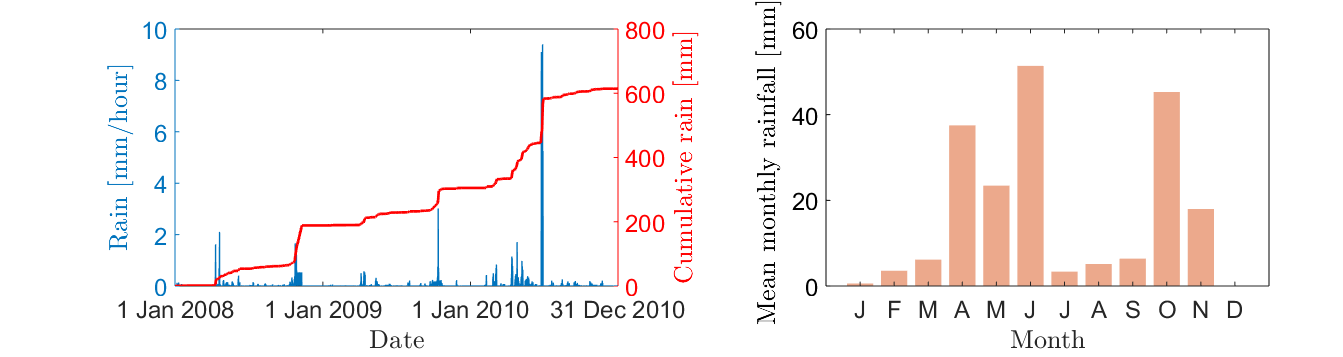}}
    \put(0.01,0.52){(a)}
    \put(0.01,0.26){(b)}
    \end{picture}
        \caption{(a) Satellite image of banded vegetation patterns in the Haud region of Africa (8$^\circ$00'00.0"N, 47$^\circ$30'00.0"E). Dark is vegetation and orange is bare soil. Sentinel-2A satellite image~\cite{DRUSCH201225} was taken in 2016. 
        (b) Hourly rainfall data for the same region, along with monthly averages. Note the two rainy seasons in this region. Rainfall data retrieved from the Modern Era Retrospective--Analysis for Research and Applications (MERRA) ~\cite{gelaro2017modern} and downloaded from {https://disc.gsfc.nasa.gov/daac-bin/FTPSubset2.pl}. 
            \label{fig:rain}
}
\end{figure*}

Figure~\ref{fig:rain}(a) shows a Sentinel-2A satellite image~\cite{DRUSCH201225} of banded vegetation patterns from the Haud region of the Horn of Africa taken in 2016.   A time series of rainfall~\cite{gelaro2017modern} at the location  is shown in \cref{fig:rain}(b). Some of the earliest field studies of banded patterns were carried out within this region, and postulated a connection between the formation of bands and the hydrology of the area~\cite{macfadyen1950vegetation,greenwood1957development}. For example, early observations describe sheet flow in the interband region during rain events, with the rain ``arrested by the next vegetation arc down the slope'' ~\cite{hemming1965vegetation}.  The presence of vegetation promotes water infiltration into the soil both by intercepting runoff and enhancing infiltration as a consequence of the increased biologically-induced soil macroporosities  ~\cite{boaler1964observations,cornet1988dynamics,dunkerley2002infiltration,casenave1992runoff}.  Such feedbacks account for the significantly higher levels of soil moisture observed within the vegetation bands than in the bare soil regions after rain events~\cite{worrall1959butana}. Such interactions between vegetation and hydrological processes appear as nonlinear feedbacks in mathematical modeling studies, where they act as mechanisms underlying the formation of regular vegetation patterns~\cite{meron2015nonlinear,meron2016pattern,meron2018patterns}. Vegetation pattern models often also rely on a difference in timescales associated with fast water transport and slow biomass dispersal as a key ingredient for generating pattern-forming instabilities. This separation of scales has been exploited in analysis of the simplest class of conceptual reaction-advection-diffusion models~\cite{bastiaansen2019stable,bastiaansen2019dynamics,bastiaansen2020effect}.

Hydrology-vegetation feedbacks, where the presence of vegetation increases infiltration and reduces evaporation, have been incorporated at an annually-averaged timescale into contemporary reaction-advection-diffusion  models for vegetation patterns~\cite{gilad2004ecosystem,rietkerk2002self}.  
We build on and extend conceptual models due to Rieterk et al.~\cite{rietkerk2002self} and Gilad et al.~\cite{gilad2004ecosystem}, by developing a fast-slow mathematical framework to capture these hydrological and ecological processes on their appropriate timescales.  We also add an additional feedback to the conceptual modeling framework: the speed of the downhill surface water flow is reduced by surface roughness effects in the vegetated zones~\cite{istanbulluoglu2005vegetation,thompson2011vegetation}. This new fast-slow switching framework provides an alternative both to detailed mechanistic models and to conceptual models that do not resolve the rainstorm timescale. Mechanistic models encode many processes on the timescales on which they occur~\cite{fatichi2012mechanistic,paschalis2016matching} but often require a large number of parameters, making it difficult to identify key underlying mechanisms.  Conceptual models formulated on a long, annually averaged,  timescale~\cite{gilad2004ecosystem,rietkerk2002self,Klausmeier1999} require ``effective'' parameter values to capture faster processes~\cite{ursino2005influence}.  For example, one way to encode water infiltration and biomass evolution on the same timescale is to reduce the rate at which water infiltrates from the surface into the soil by a factor of 1000 or more~\cite{rietkerk2002self}.  This reduction of the infiltration rate then leads surface water to be present year-round at a height that would be observed only during rainstorms.  In the framework we use here, we take an intermediate path.  Fast hydrological processes evolve on the timescale set by a rain event, with slow ecohydrological processes evolving in between rain events with a timescale set by the vegetation dynamics.  This results in a conceptual model, but one that is designed to operate on the timescales of key processes in the system.

By capturing processes  on the timescales on which they occur, we can use infiltration and water transport parameters consistent with the ecohydrology literature. 
The model proposed in this study is able to form patterns and capture certain pattern characteristics, with ecohydrologically consistent parameters, and without additional parameter fitting.  The natural separation of the processes by their timescales admits a significant simplification within each of the two systems, while still capturing the influence of both fast and slow processes on dynamics. 
Biomass-water feedbacks captured by the model, which act on the fast rain-event timescale, include increased infiltration of surface water into the soil in the bands, and a slower overland surface water flow in the vegetated zones.

There have been other notable efforts to extend conceptual modeling frameworks to capture the effects that seasonality~\cite{guttal2007self}, pulse intermittency and intensity~\cite{ursino2006stability,kletter2009patterned,baudena2013vegetation}, and stochasticity~\cite{Dodorico2006,siteur2014will} of rainfall have on vegetation patterns. However, many of these studies fail to appropriately adjust effective parameters when additional timescales are captured within the model. 
A few of the studies are similar in philosophy to the fast-slow framework we propose in that they handle the fast hydrological processes associated with rain events separately from the rest of the slower processes.  The work of Siteur et al.~\cite{siteur2014will} employs simplified infiltration and soil moisture dynamics, compared to those in the model we are presenting.  They treat rain events as delta-function pulses of surface water, instead of resolving surface water dynamics during rain events.  The benefit of their extreme simplification is the ability to  make predictions about annually averaged quantities for the case that rain events are stochastic in time and intensity.  Other work~\cite{kletter2009patterned,baudena2013vegetation} implements a continuous-time model, based on~\cite{gilad2004ecosystem}, using a numerical scheme that also simplifies the computation of the fast hydrological processes during rain events.  In particular, ~\cite{kletter2009patterned,baudena2013vegetation} use a steady-state relationship between surface water height and biomass while it is raining, instead of resolving the time dynamics of the surface water.

The paper is organized as follows.  In \cref{sec:soilmoisture} we first discuss some existing  model predictions and ground-based observations regarding the distribution of soil moisture relative to the vegetated and bare regions.  There are significant qualitative differences in predictions between various classes of models, but available observations suggest that the soil moisture is dramatically increased within vegetation zones relative to the bare soil during rain events and also depleted at a faster rate via transpiration during dry periods.  We then present a reaction-advection-diffusion model for banded vegetation patterns that couples processes operating on both fast and slow timescales in \cref{sec:seasonal}.  We use this coupled model as motivation for the fast-slow switching model presented in \cref{sec:fastslow}. The precipitation in the coupled model is assumed to vary in time, parameters are chosen to be consistent with the hydrological literature when available, and we explore predictions about pattern formation within the model based on linear stability analysis of spatially-uniform states. The time-dependence of precipitation has a substantial impact on pattern characteristics such as precipitation level at onset, band spacing and upslope colonization rate. 

We use the structure of the coupled model, along with chosen parameters, to identify a timescale separation in the system: the ratio of the timescale for surface water to infiltrate into the soil to the timescale for biomass growth is small.  This small ratio suggests a fast-slow switching model, which we describe in \cref{sec:fastslow}, and explore through numerical simulation in \cref{sec:sim}.  The simulation results of the fast-slow model are consistent with linear predictions from the coupled model from \cref{sec:seasonal} with the same temporal rain input profile, and qualitatively match the observed soil moisture dynamics from the literature that is presented in \cref{sec:soilmoisture}.   Finally, in \cref{sec:discussion}, we discuss key predictions of the fast-slow model in the context of other existing modeling frameworks, highlight advantages of the fast-slow model, and suggest potential directions for future research.  Additional simulation results for the fast-slow switching model are presented in \ref{app:sim} of the online supplement and the linear stability calculations of the coupled model, which are based on Floquet theory, are detailed in \ref{app:linstab} of the online supplement.

\section{Soil moisture: model predictions and data from literature }
\label{sec:soilmoisture}

Surface water dynamics occur on a minute to hour timescale, while vegetation growth and death occurs on a week to month timescale.  Because soil moisture is replenished by fast infiltration of surface water and depleted by slow evapotranspiration, its dynamics occur on multiple timescales.   Models of vegetation patterns vary widely in both how they capture these multiscale dynamics, and in their predictions about the soil moisture distribution in the vegetation bands relative to the bare soil regions.    In this section, we discuss differences in predictions among different classes of models ranging from  conceptual, annually averaged ones to very detailed mechanistic ones. We start the discussion by presenting ground measurements of soil moisture at a banded vegetation site in Mexico~\cite{cornet1988dynamics}. 

\begin{figure}
    \centering
    \includegraphics[width=\linewidth]{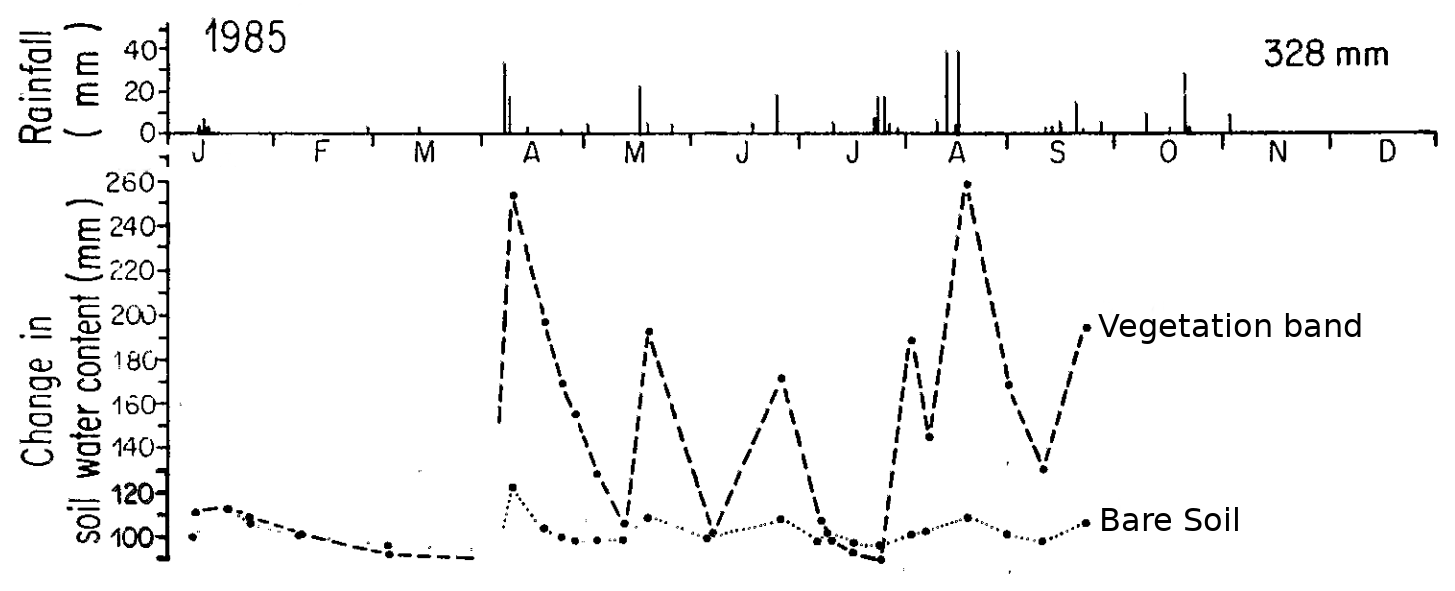}
    \caption{Adapted from Cornet et al. (1988)~\cite{cornet1988dynamics}.  (Top) Rainfall data in 1985 at a site in the Chihuahuan Desert of Mexico exhibiting banded vegetation patterns.  (Bottom) Dashed line indicate measurements of soil water content in the vegetation band and dotted line indicates the  inbetween  bare soil region.}
    \label{fig:cornett}
\end{figure}

During rainfall events biomass feedback on infiltration acts  to increase soil moisture where plants are by allowing surface water to enter the soil at a higher rate relative to the bare soil regions~\cite{worrall1959butana,boaler1964observations}. However, soil moisture may be lost faster where biomass is present because of transpiration.  
While time-resolved soil moisture data at sites that exhibit banded vegetation is limited, the work of Cornet et al.~\cite{cornet1988dynamics} provides a time series of soil water content for a vegetation band and the bare soil region just uphill at a site in the Chihauahuan Desert for 1985.   In data from~\cite{cornet1988dynamics}, after rain events, the soil water content within the band increases dramatically relative to that of the bare soil, as shown in ~\cref{fig:cornett}.  It is only after extended periods without rain that the soil water content within the vegetation band becomes comparable, or even slightly below, the soil moisture of the bare ground.  Even with these fluctuations, it appears from \cref{fig:cornett} that the soil moisture is, on average, more concentrated in the vegetation bands than in the bare soil regions. Qualitatively similar features occur for a site with banded vegetation patterns in Niger~\cite{seghieri1997relationships}.   Data on gapped patterns collected by Barbier et al.~\cite{barbier2008spatial} also indicates increased soil moisture levels in vegetated regions, although the shorter-timescale dynamics are qualitatively different. 
More comprehensive field studies would determine if similar observations occur at additional sites beyond these observational studies, and over longer time-periods. Many predictions from models are inconsistent with these observations of soil moisture distribution.

Conceptual reaction-advection-diffusion models can make qualitative predictions about the system over century timescales or longer. Such long-time simulation results ensure that the asymptotic behavior of the model is captured, and not just transient behavior that may be sensitive to the choice of initial conditions on biomass distribution.  While these conceptual models typically encode a separation of scales into the advection and diffusion terms to capture slow biomass dispersal and fast water transport, the reaction terms associated with local ecohydrological processes are typically formulated on the slow biomass timescale. Because these models do not resolve individual rain events, the resulting predictions are interpreted as those of the annually averaged system, and therefore miss important details about soil moisture on shorter timescales.

The simplest of these conceptual models lump processes associated with surface and subsurface hydrology  
into a single ``water" field, leading to a pair of reaction-advection-diffusion equations that captures interactions between biomass and water. Such ``two-field'' models, exemplified by Klausmeier~\cite{Klausmeier1999}, predict the highest concentration of water 
in the regions of bare soil between the vegetation bands.        
So-called ``three-field" reaction-advection-diffusion models separately track the dynamics of surface and soil water, in addition to the biomass field.  These models predict the soil moisture, on average, will peak within the biomass for typical parameter choices~\cite{gilad2004ecosystem,rietkerk2002self}.  
However, it has also been shown, in the context of the three-field model by Gilad et al.~\cite{gilad2004ecosystem}, that the soil moisture spatial profile can be either in or out of phase with the spatial profile of biomass~\cite{kinast2014interplay}, depending on which mechanism for pattern formation dominates. 
While such models are capable of qualitatively capturing annually averaged dynamics observed in Fig.~\ref{fig:cornett}, they fail to provide useful predictions on shorter timescales.

Detailed mechanistic ecohydrological models for general water-limited environments~\cite{fatichi2012mechanistic}, which capture  processes on the fast time scales of individual rain events, have also been used to study the banded vegetation patterns.
Such models predict 
that soil moisture can switch between being more highly concentrated under the vegetation bands to being more highly concentrated in the bare soil region for extended periods of time (a year or more).  In these models, decadal time-averaged soil moisture is more highly concentrated in the bare soil region~\cite{paschalis2016matching}.  The annually averaged predictions of these models, like those of the Klausmeier model, are in contrast with 
the measurements shown in Fig.~\ref{fig:cornett}.  Moreover, the complexity of such models limits numerical simulations to timescales of a few decades or less, and prohibits detailed exploration of the extensive parameter space.

Any prediction that soil moisture is, on annual average, more concentrated in the bare soil region, is 
inconsistent with the limited time-series data that is available for vegetation patterns 
~\cite{cornet1988dynamics,seghieri1997relationships,barbier2008spatial}.   
Because of competition between increased infiltration and increased transpiration at vegetation bands relative to the bare soil regions, modeling choices about biomass growth based on plant physiology may play an important role in predictions about soil moisture distribution~\cite{ursino2007modeling}.  While we focus on the modeling of hydrological processes at appropriate timescales in this work, we expect that a critical analysis of model details of biomass dynamics, especially combined with more field measurements, could provide additional insights.

\section{A three-field coupled timescale model}
\label{sec:seasonal}

Before introducing the fast-slow switching model, the main focus of this study, we first present a reaction-advection-diffusion model in \cref{sec:model:3field} with a seasonally-varying precipitation input that captures processes across the relevant fast and slow timescales. This `three-field coupled timescale model'   provides motivation for the switching model, and we  discuss how parameters are chosen in \cref{sec:model:par}. In \cref{sec:linearstab}, we explore the linear stability of the spatially uniform states within the seasonal model.  Using a dimensionless version of the coupled model (\cref{sec:model:nondim}) we demonstrate that the separation of timescales leads to a small parameter.   The fast and slow systems that comprise the switching model are introduced in \cref{sec:fastslow} by considering the limit in which this small parameter approaches zero.  While the coupled model motivates the switching model,  we  treat  the switching model as distinct from the coupled model instead of considering it as a means to approximate solutions of the coupled model. 

\subsection{Motivation: A three-field coupled timescale model }
\label{sec:model:3field}
We first consider a model in which we input a time-dependent precipitation $P(T)$ uniformly on a one-dimensional spatial domain. The model evolves a biomass field $B(X,T)$ ($kg/m^2$), a soil moisture field $s(X,T)\in[0,1]$, and a surface water height $H(X,T)$ ($cm$).  In restricting the description to these three dynamical variables, the model has its roots in the  three-field conceptual reaction-advection-diffusion models of Rietkerk et al.~\cite{rietkerk2002self} and Gilad et al.~\cite{gilad2004ecosystem}.
However, those models typically  evolve the system on the timescale of biomass and input the rain at a constant rate given by the mean annual value.
 Although the basic structure of those models is a starting point for ours, we make modifications to capture details of the hydrological processes on the timescales at which they occur.  

We will refer to our version of the three-field model as a ``three-field coupled timescale model" or simply ``coupled model" in this paper.
It takes the  following  form:
\begin{subequations}
\label{eq:3field}
 \begin{align}
 \frac{\partial H}{\partial T} &= P(T) - \II(H,s,B)   + K_V\frac{\partial}{\partial X} \left( \frac{ \sqrt{{\cal \zeta}}\ H^{\delta}}{1+N B} \right)   
 \label{eq:3field:H}\\
\phi Z_r \frac{\partial s}{\partial T}  &= \II(H,s,B) - L s   - \Gamma Bs  
\label{eq:3field:s}\\ 
\frac{\partial B}{\partial T}  &=   C  \Bigl(1-\frac{B}{K_B}\Bigr)\Gamma Bs- M B+D_B \frac{\partial^2 B}{ \partial X^2}  
\label{eq:3field:B}
 \end{align}
\end{subequations}
where the infiltration rate of water from the surface into the soil is given by
\begin{equation}\label{eq:3field:infl}
\II(H,s,B)=K_{I}\Bigl(\frac{B+fQ}{B+Q}\Bigr)\Bigl(\frac{H}{H+A}\Bigr)(1-s)^{\beta}.
\end{equation}
This empirical infiltration model captures feedback from biomass, surface water and soil saturation in a highly simplified way as compared to more physical-based approaches that rely on modeling the vertical distribution of soil moisture~\cite{vereecken2019infiltration}. Such simplified approaches have proven successful in other contexts~\cite{rigby2006simplified}.

 In the next  section, we describe each of the terms in the coupled model~\eqref{eq:3field} to highlight differences from previous models and to justify our parameter choices which are summarized in \cref{tab:dim}.  
A summary of some of the most significant differences between the coupled model~\eqref{eq:3field} and, for example, the Gilad et al. model~\cite{gilad2004ecosystem,gilad2007mathematical} are:  
  Precipitation is time-dependent,
 there is no soil diffusion in the coupled model,
 there is no root augmentation feedback in the coupled model,
 there is biomass feedback on surface water transport in the coupled model and,
 in the coupled model, the infiltration slows as soil moisture saturates and becomes independent of surface water height for large enough surface water height.

\subsection{Choosing Parameters}
\label{sec:model:par}

\begin{table}[tbp]
	\renewcommand{\arraystretch}{1.55}
	\centering
	\begin{tabular}{|c|c|c|l|} 
		\hline
		parameter & units& default value(s) & description \\
		\hline
		\hline
		$K_I$ & $ cm/day$ & 500 & infiltration rate coefficient  \\
		\hline
		$f$ & -- & 0.1 & bare/vegetated infiltration contrast\\
		\hline
		$Q$ & $kg/m^2$ & 0.1 & Biomass level for infiltration enhancement \\
		\hline
		$A$& $cm$& 1 &infiltration rate $H$-independent for $H\gg A$\\
		\hline
		$\beta$ & -- &  4 &infiltration exponent (soil moisture)\\
		\hline
		$\zeta$ & -- & 0.005 & elevation grade\\
		\hline
		$\delta$ & -- & $5/3$ (or 1)$^*$ & surface water transport exponent \\
		\hline	
		$K_V$&$m/day/cm^{2/3}$ (or $m/day$)$^*$& $2\times 10^5$ & surface water transport coefficient\\
		\hline
		$N$ & $m^2/kg$& 20 & surface roughness coefficient\\
		\hline
		$D_B$& $m^2/day$& $0.01$ &  biomass diffusion\\
		\hline
		$\phi Z_r$&  $cm$&  $27^{**}$ & soil water capacity\\
		\hline
		$L$ & $cm/day$ & 0.2& evaporation rate\\
		\hline
		$\Gamma$ & $(cm/day)/(kg/m^2)$ & 0.67 & transpiration coefficient\\
		\hline
		$K_B$ & $kg/m^2$ & 4 & biomass carrying capacity\\
		\hline
		$C$ & $(kg/m^2)/cm$ &0.1 & water use efficiency coefficient\\
		\hline
		$M$& $day^{-1}$& 0.01 & biomass mortality rate\\
		\hline
	\end{tabular}
	\caption{\label{tab:dim} Summary of the coupled model~\eqref{eq:3field} default parameter values, as described in Section~\ref{sec:model:par}. $^*$ in numerical simulations, we often use $\delta=1$, which changes the units of $K_V$, although not the default numerical value we use for it. $^{**}$ This value is based on typical soil porosity $\phi=0.45$ and a root depth  $Z_r=60\ cm$. 
	}
\end{table}%

\paragraph{Surface Water Transport}  We take the Manning formula for gravity-driven open channel flow~\cite{chow1959open,bonetti2017manning} as a basis for our surface water transport model. According to this formula, the average velocity of the fluid in a wide channel is empirically determined to be $\mathcal{V}\propto n^{-1}\zeta^{1/2} H^{2/3}$, where $n$ characterizes the surface roughness, $\zeta$ is the local elevation grade and $H$ is the height of the water. By modifying the surface roughness term based on literature values for vegetated surfaces and bare soil, we include the influence of biomass on runoff interception leading to a slow-down in the surface water speed within vegetated regions.
In particular, we take the water flow speed to be given by 
\begin{equation*}
\mathcal{V}(H,B)=K_V \frac{\sqrt{\zeta}\  H^{\delta-1}}{(1+NB)}
\end{equation*}
where the term $1+NB$ allows us to describe the change in surface roughness from bare to vegetated soil. 
The value of the exponent that is consistent with the Manning formula corresponds to $\delta=5/3$. With $\delta=5/3$ and a typical value for the elevation grade of $\zeta=0.005$ (i.e. an $0.5\%$ grade) we estimate the remaining parameters using  Manning roughness coefficients $n$~\cite{chow1959open,bonetti2017manning}. 
Typical values of $n$ for bare soil ($n_g=0.02$~$s/m^{1/3}$) and dense vegetation ($n_v=0.1$~$s/m^{1/3}$) lead to $K_V=2\times 10^5$~$(m/day)/(cm^{2/3})$ and $N=20$~$m^2/kg$.  Estimations using a typical surface water height around $H=1$~$cm$ result in the same values of $K_V=2\times 10^5$~$m/day$  (note different units)
 and $N=20$~$m^2/kg$ when using $\delta=1$.  These values and $1$~$cm$ surface water when $\delta=5/3$ lead to a water velocity of around $\mathcal{V}\approx 16$~$cm/s$ on bare soil and just over a factor of 10 slow down through vegetation with biomass value $B=0.5$~$kg/m^2$.  
While the value of the exponent that is consistent with the Manning formula is $\delta=5/3$, we take $\delta=1$ for most of the results in \cref{sec:sim} for computational convenience. This makes the water flow speed independent of $H$ and the advective transport term in~\cref{eq:3field:H} linear in the surface water height $H$. (We present some numerical calculations and simulations with $\delta=5/3$ in \cref{fig:uv:floquet}(a), \cref{fig:nonlinh:kn5},
and \cref{fig:nonlinh:spatialslice}; our limited exploration shows results that are similar when  $\delta=5/3$ is replaced by $\delta=1$.)   

\paragraph{Soil Moisture Dynamics} 
 
 We use a ``bucket model" that tracks a depth-averaged soil moisture within the plant rooting zone according to the soil moisture balance equation ~\eqref{eq:3field:s}, where $s$ is the relative soil moisture ~\cite{rodriguez2007ecohydrology}. Denoting the rooting depth by $Z_r$ and the soil porosity by $\phi$, the quantity $s \phi Z_r$ represents the volume of water contained in the root zone per unit ground area, and has units of column height. In the model here, we assumed typical values of $Z_r=60$~$cm$ and $\phi=0.45$ ~\cite{rodriguez2007ecohydrology}.
Water enters the soil from the surface via the infiltration model~\eqref{eq:3field:infl} and is lost through evapotranspiration as $(L+\Gamma B)s$.  The linear evaporation rate coefficient $L=0.2$~$cm/day$ is set so that the evaporation rate is approximately $0.1$~$cm/day$ when soil moisture is at $s=0.5$, which is approximately the field capacity~\cite{rodriguez2007ecohydrology}.  We neglect the shading effects on evaporation by making the evaporation rate independent of biomass $B$. We set the transpiration parameter to be $\Gamma=0.67$~$(cm/day)/(kg/m^2)$ in order that the transpiration rate with biomass density $B=1$~$kg/m^2$ reaches approximately $0.4$~$cm/day$ when soil moisture is $s=0.5$.

We take a heuristic approach to modeling infiltration, similar to those of previous conceptual models~\cite{hillerislambers2001vegetation,gilad2004ecosystem}.  The parameter $K_I$ in \cref{eq:3field:infl} represents an effective infiltration rate, which was here assumed equal to $K_I=500$~$cm/day$ in order to ensure typical values of the infiltration rates,  e.g.$\sim 100\, cm/day$, in the numerical simulation. We note, however, that the value of $K_I$ depends on the soil type. Similar to previous models, we assume  a factor $f=0.1$ reduction in infiltration when biomass levels fall far below the threshold value of $Q=0.1$~$kg/m^2$.   
This biomass feedback models the presence of a soil crust that reduces infiltration in bare soil and root systems that enhance infiltration in vegetated areas. We additionally assume that the infiltration rate becomes independent of surface water height for $H$ significantly greater than $A$, where $A=1$~$cm$ (many descriptions of infiltration have this $H$-independence, see e.g.~\cite{thompson2011vegetation}).  . 
Finally we include a factor of $(1-s)^\beta$ with $\beta=4$ to account for the  reduced infiltration values as the moisture content reaches saturation.  While soil saturation effects are typically negligible for the case of constant precipitation, we are interested in considering large, concentrated rain events where the soil moisture can increase dramatically in a very short time.

We neglect subsurface water flow, under the assumption that surface water flow will be the dominant mode of water transport. 
In other models, subsurface transport has been included as linear diffusion but we do not notice significant differences in the limited simulations where we have included it (see \ref{app:sim} of the online supplement for more details).  
We also neglect leakage of water from the root zone to deeper areas for similar reasons: it did not have a large impact on the limited simulation results where we have included it.  However, a more comprehensive exploration of the role of these two subsurface soil processes, and how they are modeled, could provide a more detailed understanding of when they can and cannot be neglected.

\paragraph{Biomass Dynamics}  We leave the biomass dynamics largely unchanged in form from a simplified version of the model by Gilad et al.~\cite{gilad2007mathematical}.   The plant growth rate is taken to be proportional to transpiration rate $\Gamma Bs$, except with a logistic term that can be thought of as capturing a decrease in water use efficiency as biomass increases. Specifically, the growth term is $C(1-B/K_B)\Gamma B s$, where  the proportionality constant $C=0.1$~$(kg/m^2)/cm$ is set by the typical water-use efficiency for plants. Increased biomass begins to significantly decrease growth rates when $B\sim K_B$, where we take $K_B=4$~$kg/m^2$ (however, biomass levels stay well below $K_B$ in all simulations presented here).   
One notable difference from the biomass growth term of the Gilad et al.~\cite{gilad2007mathematical} model is that we do not include a root augmentation feedback associated with the lateral spreading of roots.  Mortality is modeled by linear loss with coefficient $M=0.01$~$day^{-1}$, which was  estimated by Mauchamp et al.~\cite{mauchamp1994simulating} using typical carbon maintenance costs. 
  
We model seed dispersal by linear biomass diffusion with rate $D_B=0.01$~$m^2/day$.  This parameter is often chosen in models of this type without clear justification; $D_B$ ranges from  $\sim 10^{-6}$~$m^2/day$ in Gilad et al.~\cite{gilad2007mathematical} to $\sim 10^{-3}$~$m^2/day$ in Klausmeier~\cite{Klausmeier1999} to $0.1$~$m^2/day$ in Rietkerk et al.~\cite{rietkerk2002self}. (It may ultimately have been chosen to help the model match pattern characteristics; see discussion in Gandhi et al.~\cite{gandhi2019vegetation}.)
Our linear stability results, presented in \cref{fig:uv:floquet}, show that this parameter, for our model, controls a cut off for short-wavelength linear instabilities of the uniform state within the seasonal model. While this parameter is unconstrained and diffusion is likely a poor model of seed dispersal, we make no adjustments to $D_B$ in our simulations, leaving it fixed at a value of $0.01$~$m^2/day$.
We note that this model does not attempt to capture drought resistant behavior that plants in dryland ecosystems are known to exhibit.  Nor does it include transport of organic material or seeds via water during rain storms.  

\subsection{Spatially uniform states and their stability}
\label{sec:linearstab}

If we were to set the precipitation rate to a constant, $P(T)=P_0$, then we could explicitly determine two possible types of spatially uniform steady state solutions of~\eqref{eq:3field}.  
One is the bare soil state, defined by $B_{bs}=0$, and with 
\begin{equation}\label{eq:ss:bs}
     s_{bs}=\frac{P_0}{L},\; \mathcal{I}_{bs}=K_I f (1-s_{bs})^{\beta}, \; H_{bs}=A\frac{P_0 }{\mathcal{I}_{bs}-P_0},
\end{equation}
which exists for all $P_0$. This trivial solution becomes unstable at a transcritical bifurcation point $P_c\equiv M L/ C \Gamma$, and for $P_0>P_c$,  there also exists a uniform vegetation solution with nonzero biomass 
\begin{equation}\label{eq:ss:uv}
    B_{uv}=K_B\frac{C P_0 - M L/\Gamma}{C P_0+M K_B},\; s_{uv}=\frac{P_0}{L-\Gamma B_{uv}}, 
\end{equation}
\[H_{uv}=A\frac{P_0}{\mathcal{I}_{uv}-P_0},\; \mathcal{I}_{uv}=K_I \frac{B_{uv}+fQ}{B_{uv} + Q}(1-s_{uv})^{\beta}.\]

We find that these two uniform solutions, and the associated transcritical bifurcation at $P_0=P_c$, carry over to the case that the precipitation comes into the system during $n_r$ equally-spaced rain events per year. We consider rain events with constant precipitation rate ${\cal P}$ ($cm/day$), each of duration $T_r$. Specifically, if we denote the Heaviside unit step function by ${\cal H}$, and we let
\begin{equation}
\label{eq:pulsedrain}
    P(T)={\cal P}{\cal H}(T_r-T), \quad T\in[0,365/n_r],
\end{equation}
then a periodic seasonal rainfall is obtained by repeating this pattern, i.e. by letting $P(T+365/n_r)=P(T)$. The total annual precipitation is $n_r T_r{\cal P}$, and it plays the role of a mean annual precipitation rate $MAP$ in our investigations. 
For the default parameters given in \cref{tab:dim}, the transcritical bifurcation $P_c$ occurs when the total annual precipitation is $\sim 110\ mm/yr$, and we find that this value does not change when we use the pulsed, periodic rain input~\eqref{eq:pulsedrain} rather than constant precipitation $P_0$.

The stability of the spatially-uniform vegetation state to heterogeneous perturbations is determined by computing  Floquet multipliers associated with the 
linear variational equations, given perturbations
proportional to
$e^{ikX}$ (See, e.g., the book by Meiss~\cite{meiss2017}). Specifically, we let 
\begin{eqnarray*}
H_k(T)&=&H_0(T)+H_{1,k}(T)e^{ikX},\\
s_k(T)&=&s_0(T)+s_{1,k}(T)e^{ikX},\\
B_k(T)&=&B_0(T)+B_{1,k}(T)e^{ikX},
\end{eqnarray*}
where $(H_0(T),s_0(T),B_0(T))$ is a spatially-uniform, temporally periodic solution of~\eqref{eq:3field} for the periodic rain input~\eqref{eq:pulsedrain}.  In practice, we compute $(H_0(T),s_0(T),B_0(T))$ by integrating the ODEs associated with spatially uniform states of \eqref{eq:3field} forward in time until the system converges sufficiently close to this stable periodic orbit.   For the linear stability calculation, we linearize in the perturbations $(H_{1,k},s_{1,k},B_{1,k})$, and  (numerically) compute the Floquet multipliers as eigenvalues of the associated Monodromy matrix ${\cal M}$.
This calculation is carried out on a two-dimensional grid of values for mean annual precipitation ($MAP$) associated with~\cref{eq:3field} and the wavenumber $k$ of the perturbation.
 The stability boundary in the $(MAP,k)$ parameter plane is then determined by interpolating where the leading Floquet multiplier crosses the unit circle; we refer to the instability region in the $(MAP,k)$-plane as the ``Turing-Hopf bubble". (See~\ref{app:linstab} of the online supplement for more details of the linear stability computations.)

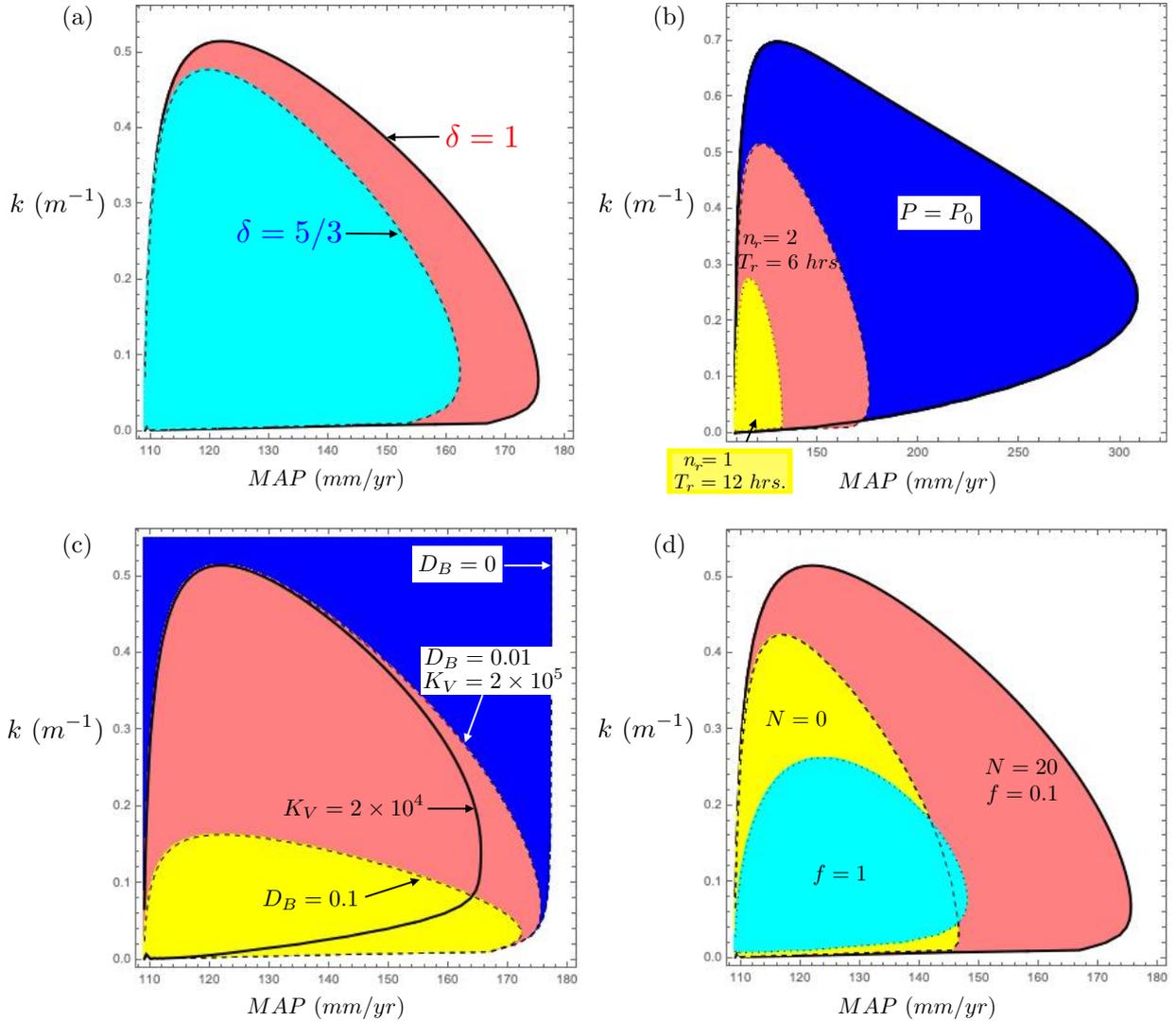
\begin{figure*}
    \centering
    \setlength{\unitlength}{\textwidth}
    \begin{picture}(1,0.9)
    \put(0,0.45){\includegraphics[width=0.49\textwidth]{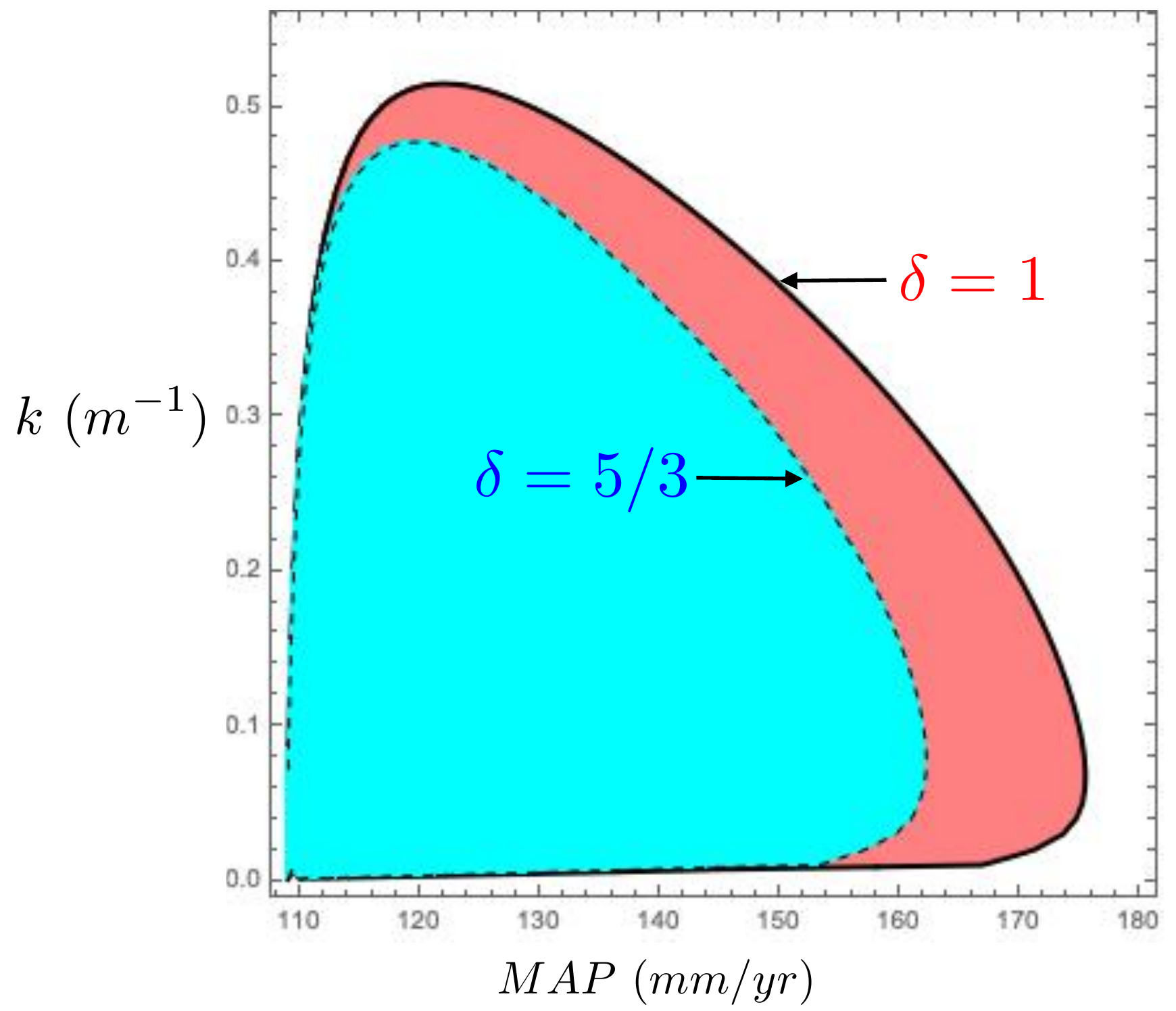}}    
     \put(0.5,.45){\includegraphics[width=0.49\textwidth]{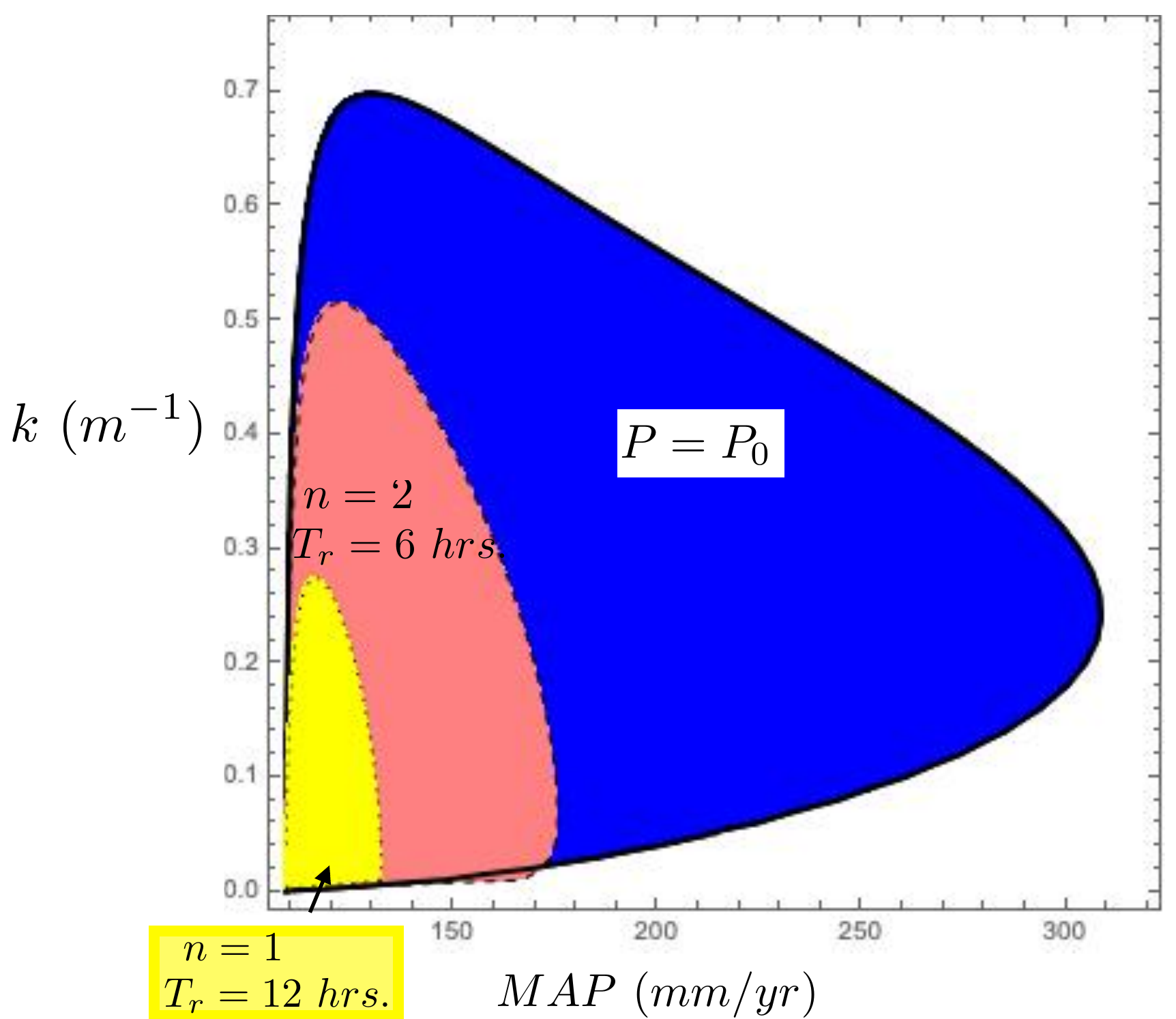}}
     \put(0,0){\includegraphics[width=0.49\textwidth]{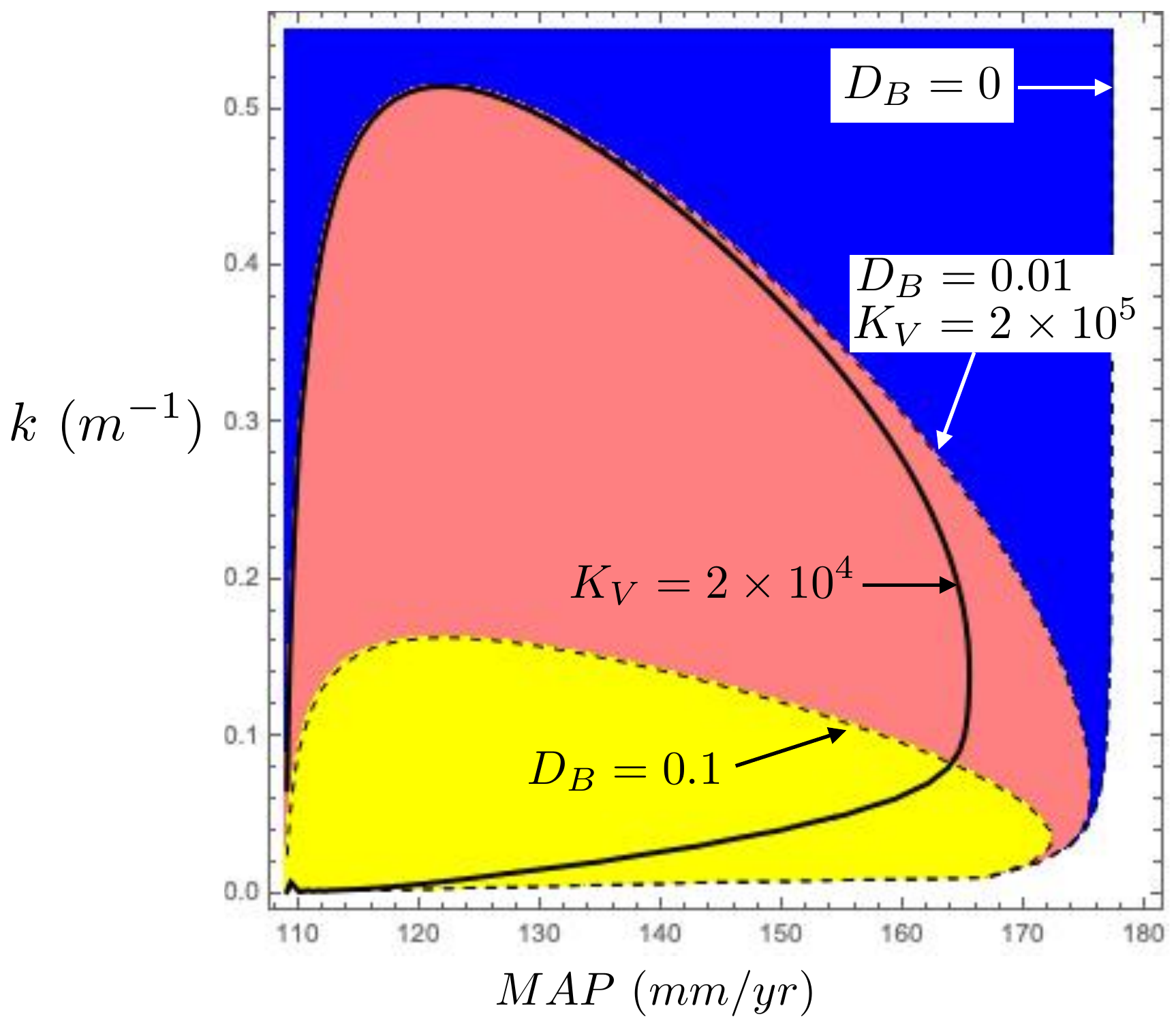}}
     \put(.5,0){\includegraphics[width=0.49\textwidth]{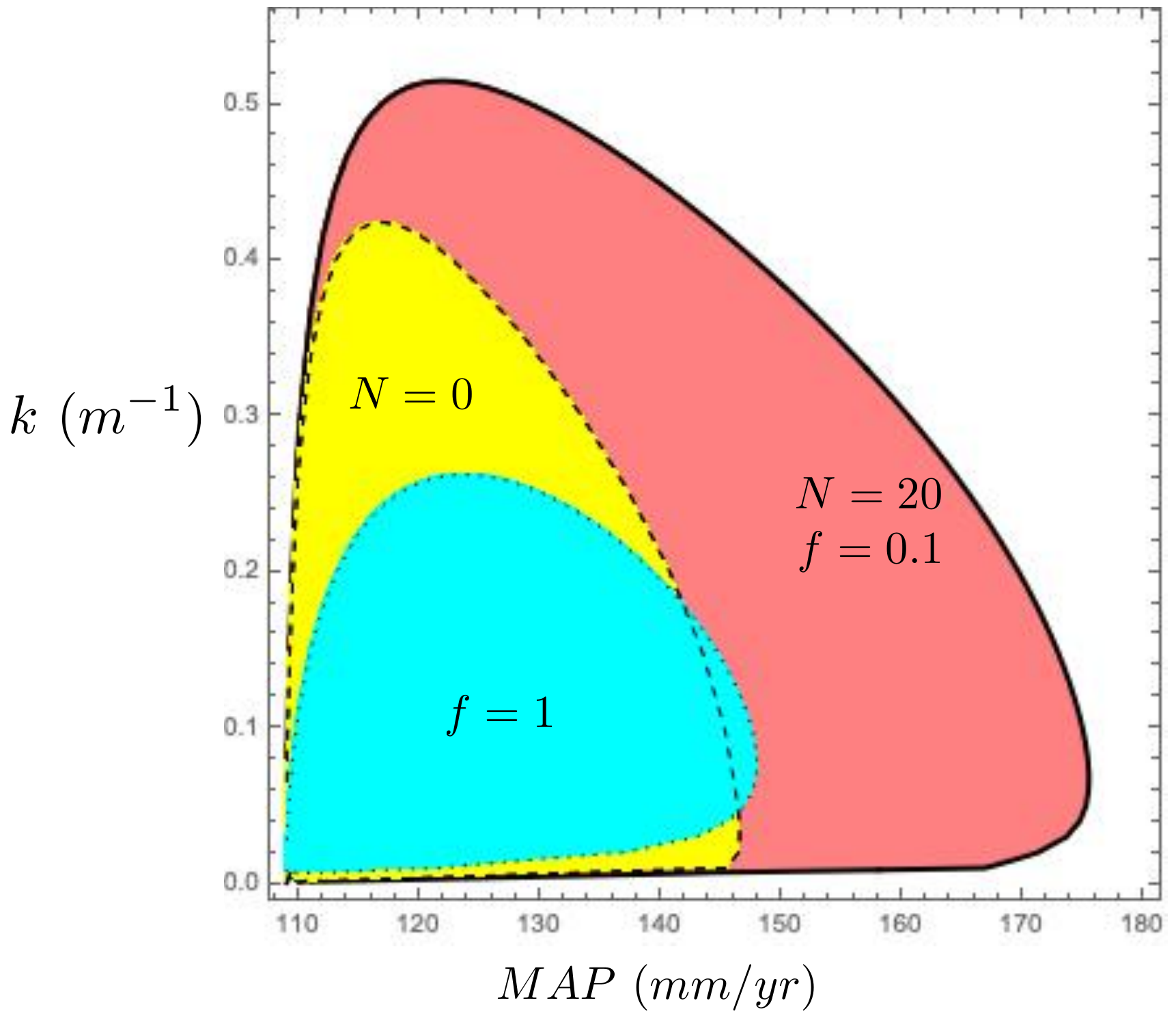}}
     
    \put(0.05,0.85){(a)}
    \put(0.55,0.85){(b)}
    \put(0.05,0.4){(c)}
    \put(0.55,0.4){(d)}
    
    \put(0.585,0.47){\tiny $r$}
    \put(0.635,0.66){\tiny $r$}
    
    \end{picture}
    
    \caption{   \label{fig:uv:floquet}
(a) Turing-Hopf bubble  in the ($MAP,k)$-parameter plane for default parameters of \cref{tab:dim} in the case that precipitation consists of a six-hour long rain event every six months. Computations were done with transport exponents $\delta=1$ (larger red bubble) and $\delta=5/3$ (smaller cyan bubble). (b) Impact on Turing-Hopf bubble under changes to the precipitation \eqref{eq:pulsedrain} for default parameters ($\delta=1$ case). The large blue bubble has constant precipitation $P_0$, the red has two equally-spaced rain events of 6 hour duration, and the small yellow bubble is associated with one rainy season of $12$ hour duration. Note that the scale has changed for this panel; the red bubble is the same as in figures (a), (c) and (d). (c) Impact of varying the transport parameters $D_B$, associated with biomass diffusion, and $K_V$, associated with overland surface water flow in the case that $\delta=1$ in \cref{eq:3field}. The bubble associated with default parameters of \cref{tab:dim}  is in red, and unless otherwise  indicated $D_B=0.01\ m^2/day$ and $K_V=2\times 10^{5}\ m/day$.  The short-wavelength (high $k$) cut-off vanishes if there is no biomass diffusion (blue region). The solid black line indicates the impact on the Turing-Hopf bubble of decreasing $K_V$ by an order of magnitude; this suppresses the longest-wave (small $k$) instabilities. (d) Comparing the Turing-Hopf bubble for default parameters in red ($\delta=1$) with those obtained when infiltration feedback is removed (cyan, $f=1$), or biomass feedback on transport is removed (yellow, $N=0\ m^2/kg$).
} 
\end{figure*}

Results of some of our numerical computations of the Turing-Hopf bubble are presented in \cref{fig:uv:floquet}. In each of figures (a)-(d) the red Turing-Hopf bubble is the reference one. It was obtained using the default parameters of \cref{tab:dim} with overland water transport exponent $\delta=1$; the precipitation function \cref{eq:pulsedrain} had $n_r=2$ and $T_r=0.25\ days$. This means that there were two big rain events
per year, separated by six months (two rainy seasons), and each lasted for six hours.  Here we summarize the findings from our linear stability calculations. 

\begin{itemize}
    \item 
\Cref{fig:uv:floquet}(a) shows that increasing the exponent from $\delta=1$ to $\delta=5/3$ leads to a slight decrease in the size of the Turing-Hopf bubble. Most of our simulation-based investigations incorporate $\delta=1$ as taking $\delta=5/3$ significantly increases computation time.  

\item
\Cref{fig:uv:floquet}(b) shows the impact on the instability region of changing the duration or frequency of the rain events. With constant rain input $P_0$ the instability region is significantly greater as shown by the large blue Turing-Hopf bubble that extends all the way to $MAP\approx 310\ mm/yr$. Alternatively, if we keep the intensity of rain the same but decrease the frequency from two rain events per year to just one rain storm, the instability region shrinks quite significantly to the small yellow Turing-Hopf bubble. This suggests that the MAP level at onset of instability of the uniform state decreases as rain events become less frequent. In fact, if we decrease the frequency of large rain events to one every 2 years, then there is no pattern-forming instability of the uniform state for these parameter values. 

\item
\Cref{fig:uv:floquet}(c) shows how the  transport parameters $K_V$ and $D_B$ in \cref{eq:3field} impact the shape of the Turing-Hopf bubble. While there is an instability of the uniform state even when we neglect biomass diffusion ($D_B=0\ m^2/day$), there is no short-wavelength (high $k$) cut-off as indicated by the asymptotically vertical boundary to the blue region at $MAP\approx 178\ mm/yr$. As $D_B$ increases to $D_B=0.1\ m^2/day$ we find that the short-wavelength modes are stabilized. The long-wavelength boundary is apparently set by the overland transport parameter $K_V$, as suggested by the change of shape of the Turing-Hopf bubble when $K_V$ is decreased by an order of magnitude to $2\times 10^4\ m/day$.

\item
\Cref{fig:uv:floquet}(d) shows how the Turing-Hopf bubble changes when one of the two biomass-hydrology feedbacks is shut off. The $N=0\ m^2/kg$ ($f=0.1$) Turing-Hopf bubble is quite similar in shape, although smaller, to that for $N=20\ m^2/kg$ ($f=0.1$). By way of contrast, the shape of the Turing-Hopf bubble with {\it only} the transport feedback (i.e. when $f=1$) is more circular, with an expected wavelength of pattern that is smaller than that when the infiltration feedback is turned on.

\end{itemize}
The Turing-Hopf bubbles in \cref{fig:uv:floquet} indicate regions of linear instability of the uniform vegetation state.  They are \textit{not} equivalent to the regions of stability of nonlinear patterns which are typically referred to as ``Busse balloons"~\cite{busse1978non}.  Even still, the numerical simulation results presented in \cref{sec:sim} suggest that these linear calculations provide some insight into characteristics of fully nonlinear states.

\subsection{Non-dimensionalization: origin of the small parameter $\epsilon$}
\label{sec:model:nondim}

We now introduce a rescaling of the three-field coupled timescale model~\cref{eq:3field} to put it into dimensionless form.  The scaling for time is chosen based on the infiltration timescale $T_I=A/K_I$ which corresponds to the characteristic time for a surface water column of height $A$ to infiltrate into the soil under optimal conditions: large reservoir of surface water, dense vegetation, and dry soil. $T_I\approx 3$ $min$ for the parameter values of \cref{tab:dim}.   In contrast the characteristic time associated with maximal biomass growth within this model is $T_G=1/C\Gamma$, which corresponds to  $\sim 15$ $days$ for our default parameters.  The spatial scale is chosen based on the characteristic advection distance of surface water before infiltrating into the soil $X_A= K_V A^\delta \sqrt{\zeta}/K_I$.  This distance is $\sim 28$~$m$ for our parameters.  Biomass is scaled by  $K_B=4$~$kg/m^2$ and surface water height by $A=1$~$cm$.  
For this choice of scalings, namely, 
\[
B=K_B b, \quad H= A h, \quad T= \frac{A}{ K_I} t, \quad X =\frac{  K_V A^{\delta}\sqrt{\zeta}} {K_I} x
\]
the coupled model~\eqref{eq:3field} becomes 
\begin{subequations}
\label{eq:nondim}
\begin{align}
\frac{\partial h}{\partial t} &= p(t)  - \iota(b,s,h) + \frac{\partial}{\partial x}  \bigg(\nu(b,h) \ h\bigg)
\label{eq:nondim:h}
\\
\frac{\partial s}{\partial t}&=\alpha \ \iota(b,s,h) - \epsilon \left( \sigma s + \gamma b s  \right) 	
\label{eq:nondim:s}
\\
\frac{\partial b}{\partial t} &=
\epsilon\left(  
 s b (1-b) -\mu b + \delta_b \frac{\partial^2 b}{\partial x^2}\right)
 \label{eq:nondim:b}
\end{align}
\end{subequations} 
where  
\[
\iota(b,s,h)=\left(\frac{b+qf}{b+q}\right)\left(\frac{h}{h+1}\right)(1-s)^\beta, \quad \nu(b,h)=\frac{h^{\delta-1}}{1+\eta b},
\]
and the dimensionless parameters are defined in \cref{tab:nondim}.  The parameter $\epsilon=T_I/T_G$ represents the ratio of the infiltration timescale to the growth timescale and is of order $10^{-4}$.  We exploit the smallness of this ratio in the following section 
to motivate the switching model that is the focus of the rest of the paper.

\begin{table}[tbp]
	\renewcommand{\arraystretch}{1.55}
	\centering
	\begin{tabular}{|c|c|c|} 
		\hline
		parameter & definition& value\\
		\hline
		\hline
		$p$ & $ P/K_I$ &  
		\\
		\hline
		$\eta$ & $ N K_B$ & 80 
		\\
		\hline
		$\delta$ & -- & 
		\\
		\hline
		$f$&--& 0.1 
		\\
		\hline
		$q$ & $Q/K_B$ &  0.025 
		\\
		\hline
		$\beta$ & -- & 4 
		\\
		\hline
		$\alpha$ & $A/\phi Z_r$ & 0.037 
		\\
		\hline	
		$\sigma$&${L}/({\phi Z_r C \Gamma})$& 0.11 
		\\
		\hline
		$\gamma$ & $K_B/( \phi Z_r C)$& 1.48 
		\\
		\hline
		$\epsilon$& ${A C\Gamma }/{K_I}$& $1.3\times 10^{-4}$  
		\\
		\hline
		$\mu$&  ${M}/({C \Gamma})$&  0.15 
		\\
		\hline
		$\delta_b$ & $D_B K_I^2 /\left(C\Gamma K_V \sqrt{\zeta}\right)^2$ &$2.8\times 10^{-3}$ 
		\\
		\hline
	\end{tabular}
	\caption{	\label{tab:nondim} Definitions and parameter values of  the dimensionless form of the coupled model given by ~\cref{eq:nondim}. 
	}
\end{table}%

\section{A fast-slow switching model}
\label{sec:fastslow}
\begin{figure*}
    \centering

\tikzstyle{decision} = [ellipse, draw, fill=red!20,  text width=10em,  node distance=4cm, minimum height=5em]

\tikzstyle{block} = [rectangle, draw, fill=blue!20, 
    text width=14em, text centered, rounded corners, minimum height=4em]
\tikzstyle{line} = [draw, -latex']
   
\begin{tikzpicture}[node distance = 9cm, auto]
    \node [block] (fast) 
        {\textbf{Fast System}: $h(x,t)$, $s(x,t)$\\ \Cref{eq:fast}:
        precipitation, surface water transport, infiltration from surface into soil   };
    
    \node [decision, below of=fast] (f2s) 
        {soil moisture from \textbf{Fast}, after $h$ is negligible, initializes \textbf{Slow}};
    
    \node [block,right of=f2s] (slow) 
        {\textbf{Slow System}: $s(x,\tau)$, $b(x,\tau)$\\
        \Cref{eq:slow}: evapotranspiration; biomass growth, mortality and dispersal };
  \node [decision, above of=slow] (s2f) 
        {Soil moisture and biomass from \textbf{Slow}, once rain starts, initialize \textbf{Fast}.};
    \path [line] (fast) -- node {$h(x,t_f)\sim \mathcal{O}(\epsilon)$ } (f2s);
    \path [line] (f2s) -- node {$s_f(x)$} (slow);
    \path [line] (slow) --node {$p(\tau_s)>0$} (s2f); 
    \path [line] (s2f) -- node {$s_s(x)$, $b_s(x)$} (fast);

\end{tikzpicture}

    \caption{The fast system is initialized using the current biomass profile and soil moisture distribution.  It evolves the surface water and soil moisture on the fast time until surface water is no longer present.  The final soil moisture distribution and unchanged biomass profile are then used to initialize the slow system.  It evolves the soil moisture and biomass until the start of the next rain event, and the cycle repeats. 
    }
    \label{fig:fastslow}
\end{figure*}
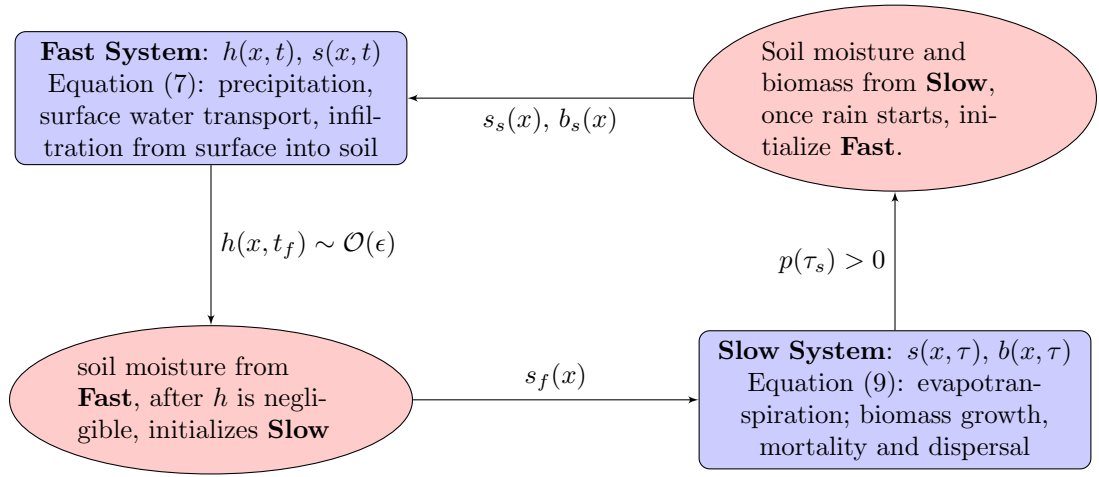

Rainfall initiates fast hydrological processes associated with overland waterflow and, with it, infiltration from the surface into the soil.  These processes occur on timescales of minutes to hours.  Evapotranspiration and plant growth occur on much longer timescales, days to months, and we assume can be neglected during rain events.  This separation of scales is evident in the coupled model presented in \cref{sec:seasonal} from the small parameter $\epsilon$ in \cref{eq:nondim} which represents the ratio of the water infiltration timescale $T_I$ to the biomass growth timescale $T_G$.   In this section, we present a switching model that captures dynamics of fast processes initiated by rainfall and switches to the dynamics of slower processes in the absence of surface water. \Cref{fig:fastslow} shows a schematic diagram of such a model, which we describe in detail in ~\cref{sec:model:fastslow}.  We then, in \cref{sec:model:bc}, discuss numerical methods employed to explore the model along with some potential issues associated with our choice of periodic boundary conditions.

\subsection{Fast-Slow Switching Model: exploiting the limit $\epsilon\to 0$}
\label{sec:model:fastslow}
We now motivate a switching model for banded vegetation patterns by leveraging the fact that $\epsilon \ll 1$ in the dimensionless model~\cref{eq:nondim}. We first consider the case that $\epsilon=0$, corresponding to the assumption that the dynamics occurring on the biomass growth timescale are completely negligible.  The $\epsilon=0$ assumption leads to a fixed biomass profile $b(x)$ since $\partial b/\partial t = 0$ from ~\cref{eq:nondim:b}.  The remaining two equations of system~\eqref{eq:nondim} become:
\begin{subequations}
\label{eq:fast}
\begin{align}
\frac{\partial h}{\partial t} &= p(t)  - \left(\frac{b+qf}{b+q}\right)\left(\frac{h}{h+1}\right)(1-s)^\beta + \frac{\partial}{\partial x}  \bigg(\frac{h^{\delta}}{1+\eta b}\bigg) 
\label{eq:fast:h}
\\
\frac{\partial s}{\partial t}&=\alpha  \left(\frac{b+qf}{b+q}\right)\left(\frac{h}{h+1}\right)(1-s)^\beta. 
\label{eq:fast:s} 	
\end{align}
\end{subequations} 
We refer to~\cref{eq:fast} as the ``fast system," which models the dynamics during and shortly after rain events.  (For convenience, we substituted in the expressions for infiltration rate $\iota(b,s,h)$ and transport speed $\nu(b,h)$ that are defined below \cref{eq:nondim}.)

We start evolution of the fast system when a rain event begins, and continue until the surface water $h$ becomes negligible, i.e. below a threshold of order $\epsilon$. We expect these fast dynamics to occur over an hours timescale, after which we obtain an updated soil moisture profile 
\begin{equation}
s \xrightarrow{\text{fast system}} s + \theta(b,s,p) 
\end{equation}
that has been replenished by the rain event.  The profile of the soil moisture replenishment term $\theta$ is determined by the fast system~\eqref{eq:fast}, and depends on the biomass $b(x)$ and soil moisture  $s(x)$ profiles prior to the rain event along with the rain event $p(t)$ itself.  

The fast dynamics of \cref{eq:nondim} are terminated once $h\sim \OO(\epsilon)$, after which the soil moisture and biomass evolution proceeds on a slow timescale given by $\tau=\epsilon t$ under the assumption that the surface water height remains fixed at $h=0$. Rescaling time in \cref{eq:nondim:s,eq:nondim:b} then leads to 
\begin{subequations}
\label{eq:slow}
\begin{align}
\frac{\partial s}{\partial \tau}&= -  \left( \sigma s + \gamma s b  \right) 	
\label{eq:slow:s}
\\
\frac{\partial b}{\partial \tau} &=
 s b (1-b) -\mu b + \delta_b \frac{\partial^2 b}{\partial x^2}.
 \label{eq:slow:b}
\end{align}
\end{subequations} 
We refer to \cref{eq:slow} as the ``slow system" and use it to evolve biomass and soil moisture in the absence of surface water.  The slow system~\eqref{eq:slow} is initialized with the biomass profile $b$ prior to the rain event and the replenished soil moisture $s+\theta$, and evolves until the next rain event occurs.   
In results and discussion sections that follow, we numerically explore this fast-slow model that has been motivated by the coupled model~\eqref{eq:3field}.

\subsection{Numerical methods and  boundary conditions}
\label{sec:model:bc}

We discretize in space with finite differences using first-order upwinding for advection and second-order centered differences for diffusion.     For time integration we employ Matlab's ode15s, an implicit variable-order solver  designed for stiff equations~\cite{shampine1997matlab}. At the start of each rain event, the fast system is initialized with the output biomass (which will remain fixed), soil moisture of the slow system and no initial surface water.  The fast system is run for twice as long as the rain event, with a constant rain input for the first half of this time.  The surface water height typically has fallen to below $10^{-3}$~$cm$ by this point, and the simulation proceeds to the slow system initialized with the updated soil moisture and previously output biomass.   However, if the surface water is not below a threshold value of $H_{thresh}=0.1$~$cm$, the fast system is run again with no rain input for fixed intervals of time until the threshold is reached and  the simulation can move on to the slow system.

\begin{figure}
    \centering
    \includegraphics[width=\linewidth]{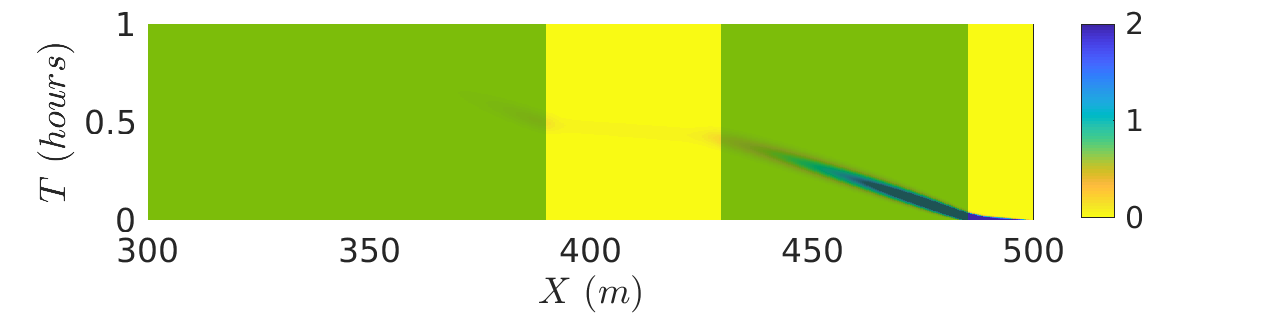}
    \caption{Spacetime plot of surface water height in units of $cm$ shown with the blue indicating 2~$cm$.  The green indicates where vegetation is above 0.01~$kg/m^2$.  The surface water, initialized at 2~$cm$ along the top 15~$m$ of the domain falls to below 0.2 $mm$ by the trailing edge of the vegetation band at $X\approx 419$~$m$. Parameter values used in \cref{eq:fast,eq:slow} are given by \cref{tab:nondim} with $\delta=1$. 
    }
    \label{fig:periodicbc}
\end{figure}

The simulations are carried out on a periodic domain that typically corresponds to a few wavelengths of the banded pattern. 
The underlying assumption for this choice of boundary conditions is that we are considering a small section in the middle of a very long hillslope.  One potential issue that may arise with periodic boundary conditions is that, during the fast system, surface water may repeatedly reach the bottom of the domain and get re-injected at the top of the domain.  
We don't expect surface water to travel more than a few vegetation bands before infiltrating into the soil, and find that our simulations are consistent with this expectation.  For example, a simulation  of the fast system~\eqref{eq:fast} with parameters from \cref{tab:dim} shows that 2~$cm$ of surface water at the top 15~$m$ of a hillslope is decreased to approximately 1\% of its original amount after traveling through an 80~$m$ wide band of vegetation  with peak biomass value of $0.18$~$kg/m^2$.  The water travels with a speed of approximately $4$~$cm/s$ through the vegetation band and  13~$cm/s$ over the bare soil.  This simulation, which is shown in \cref{fig:periodicbc}, is initialized with the state reached just prior to a rain event at 200~$years$ from the simulation shown in \cref{fig:sim:spatialslice} of Section~\ref{sec:sim:rand}.

\section{Numerical results for fast-slow model}
\label{sec:sim}
 In this section, we explore pattern formation within the fast-slow switching model, \cref{eq:fast,eq:slow}, through numerical simulation.
Inspired by the biannual rainy seasons shown in \cref{fig:rain}(b), we take as a base-case scenario two major rain events each year, spaced six months apart, with a mean annual precipitation of 160 $mm/year$.  In this case, each rain event lasts for six hours, and deposits half of the mean annual precipitation at a constant rate during that short time-frame.  Unless otherwise noted, we use parameter values from \cref{tab:nondim} with $\delta=1$, corresponding to dimensional quantities given in \cref{tab:dim}, throughout this section.  We also report all quantities in dimensional units, with the conversions described in \cref{sec:model:nondim}. (See also parameters in \cref{tab:dim}.)

We first study characteristics of stable periodic solutions obtained using sinusoidal perturbations of the uniform state as initial conditions in \cref{sec:sim:po}. 
These solutions provide a basis for interpreting the results of simulations initialized with random initial perturbations, which we discuss in \cref{sec:sim:rand}. We provide a detailed picture of transients that persist on ecologically relevant timescales, e.g. decades to centuries, for this case and describe the observed model behavior on the timescale of millennia.  We then, in \cref{sec:sim:precip}, study the influence that the amount of rainfall has on pattern characteristics.  Finally, in \cref{sec:sim:feedback}, we explore the role that biomass feedback plays in infiltration and surface water transport in the pattern forming process. Additional supporting simulation results are reported in \ref{app:sim} of the online supplement.

\subsection{Stable periodic states}
\label{sec:sim:po}

\begin{figure*}
    \centering  
        \setlength{\unitlength}{\textwidth}
    \begin{picture}(1,0.36)
    \put(0.01,0){\includegraphics[width=0.48\textwidth]{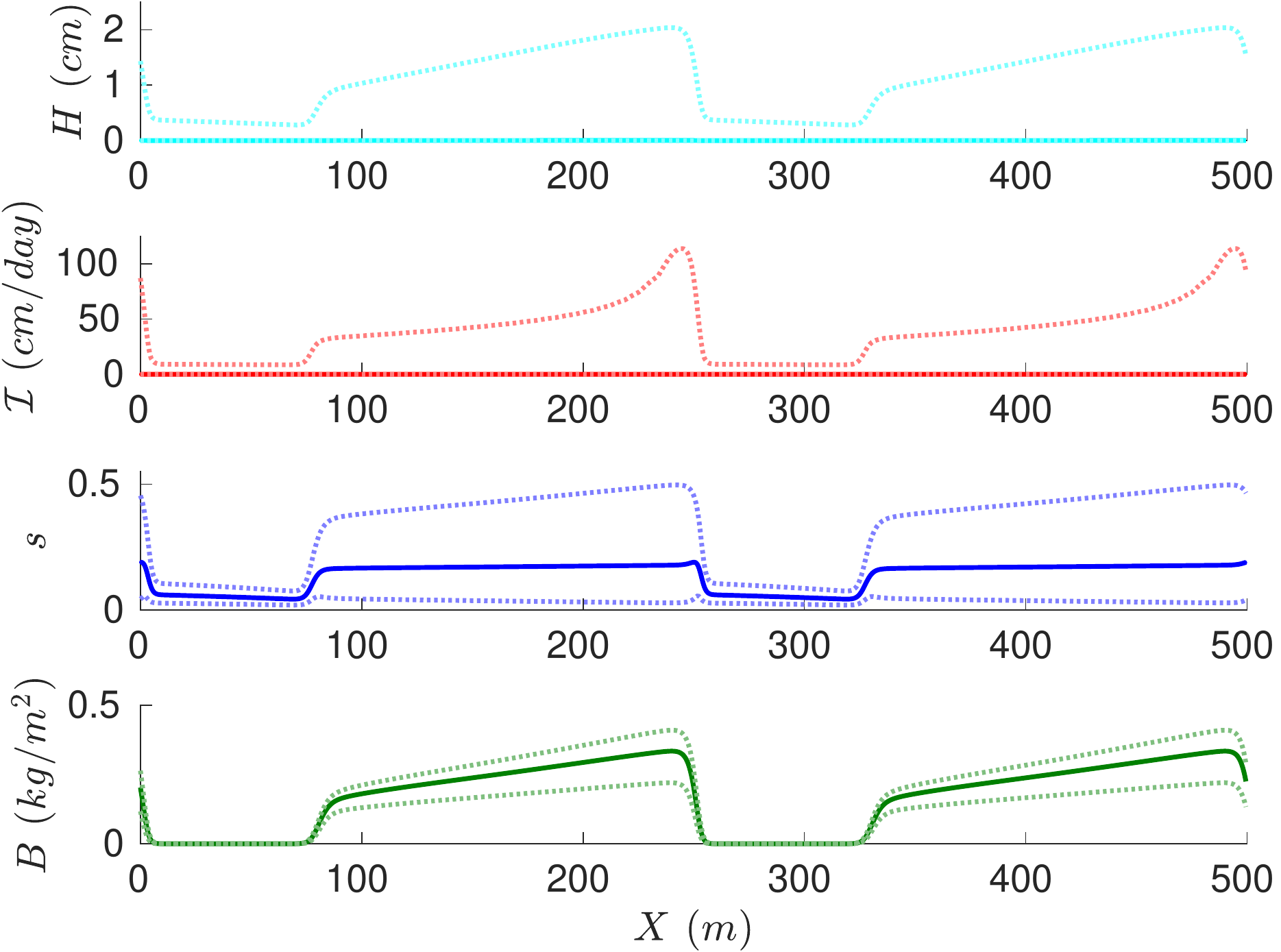}}
    \put(0.51,0){\includegraphics[width=0.48\textwidth]{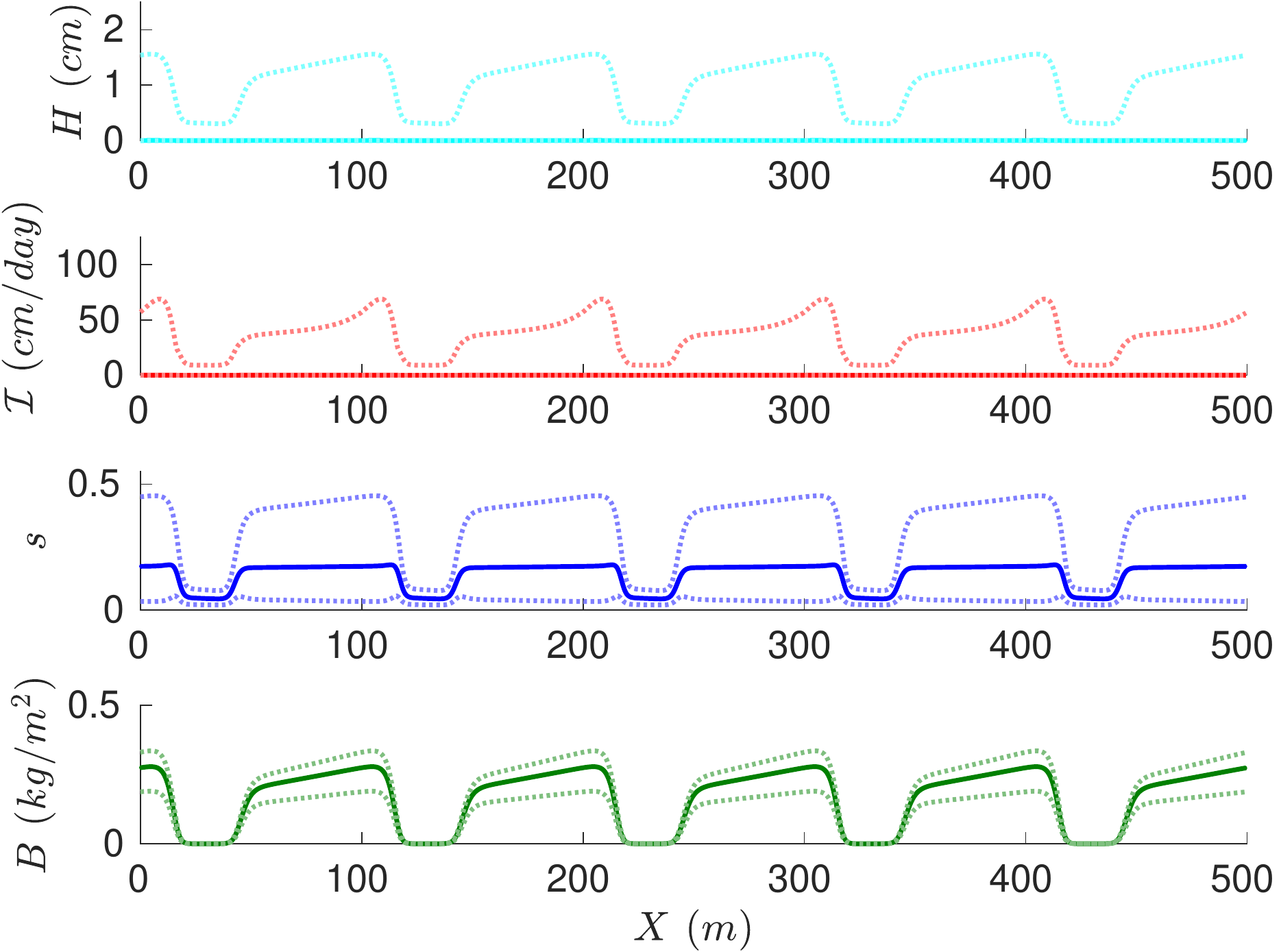}}
    
    \put(0.0,0.35){(a)}
    \put(0.5,0.35){(b)}
    \end{picture}
    \caption{Spatial profiles of patterns with initial perturbation of wavenumber (a) $k=2\pi/250$ $m^{-1}$ and (b) $k=2\pi/100$ $m^{-1}$. For each wavenumber, surface water height $H$, Infiltration rate $\II$, soil moisture $s$, and biomass density $B$ at $t=3000$ $years$ with mean annual precipitation of 160 $mm/year$ are shown.  The solid line is the annually averaged profile while the dotted lines show the pointwise minimum and maximum values over the course of the year. Parameter values used in \cref{eq:fast,eq:slow} are given by \cref{tab:nondim}, and the rain is input uniformly over two evenly-spaced six-hour rainstorms per year.}
    \label{fig:kn25profiles}
\end{figure*}

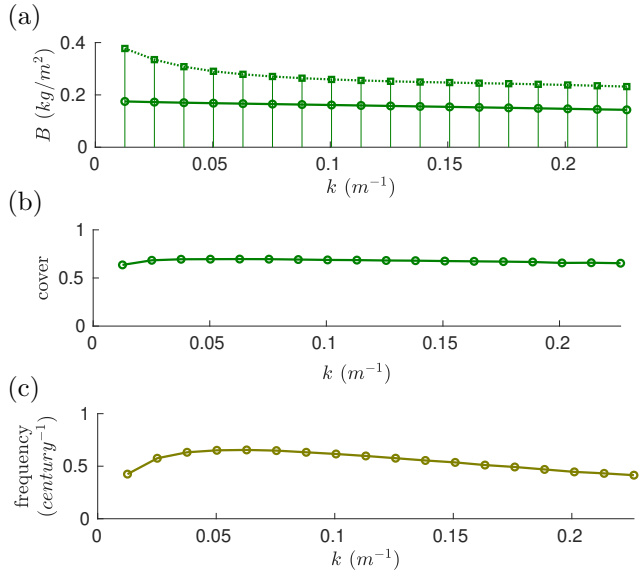
\begin{figure}
    \centering    
    \setlength{\unitlength}{0.5\textwidth}
    \begin{picture}(1,0.9)
    \put(0.04,0.6){\includegraphics[width=0.48\textwidth]{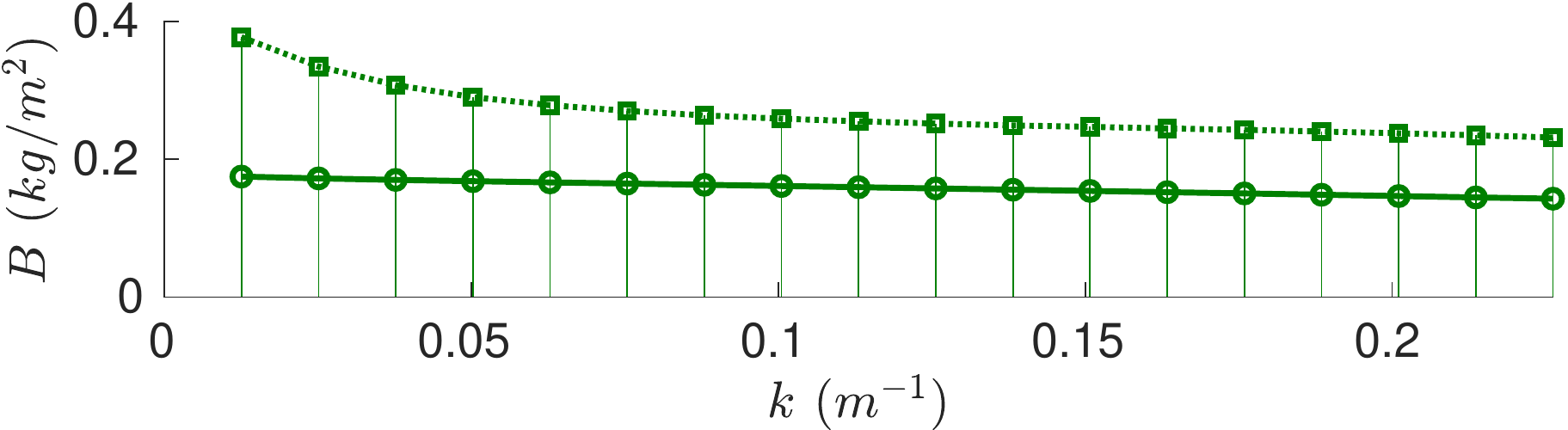}}
    \put(0.05,0.3){\includegraphics[width=0.47\textwidth]{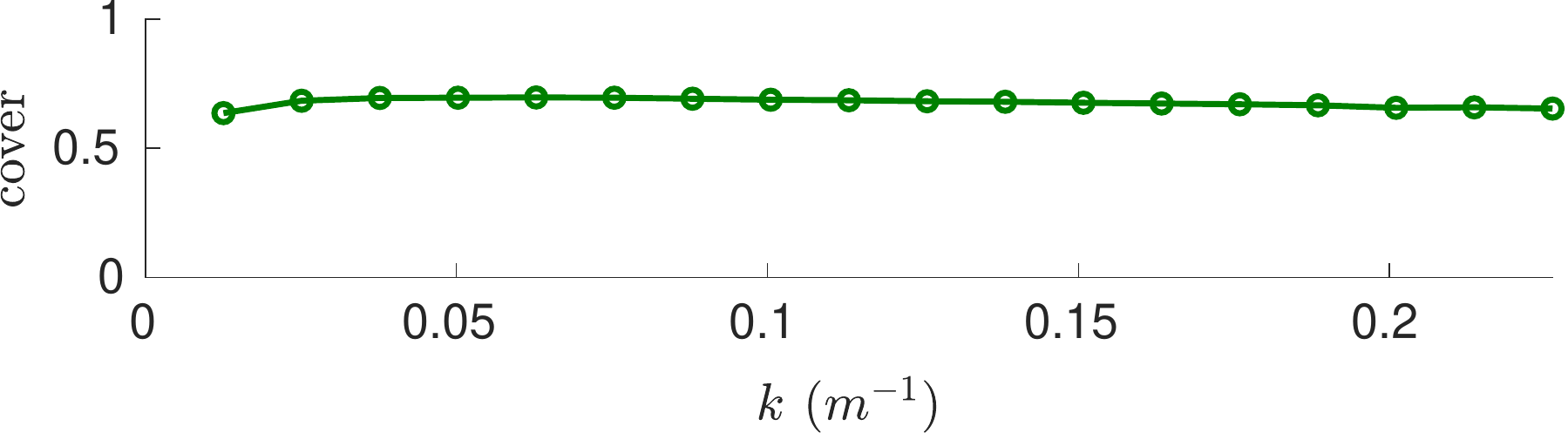}}
    \put(0.01,0){\includegraphics[width=0.5\textwidth]{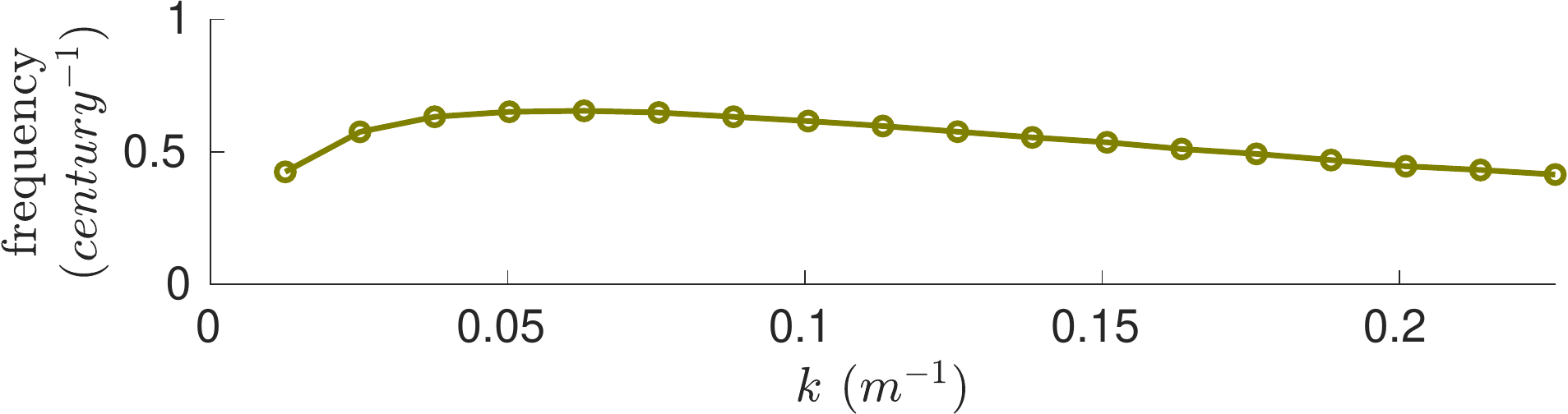}}
    \put(0,0.88){(a)}
    \put(0,0.58){(b)}
    \put(0,0.28){(c)}
    \end{picture}   
    \caption{Characteristics of solutions from simulation as a function of pattern wavenumber.  (a) Biomass averaged over time and space,  and  peak value of annually-averaged biomass are indicated by solid and dotted lines, respectively. (b) Fraction of wavelength covered by vegetation band. (c) Uphill migration frequency of biomass pulse in wavelengths of the pattern per century. Parameter values used in \cref{eq:fast,eq:slow} are given by \cref{tab:nondim}, and the rain is input uniformly over  two evenly-spaced six-hour rainstorms per year with mean annual precipitation of 160~$mm/year$. 
    }
    \label{fig:knplots}
\end{figure}

We consider a spatial domain of length $L=500$~$m$ and take the positive $x$-direction to be uphill. We initialize the fast-slow system with a 1\% sinusoidal perturbation of wavenumber $k_j=2\pi j/L$, $j=1\dots 30$, to the uniform solution with constant precipitation given by \cref{eq:ss:uv}. For $1\leq j \leq 19$, the simulation converges to a periodic traveling wave solution with wavenumber $k_j$, and for $j>19$ the simulation converges to a periodic pattern with different wavenumber (typically $k_j$ with $4\leq j \leq 6$). These stable wavenumbers correspond to a stripe spacing from $\sim 26\ m$ ($j=19$) all the way to the size of the domain $500\ m$ ($j=1$). 
\Cref{fig:kn25profiles} shows examples of spatial profiles of two different wavenumber solutions for the fixed parameter set considered here. The solid line in each plot represents the annual average while the dotted lines indicate the pointwise maximum and minimum.  The average biomass, the surface water and soil moisture all peak at the leading edge of the vegetation bands. While the surface water and infiltration rates are near zero on average, the maximum values are at around 2~$cm$ and $100$~$cm/day$, respectively,  during rain events.   The soil moisture within the vegetation bands also varies significantly in time, going from $s=0.5$ just after a rain event to below 0.05 just before the start of the next one.  The seasonal variations in biomass, on the other hand, are less dramatic. We note that, during rain events, the surface water height is expected to decrease in the vegetated regions for flat-terrain patterns because of increased infiltration~\cite{thompson2011vegetation}. However, on a hillslope, the flow induced by the elevation gradient along with the increased surface roughness of vegetation leads to an increase in predicted surface water height within the vegetation bands in the fast-slow model.        

\Cref{fig:knplots} indicates that many pattern characteristics (e.g. average biomass, fraction of wavelength covered by vegetation) show remarkably little variation with respect to wavenumber. Panel (a) indicates a monotonic decrease in the maximum biomass value as a function of wavenumber.   The fraction of the wavelength with vegetation cover, shown in Panel (b), varies little with $k$, on the order of 5\% with a maximum at around $n=5$.  To compute the fractional cover, we take the edges of the vegetation band to be the points of steepest increase/decrease in biomass.  Using a threshold value of $Q=0.1$~$kg/m^2$ to define the edges leads to nearly identical fractional cover values.

Unlike the average biomass and fractional cover, the average migration speed of the bands does vary significantly with wavenumber: it decreases monotonically from about 200 $cm/year$ for the largest band spacing of $500\ m$ ($k_1=2\pi/500$~$m^{-1}$) to about 10 $cm/year$ for a small spacing of $28\ m$  ($k_{18}=36\pi/500$~$m^{-1}$), and has a value of $65$~$cm/year$ for patterns with a wavelength of 100 $m$. As a possible reference point, observational studies of banded patterns in the Horn of Africa record typical wavelengths of $40-250$~$m$ and migration speeds measured in 10s of centimeters per year, on slopes with typical elevation grades  $0.1-1.2$\%.
~\cite{deblauwe2012determinants,gowda2018signatures,bastiaansen2018multistability}. 
The model predictions are therefore of the right order of magnitude, and this is without these characteristics being a factor in our parameter selection.   

Since there is a significant (increasing) trend in migration speed with wavelength, we choose to plot the migration frequency (wavelengths of the pattern per century) in \cref{fig:knplots}(c). For this we find a relatively modest range of $0.4-0.6$/century with a peak speed for a wavelength of 100 $m$. Deblauwe et al.~\cite{deblauwe2012determinants}, drawing on pattern observations around the globe, report slightly smaller mean migration speed values in the range of $0.1-0.3$/century, although with error bars that can bring these to our higher range.

We note that predictions of wavelength and migration speed that are based on a {\it constant} rain input to the coupled model~\eqref{eq:3field} are way off, both in comparison to the reported observations and to our fast-slow simulations that incorporate short rain events. For instance, the onset of Turing patterns for a constant rain input have  wavelength of 26 $m$, and travel rapidly at a speed of $\sim$3.5 $m/yr$ (i.e. $\sim 13$ wavelengths per century). Moreover, the associated Turing-Hopf bubble of \cref{fig:uv:floquet}(b) extends all the way to a $MAP\approx 309$ $mm/yr$, giving a rather high lower bound to where patterns would occur. Using linear predictions from the case of constant rain input, and trying to adjust the parameters to fit those observed pattern characteristics, would lead to different parameter choices. 

 \subsection{Random initial conditions and asymptotic behavior}
 \label{sec:sim:rand}
 
\begin{figure}
    \centering
      \setlength{\unitlength}{0.5\textwidth}
    \begin{picture}(1,1.5)
    \put(0.02,0.75){\includegraphics[width=0.41\textwidth]{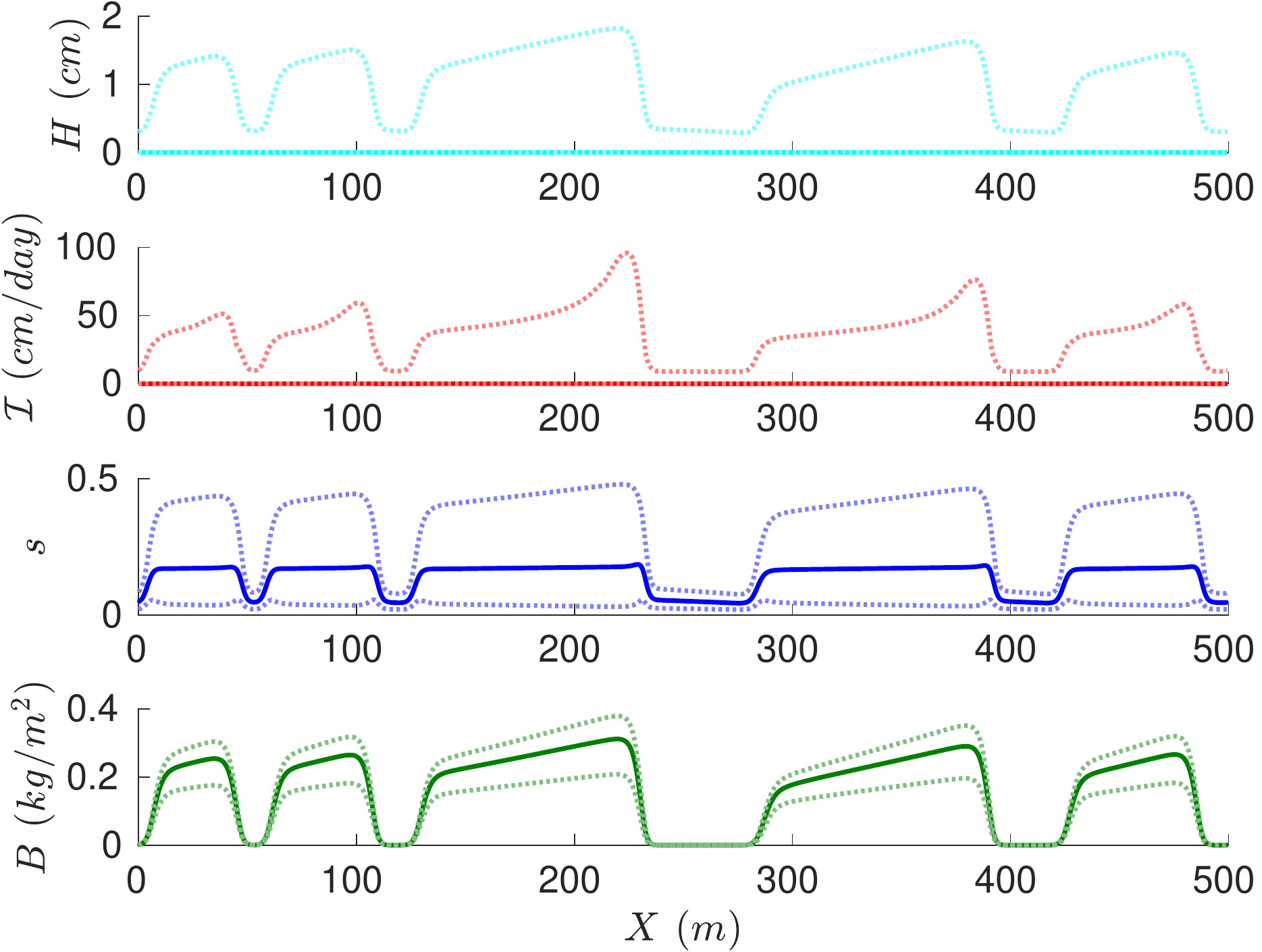}}
    \put(0.0,0.0){\includegraphics[width=0.5\textwidth]{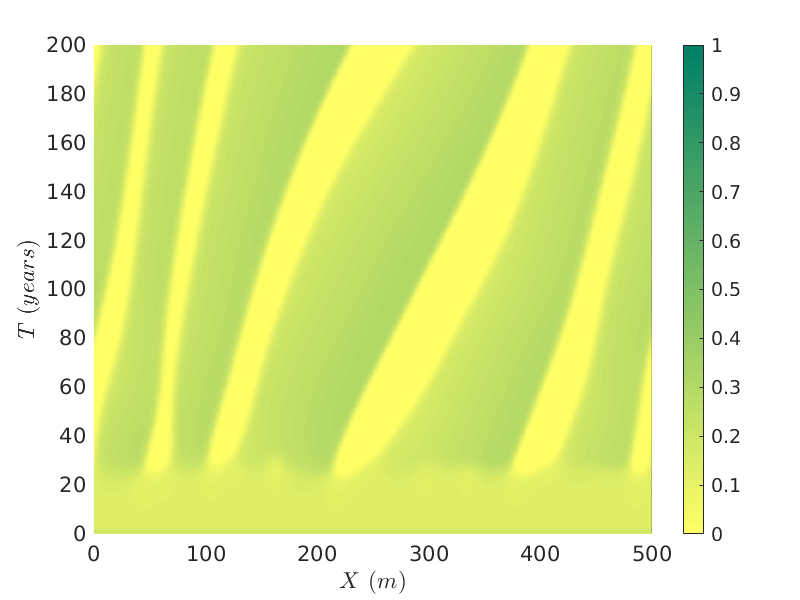}}
    \put(-.01,0.7){(b)}
    \put(-0.01,1.45){(a)}
    \end{picture}
    \caption{(a)  Spatial profile of surface water height $H$, infiltration rate $\II$, soil moisture $s$, and biomass density $B$ at $t=200$ $years$.  The solid line is the annually averaged profile while the dotted lines show the pointwise minimum and maximum values over the course of the year.   (b) Spacetime plot of annually averaged biomass $B$ in units of $kg/m^2$ over the course of the 200 $year$ simulation.  Yellow indicates low biomass while green indicates high biomass, and the color scale range is $0<B<1$~$kg/m^2$.  All subsequent biomass spacetime plots use the same color scale described here. Parameter values used in \cref{eq:fast,eq:slow} are given by \cref{tab:nondim} with $\delta=1$, and the rain is input uniformly over two two evenly-spaced six-hour rainstorms per year with mean annual precipitation of 160~$mm/year$. 
    }
    \label{fig:sim:spatialslice}
\end{figure}    

While the periodic states summarized in \cref{fig:kn25profiles,fig:knplots} persist for at least 5000 years and are stable to small perturbations, long-time simulations with random initial perturbations eventually approach the long-wavelength patterns.  We first provide a detailed account of long-lived transients that persist on ecologically relevant timescales of centuries before discussing this asymptotic behavior on the scale of millennia.  

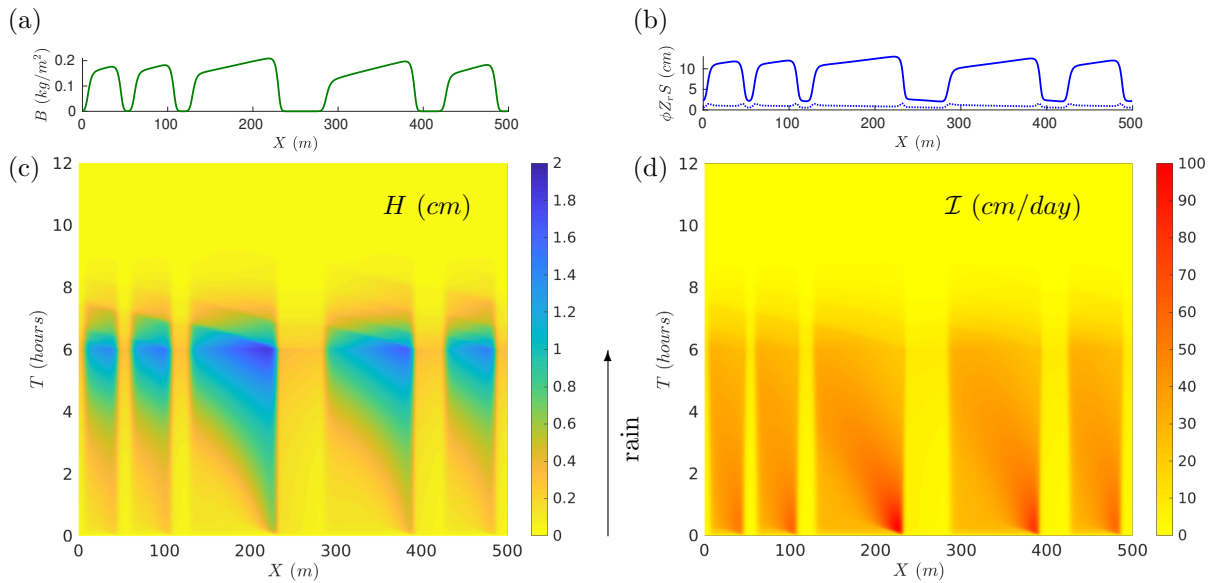
\begin{figure*}
    \centering    
          \setlength{\unitlength}{\linewidth}
    \begin{picture}(1,0.45)
    \put(0,0){\includegraphics[width=0.49\linewidth]{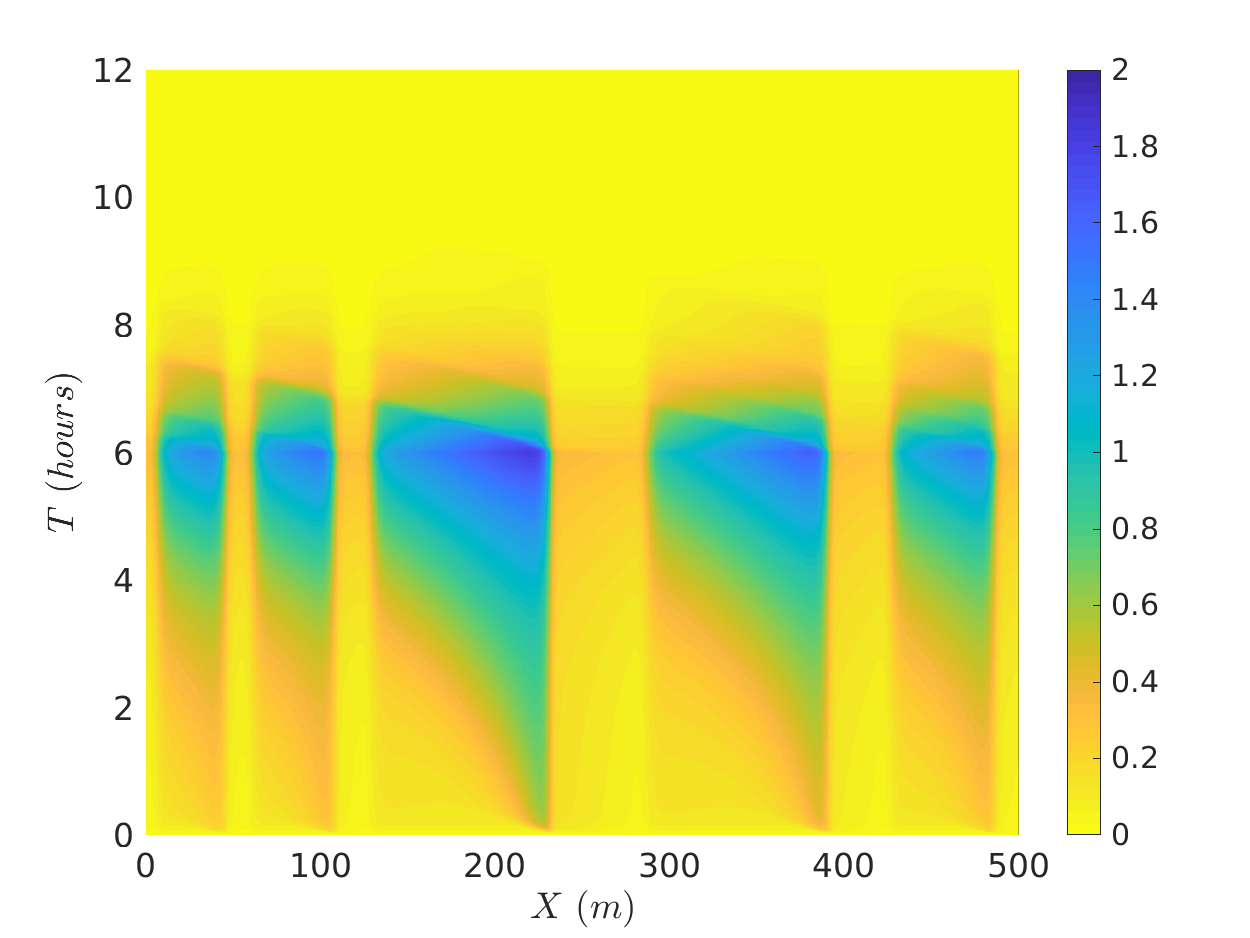}}
     \put(0.5,0){\includegraphics[width=0.49\linewidth]{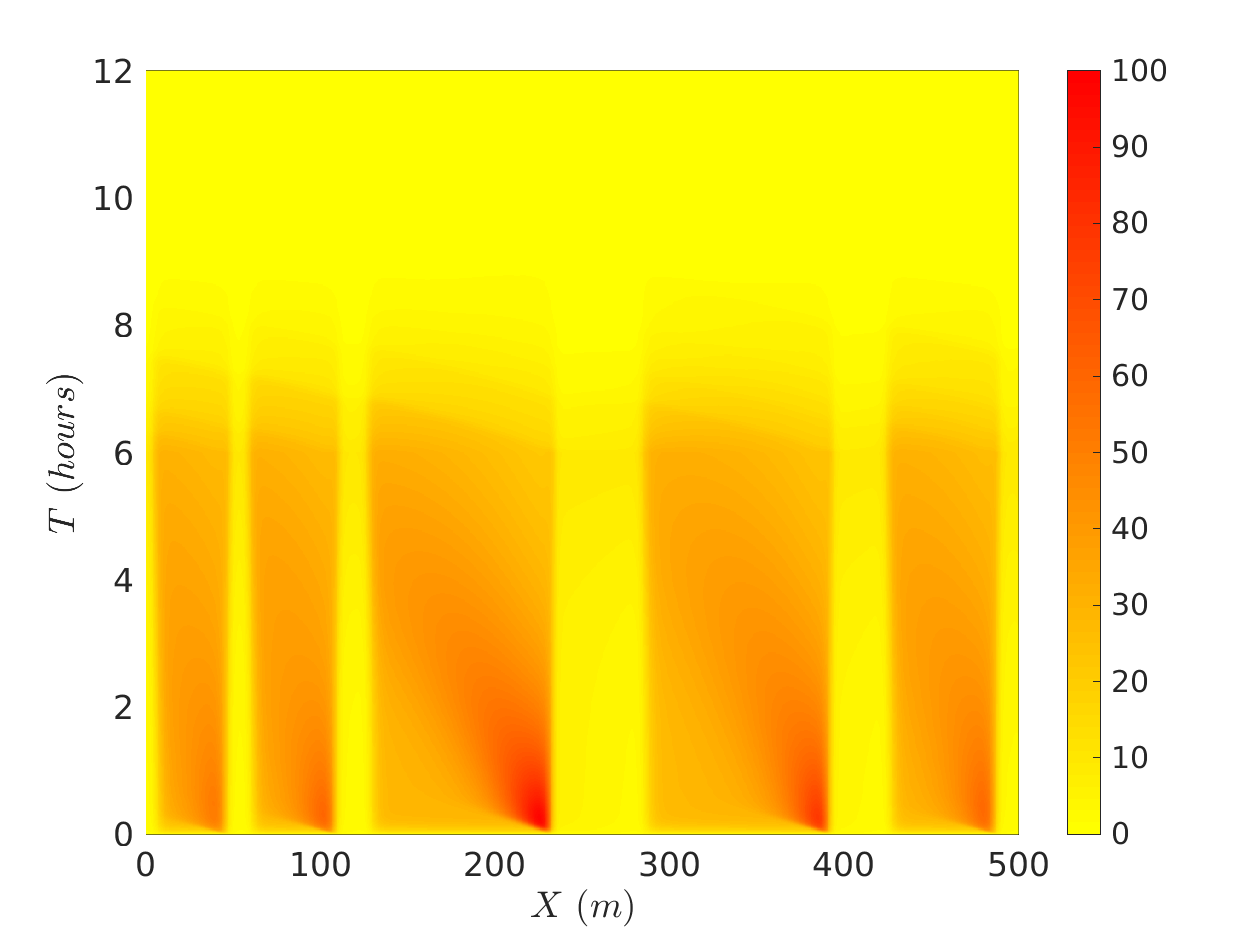}}
    \put(0.02,0.35){\includegraphics[width=0.39\linewidth]{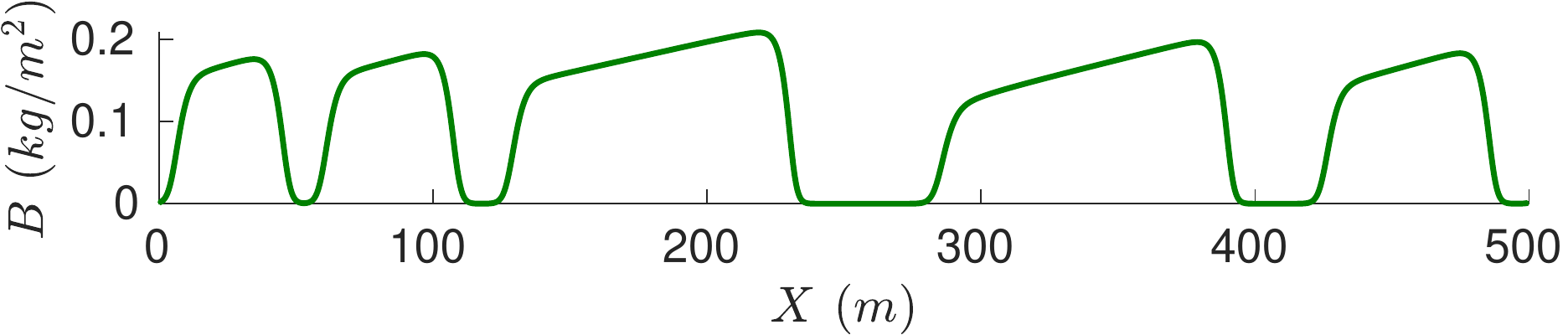}}
     \put(0.52,0.35){\includegraphics[width=0.39\linewidth]{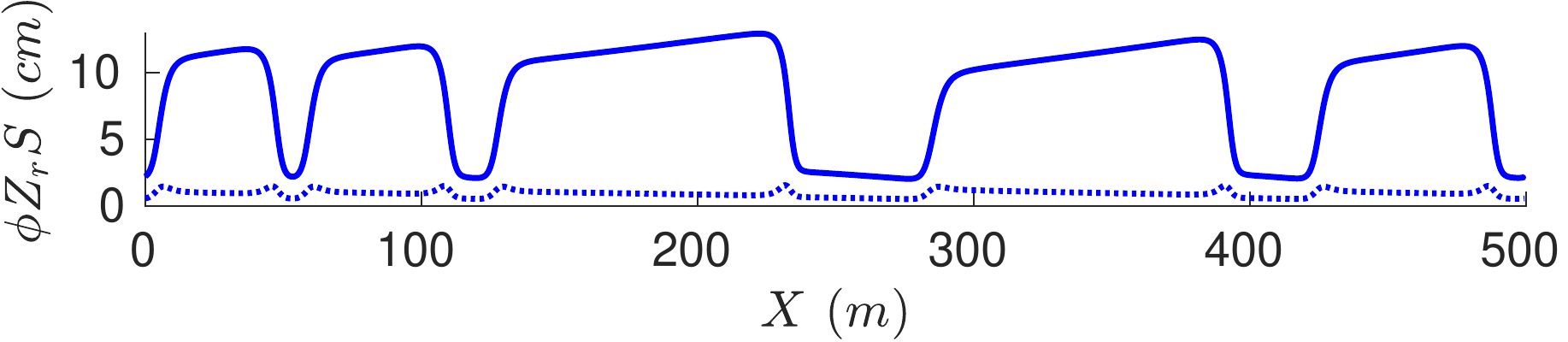}}
     \put(0,.45){(a)}
     \put(.5,.45){(b)}
     \put(0,.33){(c)}
     \put(.5,.33){(d)}
     \put(0.48,0.04){\vector(0,1){0.15}}
     \put(0.49,0.1){\rotatebox{90}{rain}}
     
     \put(0.3,0.3){$H$ ($cm$)}
     \put(0.75,0.3){$\mathcal{I}$ ($cm/day$)}
     \end{picture}
    \caption{Fast system during last year of simulation shown in \cref{fig:sim:spatialslice}. (a) Fixed biomass profile during simulation of fast system.  (b) Initial (final) soil water content ($\phi Z_r s)$ in units of $cm$ shown with dotted (solid) line. (c) Spacetime plot of surface water height $H$ in units of $cm$ during and shortly after a 6-hour storm with 80 mm of rainfall in the final year of a $200$-year simulation.     (d) Spacetime plot of instantaneous infiltration rate in units of $cm/day$ during the same time period. 
    \label{fig:sim:fast}
    }
\end{figure*}

\begin{figure}
    \centering    
    \setlength{\unitlength}{0.5\textwidth}
    \begin{picture}(1,0.6)
    \put(0.01,0.33){\includegraphics[width=0.5\textwidth]{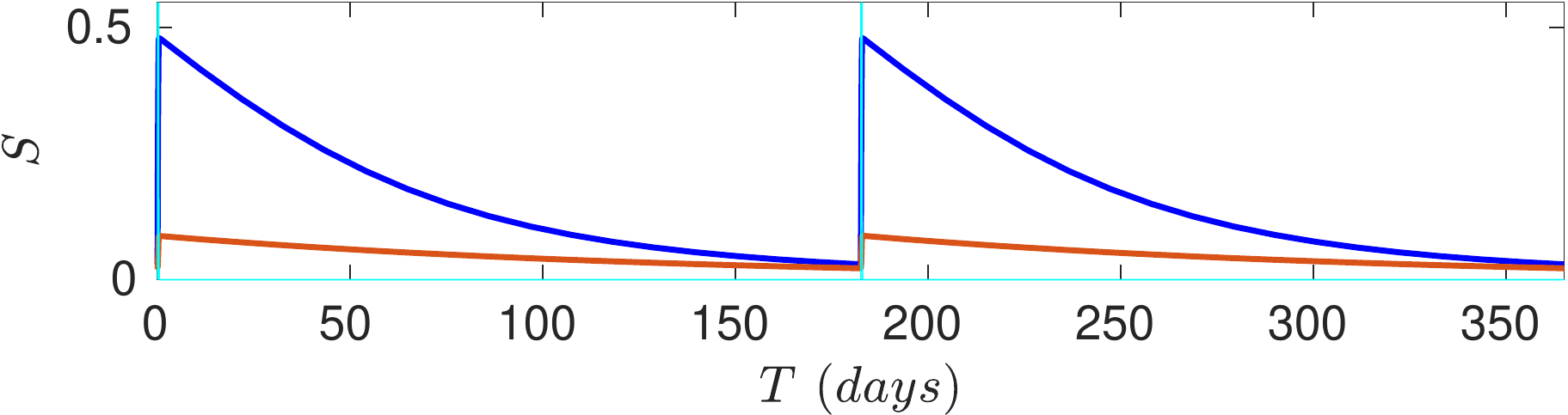}}
    \put(0.01,0){\includegraphics[width=0.5\textwidth]{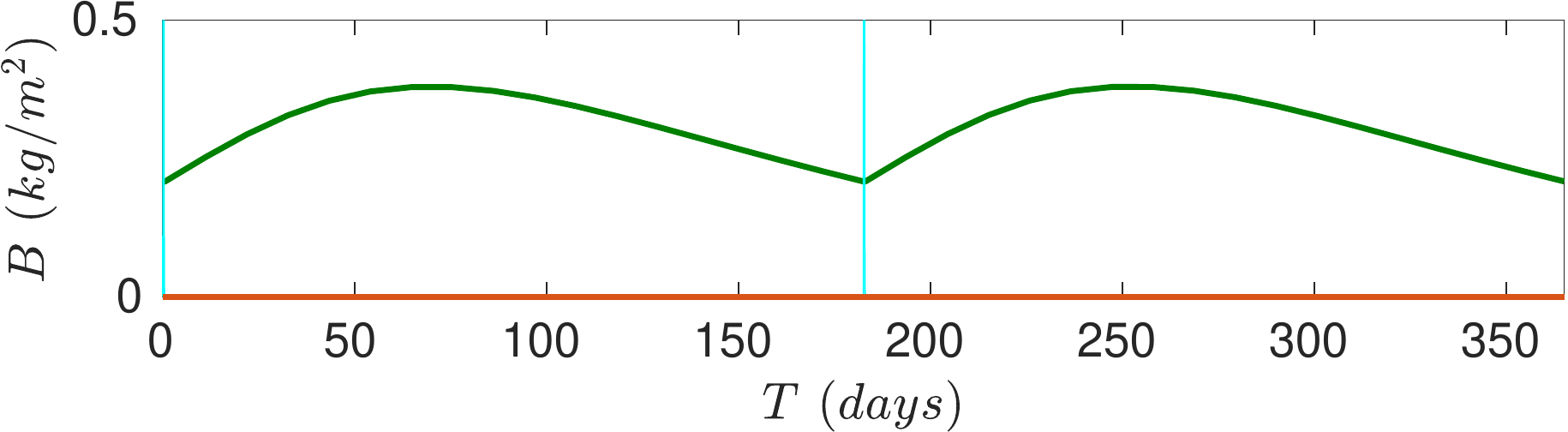}}
     \put(0,.62){(a)}
    \put(0,.3){(b)}
    \end{picture}
    \caption{Time series of (a) soil moisture and (b) biomass during the last year of the simulation shown in \cref{fig:sim:spatialslice,fig:sim:fast}. The blue (green) lines indicate  soil moisture (biomass) at the peak of the vegetation band ($x\approx 220$ $m$), and the orange lines indicate the center of the uphill bare soil region ($x\approx 255$ $m$).  The cyan vertical lines indicate the time of the rain events.  
    \label{fig:sim:slow}
    }
\end{figure}

\Cref{fig:sim:spatialslice,fig:sim:fast,fig:sim:slow} show details of a 200-year simulation on a 500 $m$ domain with mean annual precipitation of $160$ $mm/year$. These are the same parameters we used for the periodic patterns in the previous section. However, we modify the initial condition to be a 1\% random perturbation of the uniformly vegetated state that's obtained for constant precipitation, given by \cref{eq:ss:uv}. These values, $B=0.126$~$kg/m^2$ and $s=0.154$, are a good estimate for the mean of the uniform state in the seasonal case.  A spacetime diagram of the annually-averaged biomass along with spatial profiles of the surface water, infiltration rate, soil moisture and biomass during the last year of the simulation are shown in \cref{fig:sim:spatialslice}.

The simulation evolves the fast system, \cref{eq:fast}, during and shortly after each rain event. Specifically, constant precipitation, at a rate of about 13 $mm/hour$, is input during the first six hours and none during the next six hours, during which time the surface water height drops to below $10^{-3}\ cm$. The biomass profile, which remains fixed during this time, is shown in \cref{fig:sim:fast}(a).  \Cref{fig:sim:fast}(b) shows the initial and final soil water content $\Phi Z_r s$ in units of $cm$.  A total of 8~$cm$ of water is input uniformly across the domain by the rain storm and the vegetation bands collect approximately 10.6 $cm$ on average, while the bare soil regions collect approximately 1.7~$cm$ on average.  The vegetated regions therefore absorb, on average, approximately 6.2 times the amount of rainfall because of runoff from the bare soil regions. 
\Cref{fig:sim:fast}(c,d) show spacetime diagrams for surface water height $H$ and instantaneous infiltration rate ${\cal I}$ over the 12 hour period that the fast system is run.  During the rain, both the surface water height and the infiltration rate are enhanced by the presence of biomass.  The surface water is increased in the vegetation bands because increased surface roughness slows water transport.  This increased height, along with a positive feedback of biomass directly on infiltration rate, lead to the heightened accumulation of soil water content in the vegetation bands that is seen in \cref{fig:sim:fast}(b).

At the end of the 12-hour period of the fast system, we then initialize the slow system with the updated soil moisture from the fast system and the same biomass profile. \Cref{fig:sim:slow} tracks soil moisture $s$ and biomass $B$ at locations in the vegetation band (blue and green) and in bare soil (orange) over time for the slow system during the final year  of the simulation shown in \cref{fig:sim:spatialslice}.  The soil moisture in the vegetation band initially increases significantly, but eventually falls to nearly the same level as the bare soil just before the next rain event.  The biomass peak is delayed by approximately two months after the rain event. 

While the behavior in \cref{fig:sim:spatialslice,fig:sim:fast,fig:sim:slow}, initialized with random perturbation, persists for a millennium, 3500-year simulations (see \cref{fig:asymp:b}) reveal it to be transient.  This is in contrast to the cases where a periodic perturbation of specific wavenumber is introduced, as described in \cref{sec:sim:po}. The periodic states in \cref{fig:kn25profiles,fig:knplots} persist as traveling wave solutions (on annually averaged timescales) for at least 5000 years.  

The long transients persist on ecologically relevant timescales and have wavelengths consistent with predictions from linear stability analysis of the coupled model, \cref{eq:3field}, with identical rain input (i.e. six-hour constant rain intensity storms every six months).  Those calculations indicate an onset of the pattern-forming instability occurs for a total annual rain of $\sim 175\ mm$ (input in two large rain events), and are associated with a wavelength of $\sim 94$ $m$. For mean annual precipitation of $160$ $mm/year$, the ``Turing-Hopf bubble" has a width that indicates the uniform state is unstable to perturbations with wavenumbers between $k=0.006\ m^{-1}$ and $k=0.305\ m^{-1}$, corresponding to  wavelengths greater than $\sim 20\ m$ and all the way up to a kilometer scale; the most unstable mode for these conditions occurs for $k\approx 0.06\ m^{-1}$ (a wavelength of $\sim 105\ m$), where the Floquet multiplier reaches a peak value of $1.15$. Simulations of the fast-slow system on a 500 m domain, with an initial periodic perturbation of fixed wavenumber within this Turing instability range, appear to be stable, at least for wavelengths greater than $\sim 25 m$, i.e. they do not undergo band-merging events on the millennial simulation timescale. (See results presented in \cref{fig:knplots}.)  

\subsection{Nonlinear dependence of transport on surface water height}
\label{sec:sim:nonlinh}
\begin{figure}
    \centering
    \includegraphics[width=0.5\textwidth]{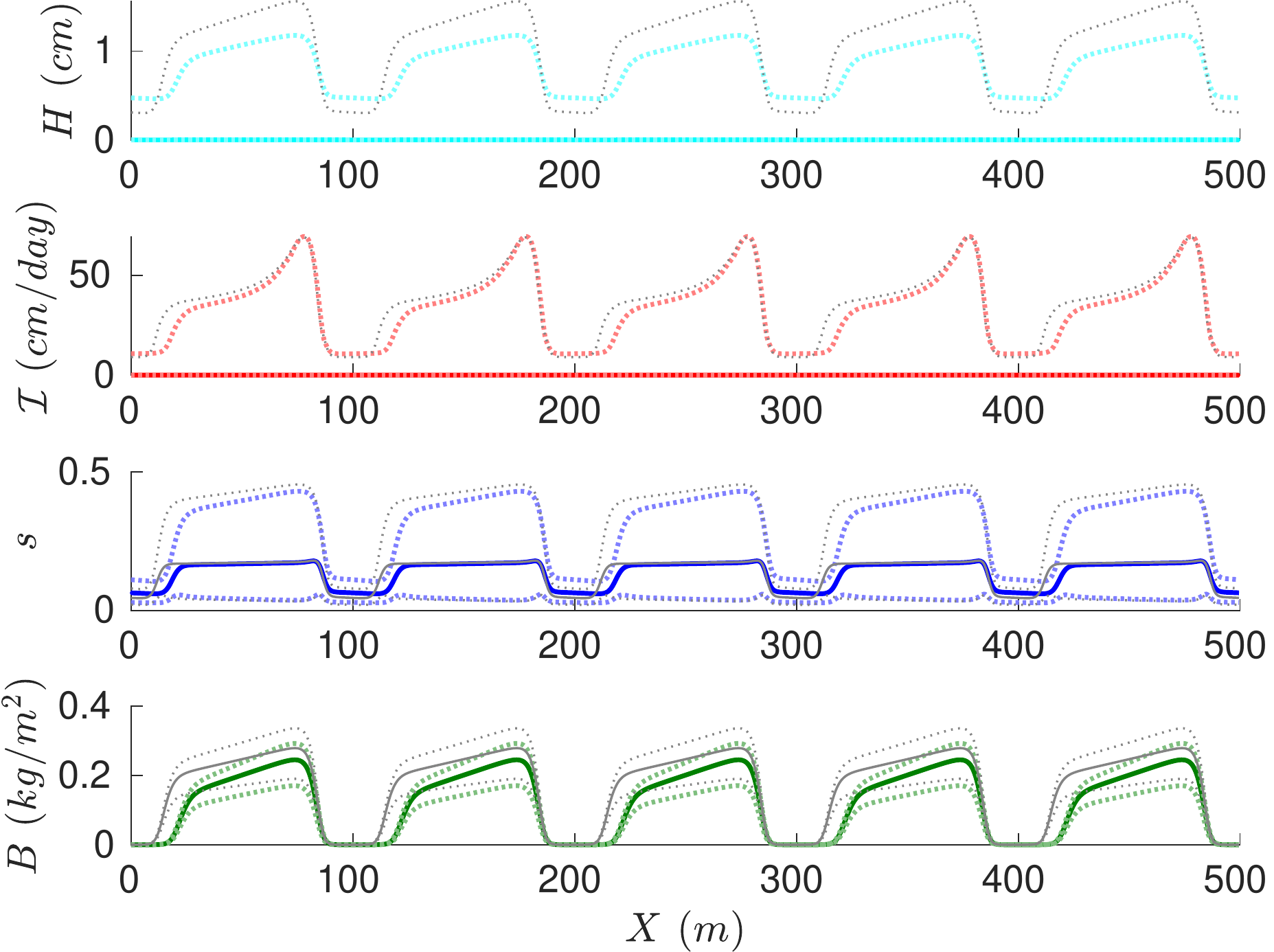}
    \caption{Nonlinear dependence of transport on surface water height ($\delta=5/3$).  Profiles at end of 200 year simulation using $\delta=5/3$ and initialized with the solution shown in \cref{fig:kn25profiles}(b) that was obtained using $\delta=1$. Colored lines show, from top to bottom: spatial profile of surface water height $H$, infiltration rate $\II$, soil moisture $s$, and biomass density $B$ at $t=200$~$years$.  The solid line is the annually averaged profile while the dotted lines show the pointwise minimum and maximum values over the course of the year. The profiles of the initial condition, a solution when $\delta=1$, are shown in light gray for reference. Parameter values used in \cref{eq:fast,eq:slow} are given by \cref{tab:nondim}, and the rain is input uniformly over two evenly-spaced six-hour rainstorms per year with mean annual precipitation of 160~$mm/year$.   }
    \label{fig:nonlinh:kn5}
\end{figure}
The theory for open channel flow suggests a nonlinear dependence of surface water height on transport, e.g. $\delta=5/3$ in \cref{eq:3field:H}.  However,  we use $\delta=1$ out of numerical convenience for the majority of simulations presented here in \cref{sec:sim}. The predictions of the linear theory for the three-field coupled model presented in \cref{fig:uv:floquet}(a) show relatively modest differences between 
the $\delta=1$ and $\delta=5/3$ cases.
The linear theory, however, is about the stability of the spatially uniform solutions and does not provide information about the fully nonlinear states.
Our limited simulations suggest that the nonlinear behavior may also be qualitatively unchanged for $\delta=5/3$, compared to $\delta=1$.
To illustrate the minor quantitative differences in periodic solutions for the two cases, we take the solution with $k=2\pi/100\ m^{-1}$ for the case of $\delta=1$, shown in \cref{fig:kn25profiles}(b), as an initial condition for a simulation with $\delta=5/3$.  A comparison of this initial condition to the final state that the simulation converges to is shown in \cref{fig:nonlinh:kn5}. The resulting periodic state exhibits slightly less biomass for $\delta=5/3$: the bands are slightly narrower with slightly lower peak values. Despite these minor quantitative differences, the overall qualitative character of the solution profiles is the same. (Another example, presented as \cref{fig:nonlinh:spatialslice} in \ref{app:sim} of the online supplement, starts with a 
random perturbation of the uniform state, and shows a much slower development of patterns for $\delta=5/3$, indicating another aspect of the numerical speed-up we achieve for $\delta=1$.)

\subsection{Dependence on mean annual precipitation}
\label{sec:sim:precip}

In this section we explore the impact on the patterns of ramping down the mean annual precipitation and then ramping it back up, with details on our protocol for this provided below. We find that the range of existence for patterns is considerably greater than suggested by the linear theory of Section~\ref{sec:linearstab}; rather than patterns only existing in the range of $109-175$ $mm/yr$, they exist as highly nonlinear states in the range of $\sim 35-201\ mm/yr$. We also find that on gradually decreasing the precipitation there are a number of discrete steps that reduce the number of bands on the $500\ m$ domain until only one band persists. However, as we then increase the precipitation, these jumps in the number of bands do not reverse themselves. Instead the single band persists but with ever increasing width until it eventually fills the domain as a spatially uniform vegetation state. We find that the average biomass decreases approximately linearly with precipitation level, but that the peak biomass values in the bands are relatively constant. 
In particular, it is the fraction coverage per wavelength that contributes most significantly to the change in average biomass on the domain, and not the peak biomass level in the bands.

\begin{figure*}
    \centering
        \includegraphics[width=0.32\textwidth]{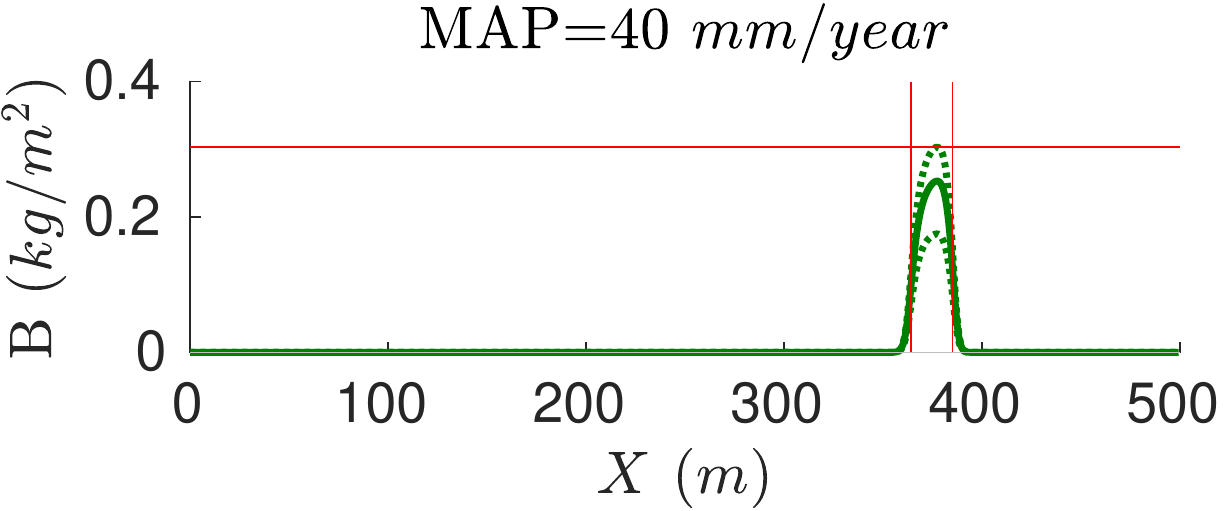}
    \includegraphics[width=0.32\textwidth]{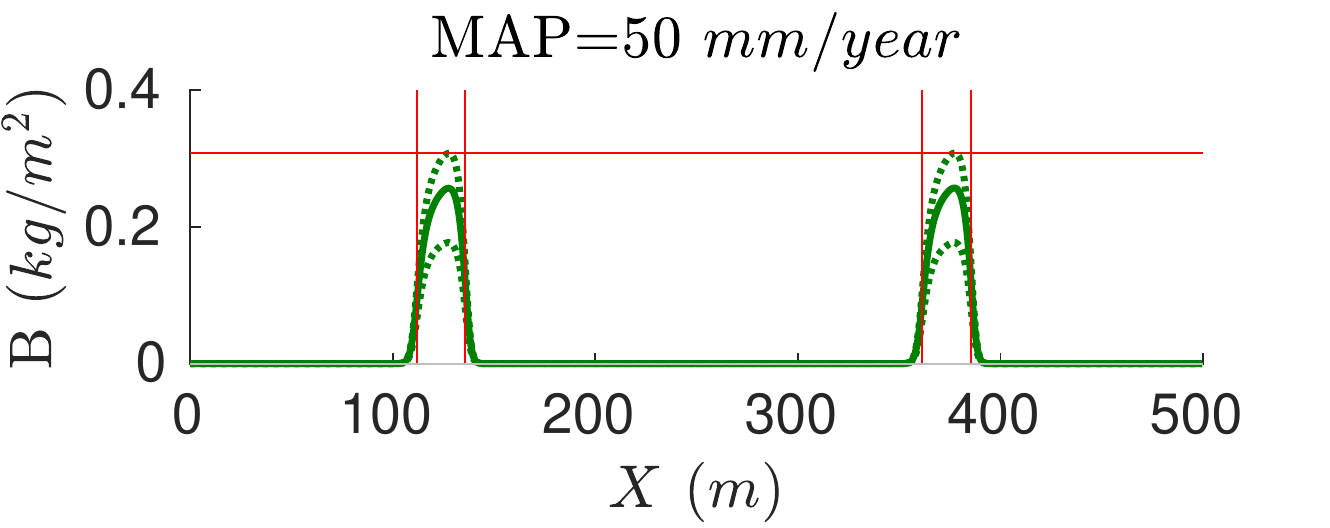}
    \includegraphics[width=0.32\textwidth]{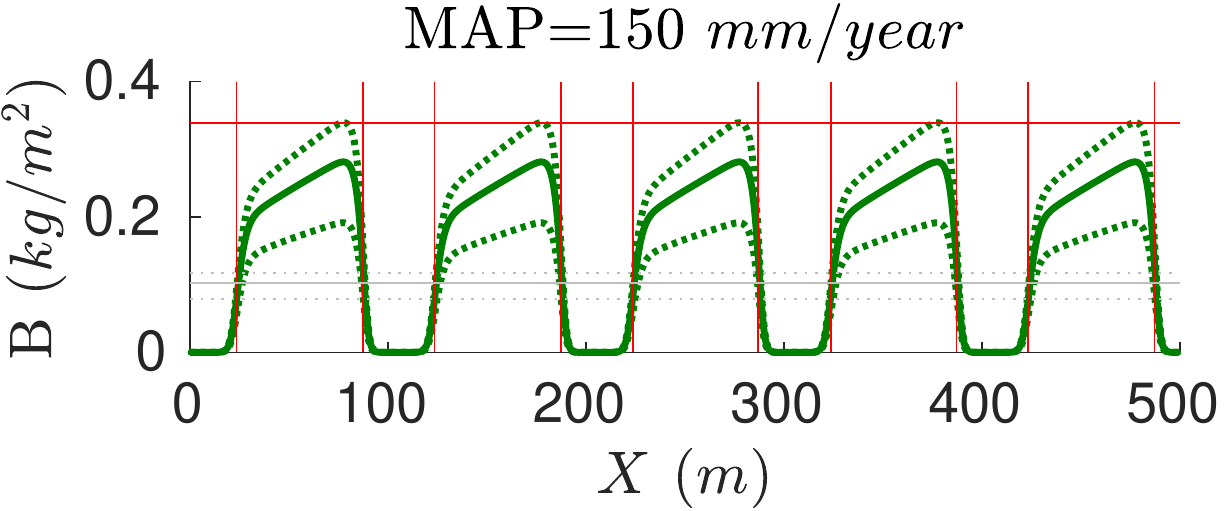}\\
    \includegraphics[width=\textwidth]{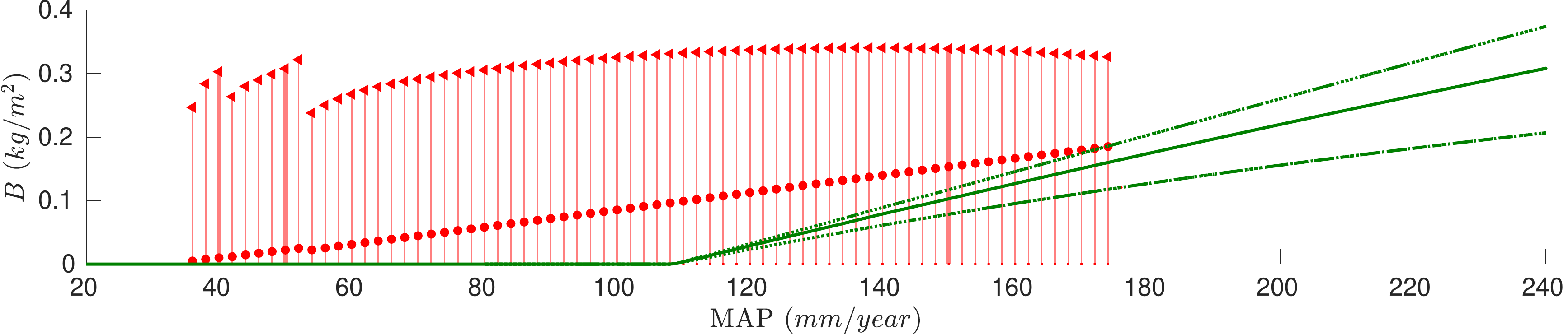}\\
    \includegraphics[width=\textwidth]{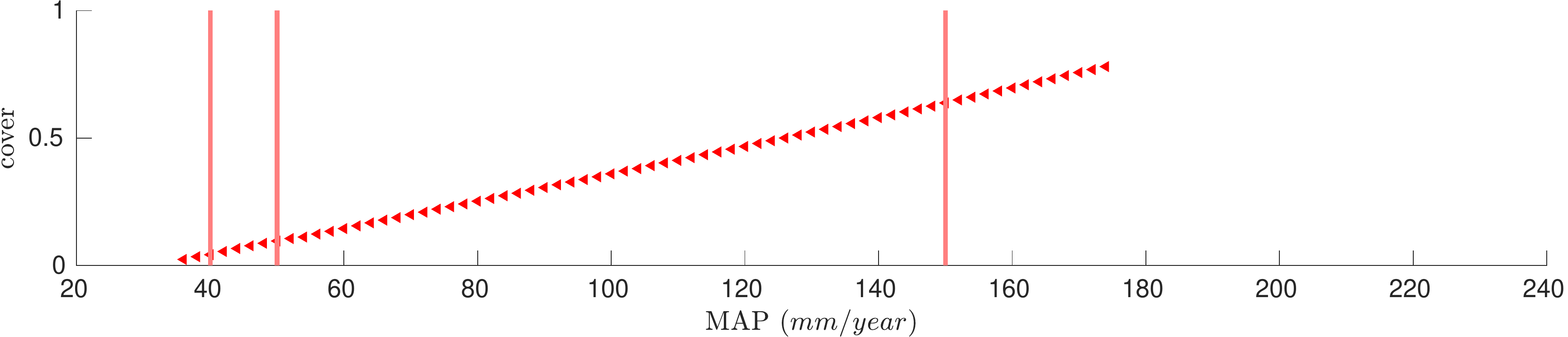}\\
    \caption{\label{fig:pscankn:down} Decreasing Precipitation. The green solid (dotted) line indicates average (minimum and maximum) values of biomass for uniformly vegetated state.  At 174 $mm/year$, this spatially uniform state is unstable to a periodic perturbation with wavenumber $k=2\pi/100$ $m^{-1}$. The red left arrow (dot) indicate the maximum (average) value of the pattern that emerges. At 52 $mm/year$ this pattern with wavelength 100 $m$ is unstable and a pattern with wavelength 250 $m$ emerges, and at 40 $mm/year$ a pattern with a single wavelength on the domain is selected.  At 34 $mm/year$ this single-wavelegnth pattern collapses to the bare soil state.  Average and min/max values of the biomass profiles are shown for the three different wavelengths. When it exists, the uniform vegetation state is included in gray for the given precipitation value. A plot of the fraction of a wavelength covered by biomass shows linearly decreasing coverage with precipitation. Parameter values used in \cref{eq:fast,eq:slow} are given by \cref{tab:nondim}, and the rain is input uniformly over two evenly-spaced six-hour rainstorms per year.  }
    
\vspace{5mm}
    \centering
     \includegraphics[width=0.32\textwidth]{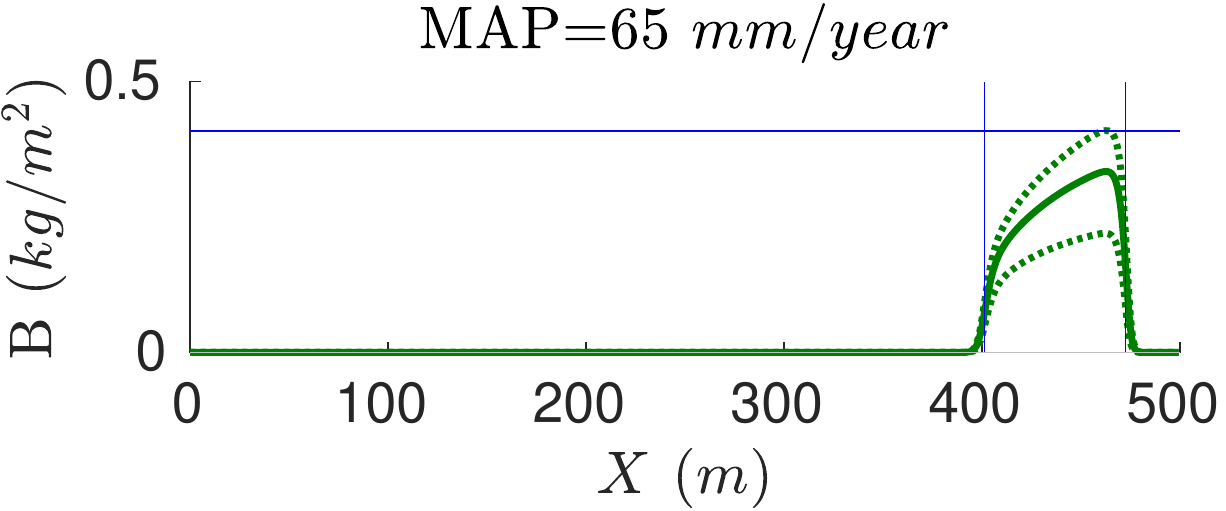}
    \includegraphics[width=0.32\textwidth]{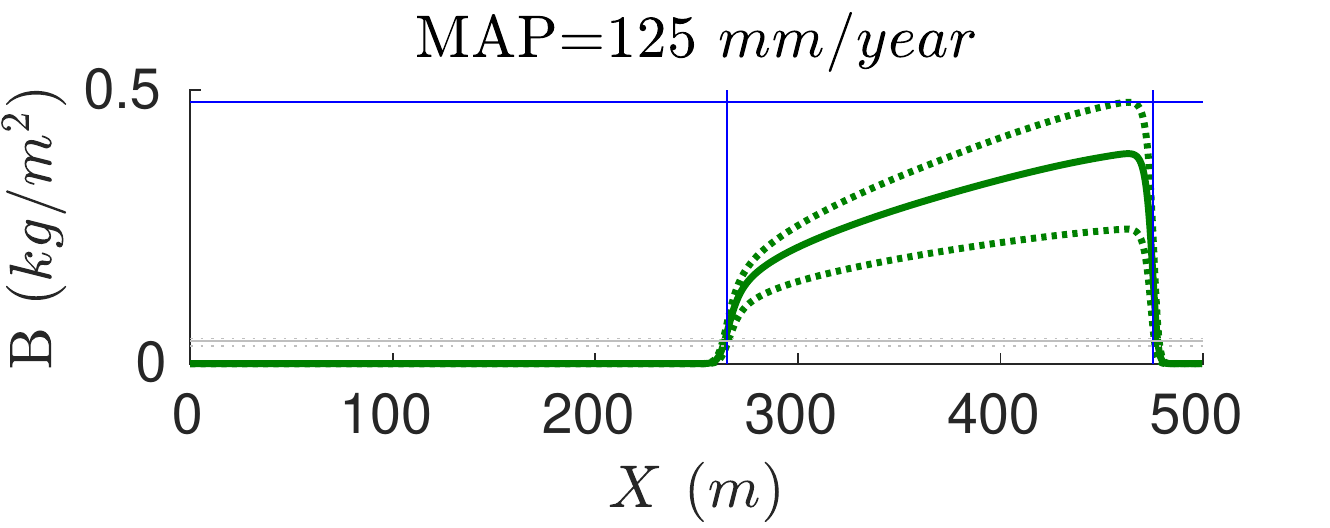}
    \includegraphics[width=0.32\textwidth]{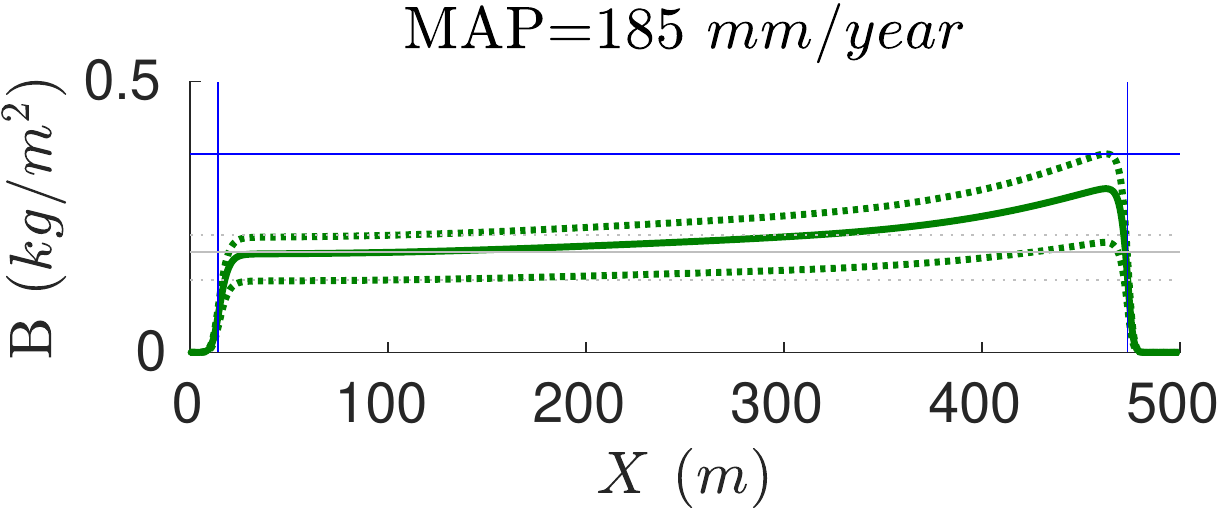}\\
    \includegraphics[width=\textwidth]{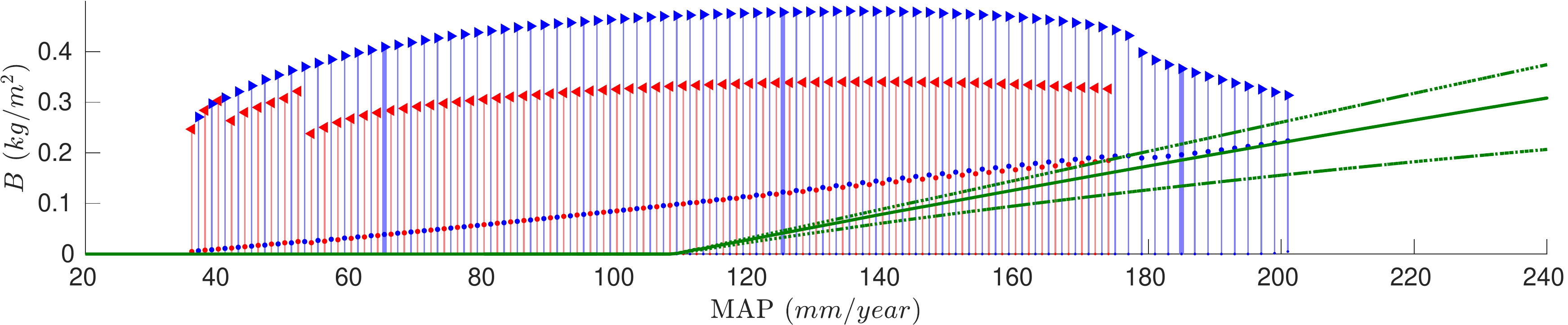}\\
   
    \caption{\label{fig:pscankn}  The green lines and red left-arrows, from \cref{fig:pscankn:down}, indicate the uniform vegetation state and the patterns obtained by decreasing precipitation with steps of 2 $mm/year$.  The blue right-arrows indicate the maximum of the biomass for increasing precipitation.  At each step of 2 $mm/year$, the state remains a single biomass pulse on the domain but with increased width. Parameter values used in \cref{eq:fast,eq:slow} are given by \cref{tab:nondim}, and the rain is input uniformly over two evenly-spaced six-hour rainstorms. }
    
\end{figure*}

Results associated with decreasing mean annual precipitation are summarized in \cref{fig:pscankn:down}.
In order to determine where the uniform state loses stability, we begin with the stable uniform vegetation state at 240 $mm/year$ and decrease the precipitation amount in steps of 2 $mm/year$. At each step, we add 1\% perturbations and allow the simulation to run for 300 years (or shorter if it converges to a uniform state more quickly).
We find that the uniform state of the fast-slow system remains stable until $174$~$mm/year$. 
Perturbing the uniform state at $174\ mm/yr$ with
1\% sinusioidal perturbations of wavenumber $k_j=2\pi j/L$ with $j=1\dots 19$, we find that the uniform state is unstable to the $3\le j\le 8$ wavenumbers, corresponding to wavelengths between about 63 and 167~$m$. This is quantitatively consistent with the predictions of linear theory for the three-field coupled model presented in \cref{sec:seasonal}.
Interestingly, the average biomass on the domain for these highly nonlinear patterned states is approximately equal to the maximum biomass obtained for the uniform vegetation state.  

We continue decreasing the mean annual precipitation  from $174\ mm/yr$, in  $2$~$mm/year$ steps, starting with the $j=5$ wavenumber pattern, corresponding to a wavelength of 100 $m$.  This state persists until the precipitation reaches 52~$mm/year$, well below the transcritical bifurcation that creates the uniform state at 109~$mm/year$.  Moreover, the patterned state sustains more biomass on average than the uniform state, for precipitation values where the uniform state does exist.  While the patterns maximum biomass value does not change very much over this range, the fraction of the domain covered by vegetation decreases linearly with precipitation, i.e. there is a linear decrease in the width of the vegetation bands with $MAP$.  At 52 $mm/year$, a pattern with 250 $m$ wavelength emerges from the simulation and transitions to a single band on the 500 $m$ domain at 40 $mm/year$.  This so-called `oasis state' finally collapses to the bare soil state at 34 $mm/year$.

As shown in Figure~\ref{fig:pscankn}, we also scan up in precipitation, starting with the oasis state at 35 $mm/year$ and moving in increments of 2 $mm/year$ as before.  This state with a single vegetation band on the domain persists all the way up to 201 $mm/year$ before loosing stability to the uniform vegetation state.  As the precipitation increases, the width of the vegetation band increases.  We note that at around 179 $mm/year$, there is a qualitative change in the shape of biomass profile, in terms of its concavity.  
Above 179 $mm/year$, the biomass values at the trailing edge of the vegetation band correspond to those of the uniform vegetation state for the given precipitation level. If we again decrease precipitation, starting now with a one-pulse solution, we find an apparent bistability of two distinct 
single-pulse  states for precipitation between approximately 175 and 177 $mm/year$, which is in a region where the uniform state is also (barely) stable. See \cref{fig:bistable1band} in \ref{app:sim} of the online supplement for more details.

\subsection{Influence of biomass feedback on infiltration and transport}
\label{sec:sim:feedback}


\begin{figure*}

    \centering
         \setlength{\unitlength}{\textheight}
    \begin{picture}(1,0.55)
    \put(-0.02,0.12){\includegraphics[width=0.25\textheight]{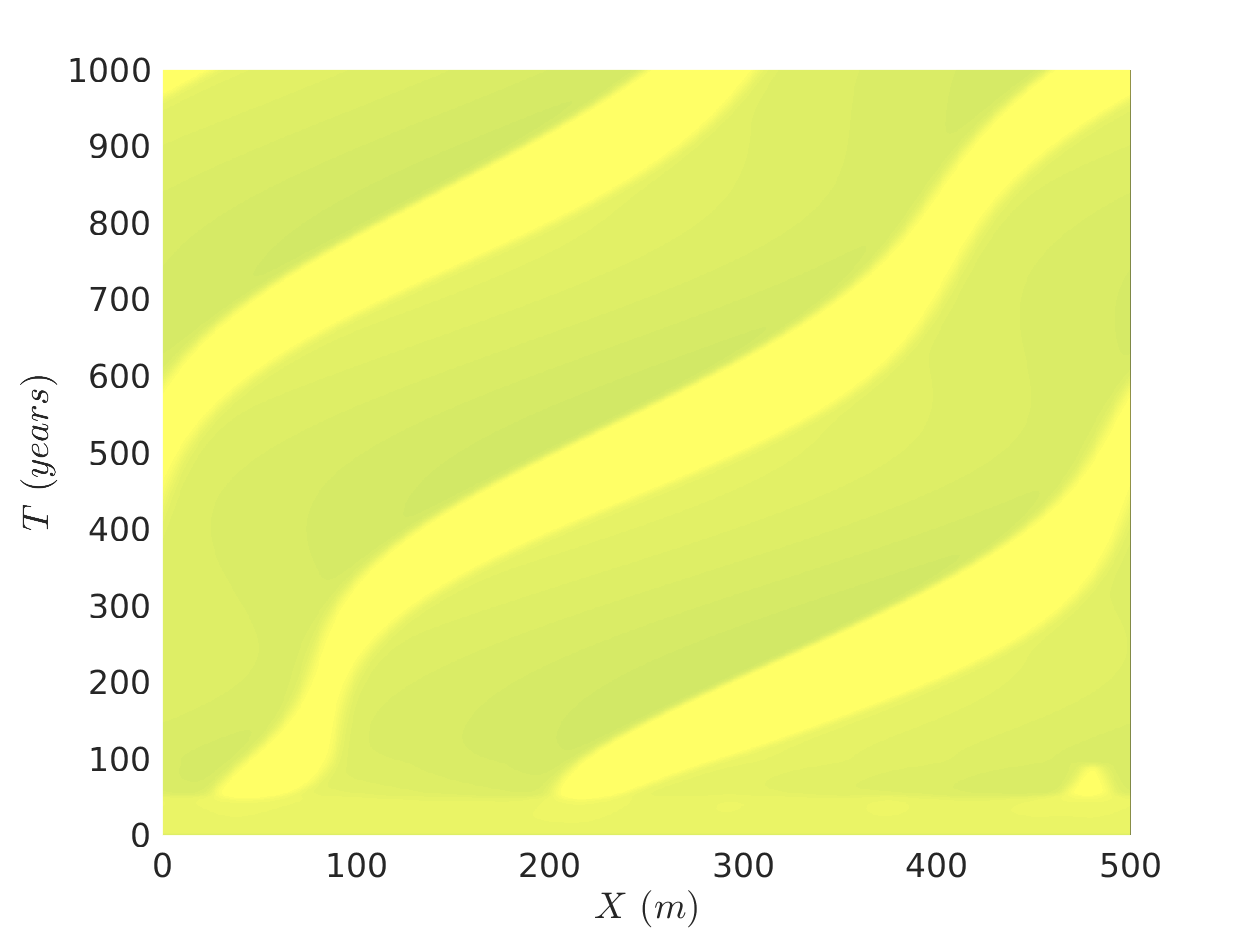}}
    \put(-0.01,0.32){\includegraphics[width=0.22\textheight]{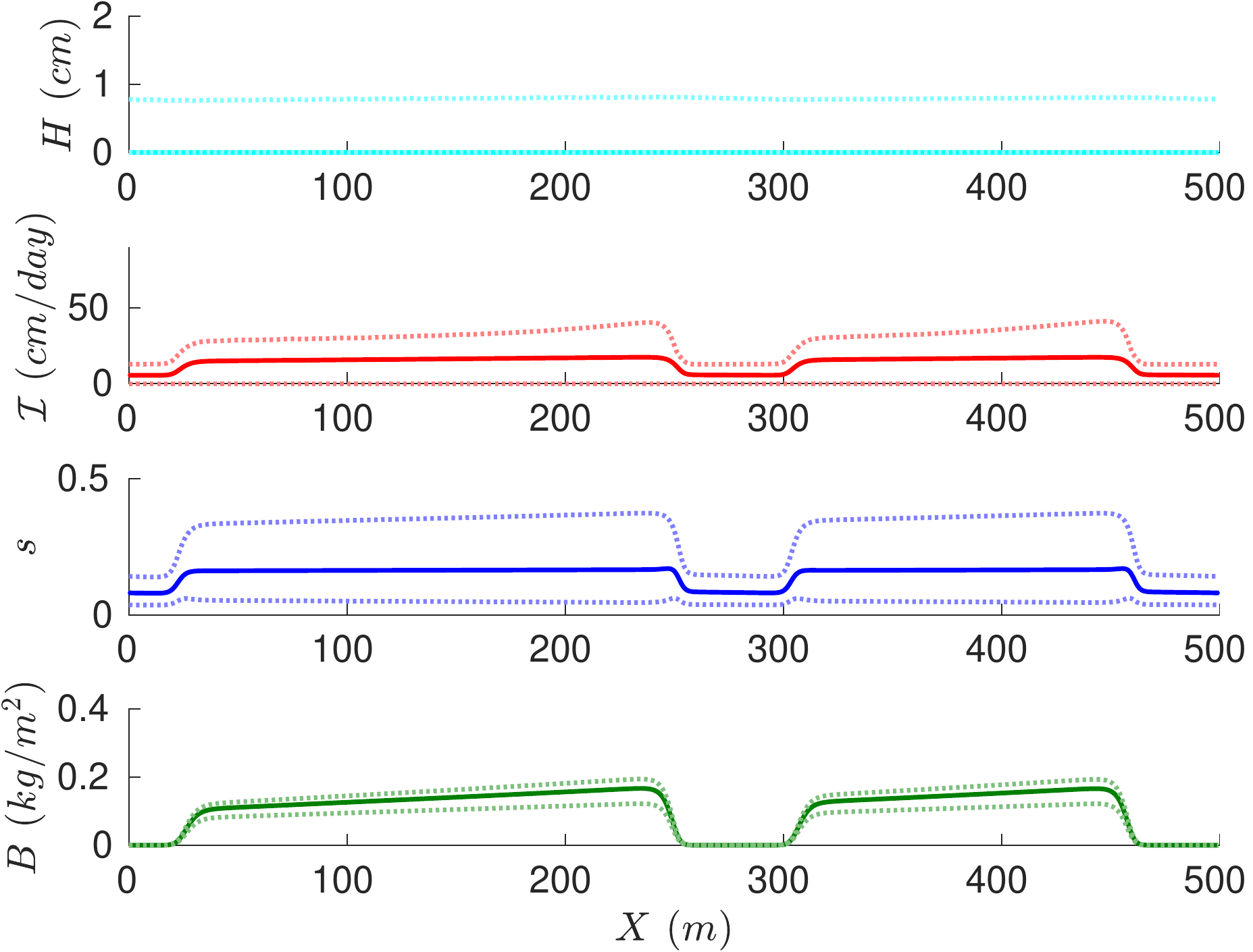}}

    \put(0.23,0.12){\includegraphics[width=0.25\textheight]{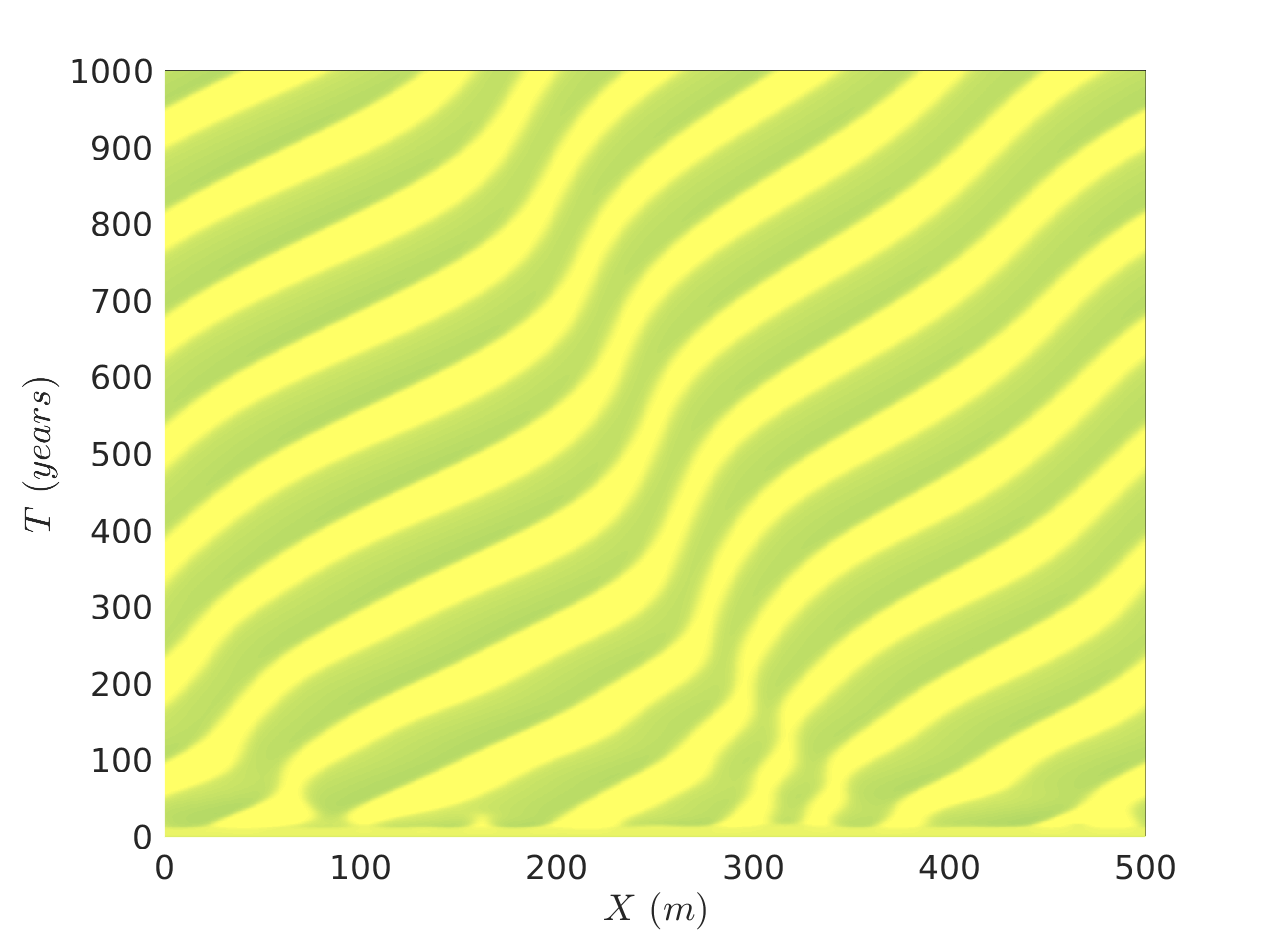}}
    \put(0.24,0.32){\includegraphics[width=0.22\textheight]{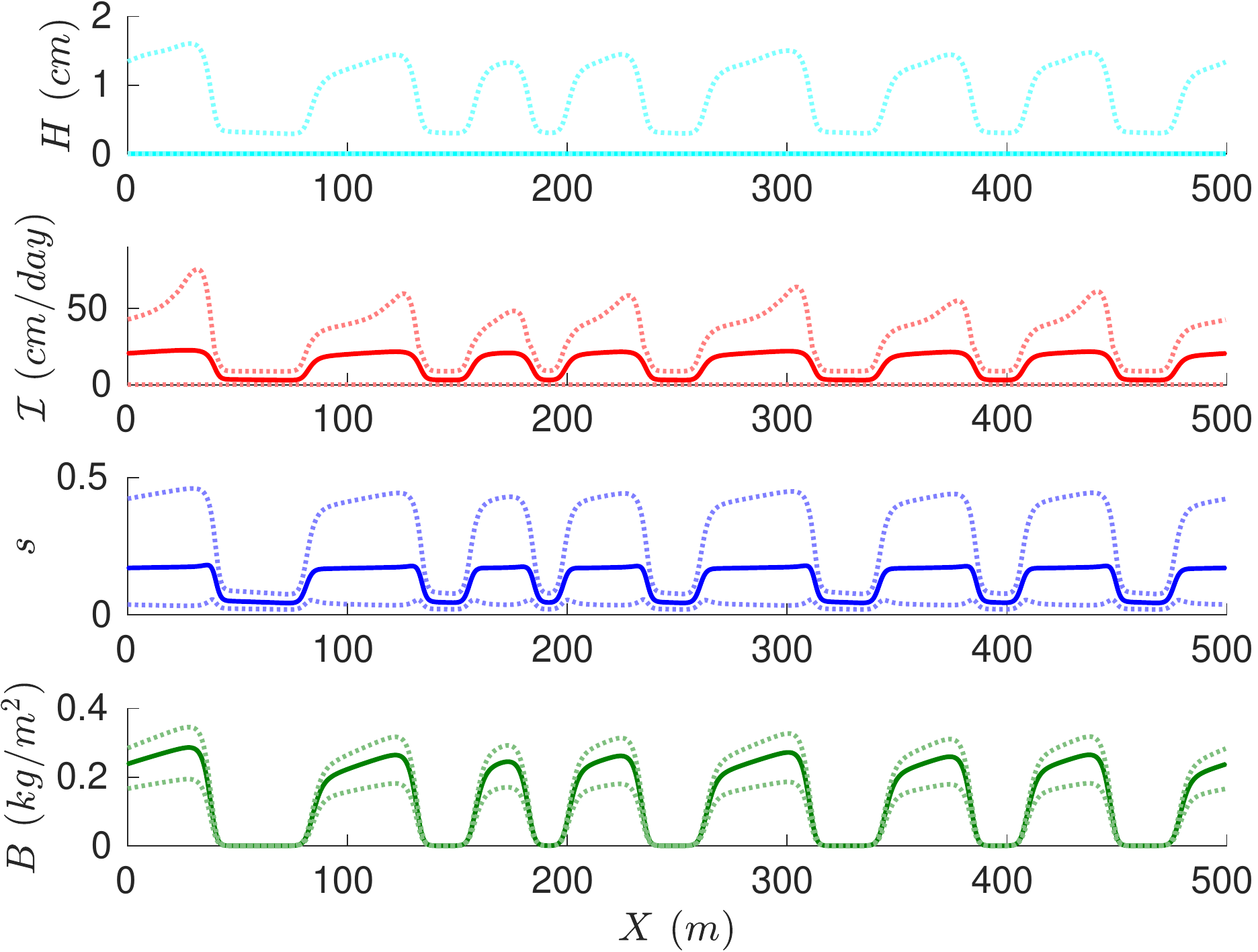}}    

    \put(0.48,0.12){\includegraphics[width=0.25\textheight]{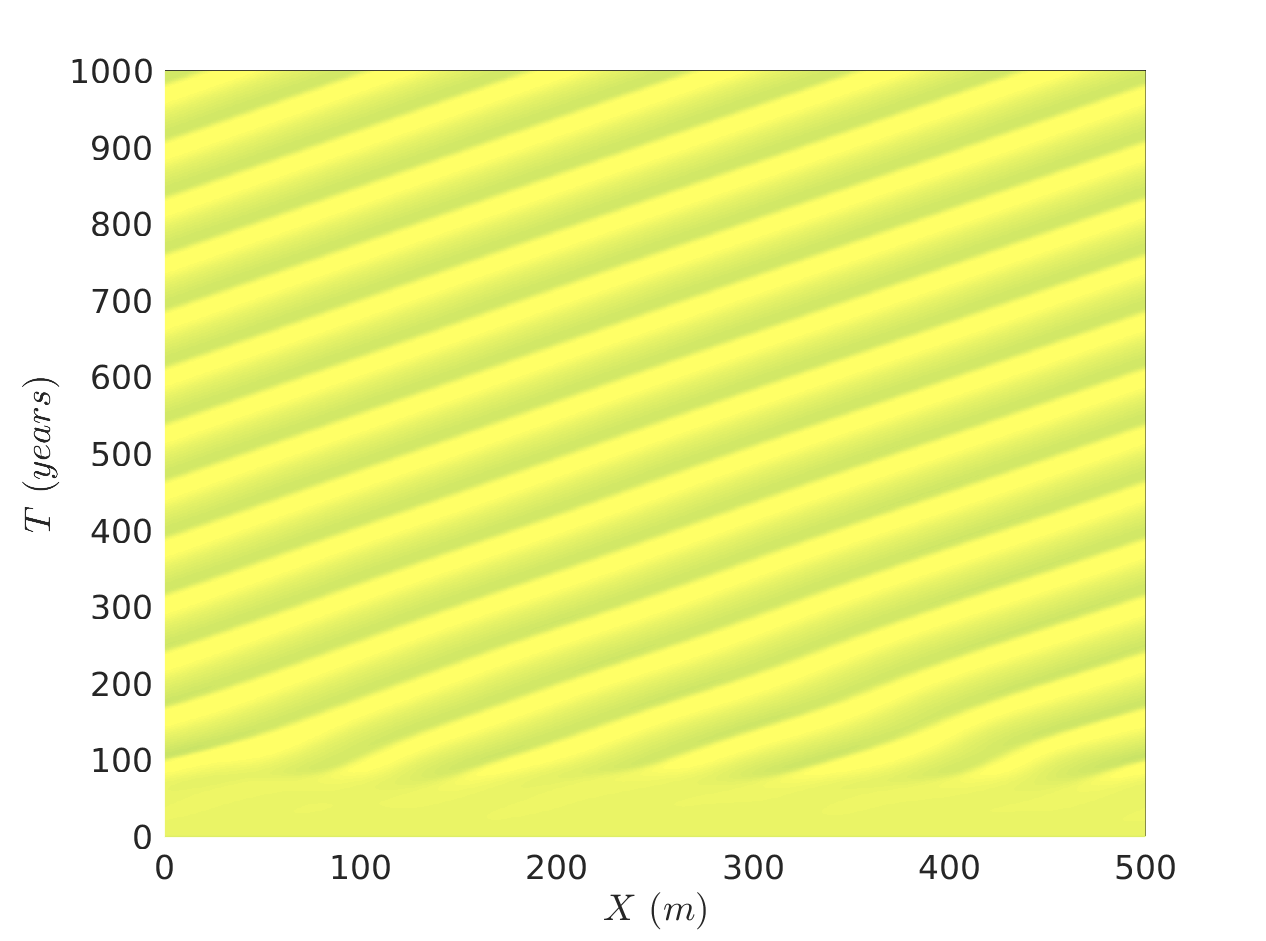}}
    \put(0.49,.32){\includegraphics[width=0.22\textheight]{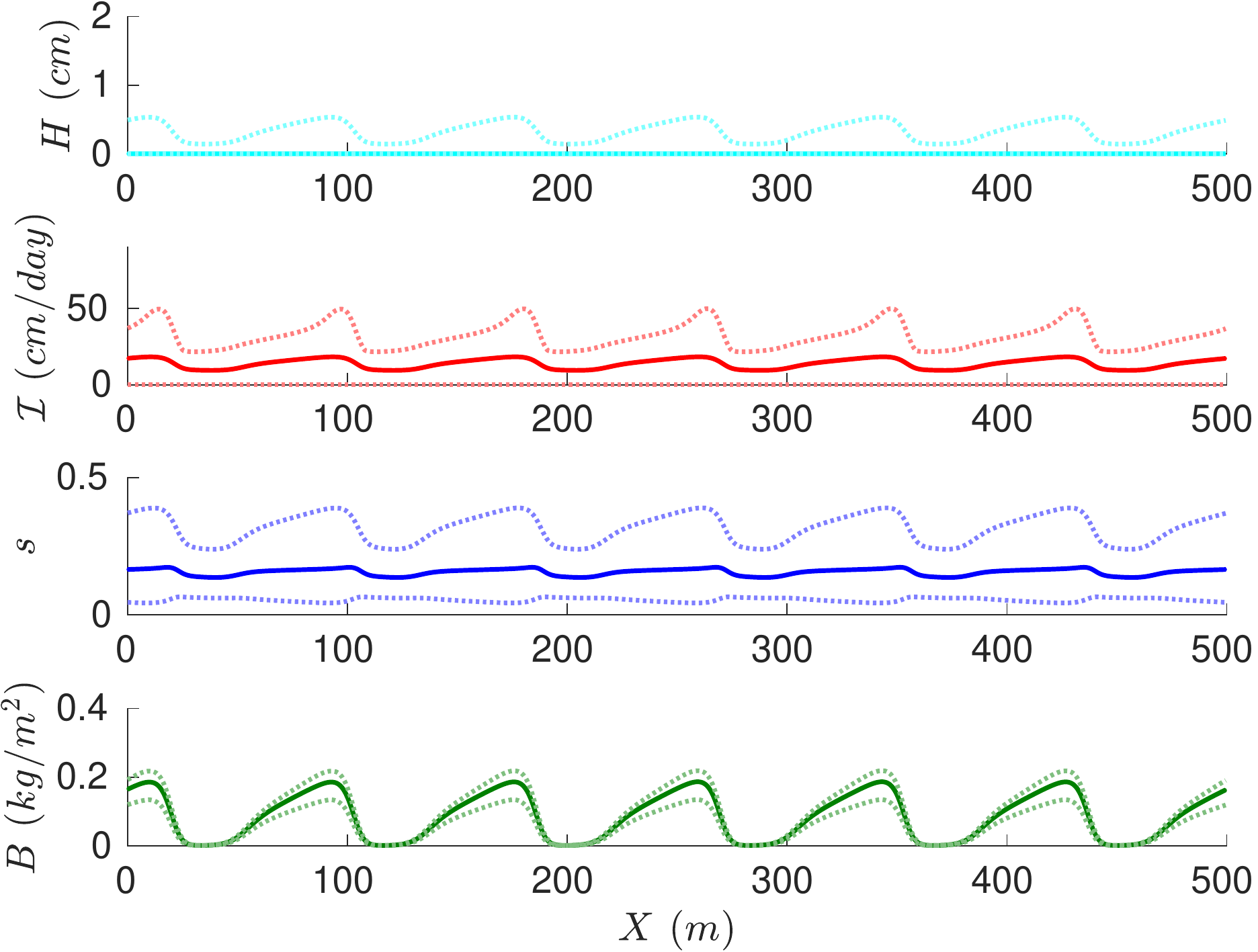}}

    \put(-0.01,0.0){\includegraphics[width=0.22\textheight]{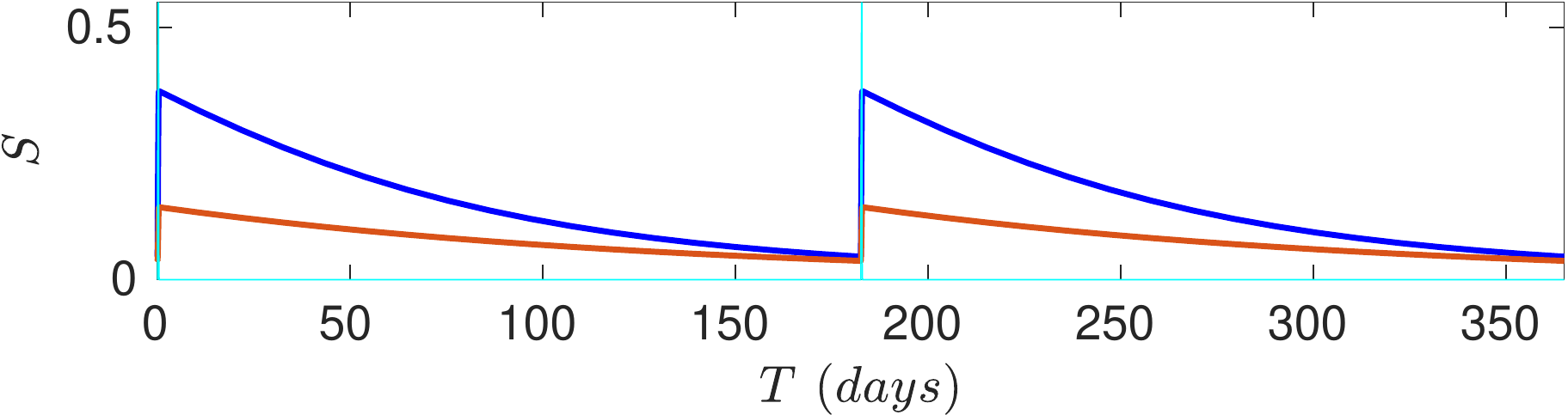}}
    
    \put(0.24,0.0){\includegraphics[width=0.22\textheight]{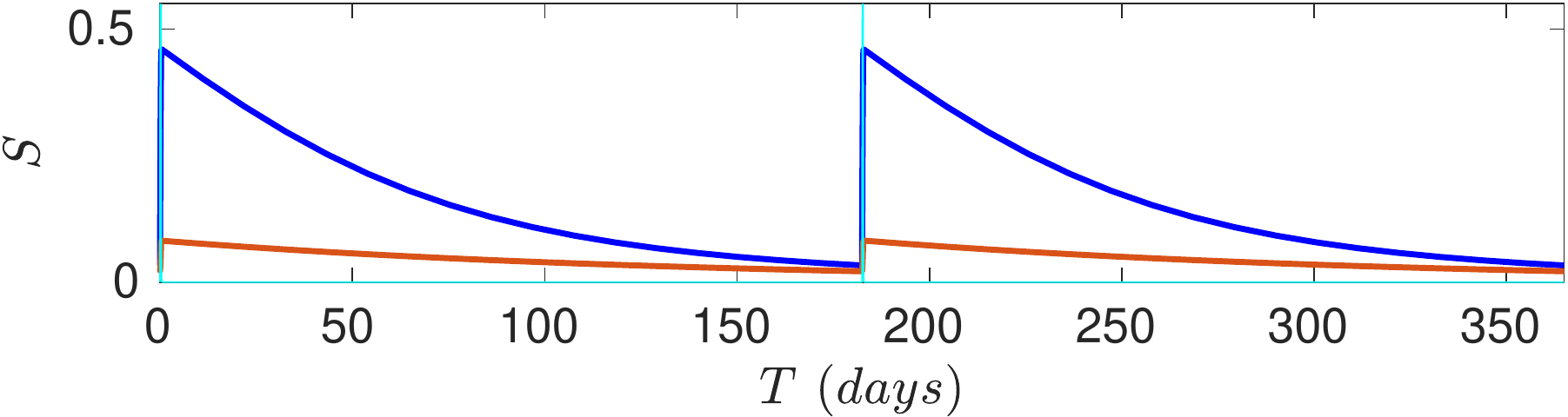}}
  
    \put(0.48,0.0){\includegraphics[width=0.22\textheight]{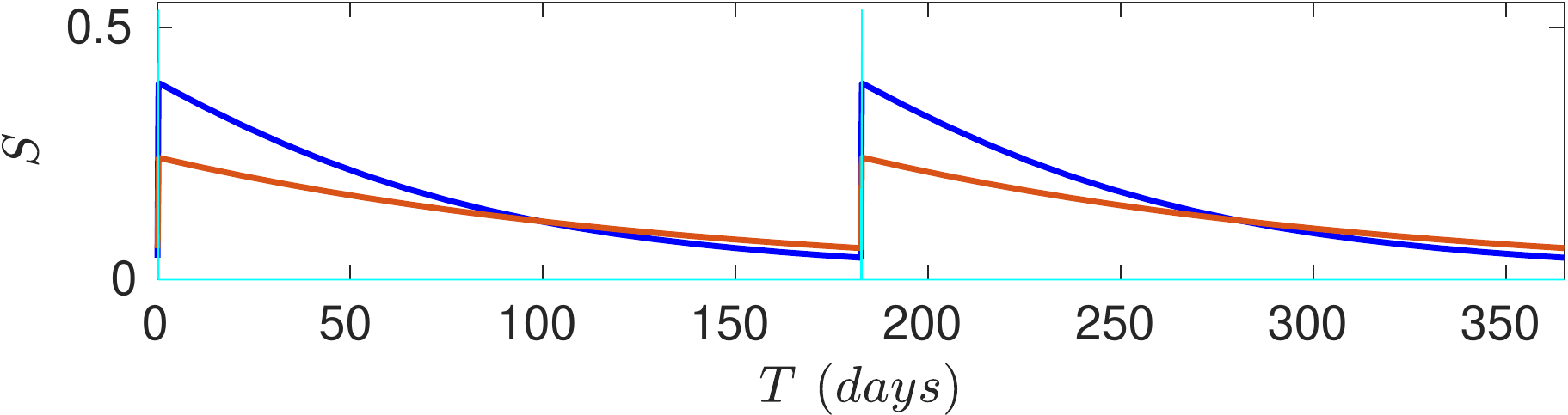}}
    
    \put(-0.02,.53){(a) transport feedback off }
    \put(-0.00,.5){infiltration feedback on}
     \put(.25,.53){(b) both feedbacks on}
     \put(0.47,.53){(c) transport feedback on   }
     \put(0.49,.5){infiltration feedback off}
     \put(0.,0.08){Corresponding soil moisture time series $s(T)$ for the bare  and vegetated soil:}
    
    \end{picture}
        \caption{Influence of biomass feedback on surface water transport and infiltration rate with mean annual precipitation of 140 $mm/year$. For each of the following choices of $(f,N)$, we show
        spatial profiles during the last year of the simulation above the spacetime plot of the biomass from a 1000-year simulation.  Below those, we present the time series of soil moisture during the last year at  the location associated with peak biomass in a vegetation band (blue) and in the center of a bare soil region (orange).  Standard parameter choices from \cref{tab:dim} are used in each case except that: (a) transport feedback is turned off, $N=0$~$m^2/kg$ and $f=0.1$ (b) standard parameters are used, $N=20$~$m^2/kg$ and $f=0.1$ (c) infiltration feedback is turned off,  $N=20$~$m^2/kg$ and $f=1$.  All other parameter values used in \cref{eq:fast,eq:slow} are given by \cref{tab:nondim} with $\delta=1$, and the rain is input uniformly over two evenly-spaced six-hour rainstorms per year.     
        }
    \label{fig:sim:infltransp}
\end{figure*}
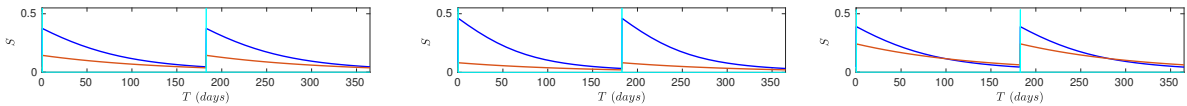

The fast component of our model~\eqref{eq:fast} incorporates two important feedbacks between the biomass and the water resource re-distribution, and in this section we explore the relative impact of those feedbacks on the pattern characteristics. The presence of biomass is known to enhance infiltration of surface water into the soil within the dryland ecosystems that exhibit banded vegetation patterns. This positive feedback plays a central role in the pattern formation process within many conceptual models~\cite{gilad2004ecosystem,Klausmeier1999,hillerislambers2001vegetation}.  The model we consider here captures an additional feedback of biomass on the hydrology, namely an increase in surface roughness that can slow down water flow where biomass is present~\cite{istanbulluoglu2005vegetation,thompson2011vegetation}. 
Our standard choice of parameters includes both kinds of biomass feedback but, as \cref{fig:sim:infltransp}  illustrates,  patterns are possible with \textit{either} feedback alone.  These results are consistent with linear predictions from the three-field coupled model shown in \cref{fig:uv:floquet}(d). 

The simulations for \cref{fig:sim:infltransp} are initialized with a random perturbation of the uniform state, and carried out with mean annual precipitation of 140 $mm/year$. The parameters characterizing the biomass feedback on surface water transport, $N$ in \cref{eq:3field:H}, and infiltration rate, $f$ in the infiltration function ${\cal I}$ in \cref{eq:3field:H,eq:3field:s}, are varied with all other parameter values taking the default parameters from  \cref{tab:dim}.  Simulation results with the standard parameters, $N=20$~$m^2/kg$ and $f=0.1$, are shown in \cref{fig:sim:infltransp}(b). These results are to be contrasted with those of columns (a) and (c) for which either the transport or the infiltration feedback is turned off by setting $N=0$ or $f=1$, respectively.  The top row shows plots of the spatial profiles of surface water, infiltration rate, soil moisture and biomass during the last year of the simulation. 
The middle row shows spacetime diagrams of the annually-averaged biomass.  The bottom plots show time series of the soil moisture at the peak biomass location 
within the vegetation band (blue) and in the center of a bare soil region (orange) during the last year of the simulation. The soil moisture plot provides a qualitative comparison for field measurements, as shown in \cref{fig:cornett}.  

We make the following observations about \cref{fig:sim:infltransp}:

\begin{itemize}
    \item{} While the width of the bare soil region is somewhat comparable between all three cases, the widths of the vegetation bands are significantly greater when there is no feedback in the transport, i.e. $N=0\ m^2/kg$. This leads to far fewer bands on the domain in this case, compared to the cases with transport feedback.
    \item{} In absence of the infiltration feedback i.e. $f=1$, the bands are more regular in spacing and travel approximately twice as fast across the domain and with uniform speed. In contrast, when there is no transport feedback, i.e. $N=0$, then we see a lag between the leading edge colonization and trailing edge death.
    \item{} The asymmetry in the profiles of the vegetation bands, between leading uphill edge and trailing downhill edge, is much more pronounced in the presence of the transport feedback. In fact, the profiles are almost triangular when the infiltration feedback is turned off, and almost rectangular when there is no transport feedback. This difference in profiles is also quite pronounced in the soil moisture profiles.
    \item{} The transport feedback is required to concentrate the surface water in the bands. Morever, the peaks of the surface water are lower
    without an infiltration feedback, and greatest when both feedbacks are present.
    \item{} The time series of the soil moisture in the lower plots shows then when there is no infiltration feedback there is a decreased  contrast in soil moisture between the bare soil and vegetation band immediately following  the rain event.  
    \item{} When there is no infiltration feedback the soil moisture at the peak of the biomass actually falls below the soil moisture in the bare soil zone. Specifically, a switching point at around 100 days after the rain event is seen in the time series of the soil moisture, where $s$ of the bare soil becomes higher than $s$ within the vegetation band.  This switching behavior is also observed in some drought phases of the field observations illustrated by \cref{fig:cornett}.
\end{itemize}

\section{Discussion}
\label{sec:discussion}

In this paper we have motivated and then investigated a fast-slow mathematical modeling framework, \cref{eq:fast,eq:slow}, for vegetation pattern formation. This fast-slow framework exploits the inherent separation of timescales between fast hydrological processes and slow biomass dynamics.  Our fast-slow model switches between a fast system, \cref{eq:fast}, that resolves aspects of the surface water and soil moisture dynamics during rain events at the fast timescale of minutes to hours, and a slow system, \cref{eq:slow}, that captures the dynamics of biomass and soil moisture interactions on the slow timescale of weeks to months when no surface water is present.  By modeling key processes on the timescales at which they occur, we are able to employ a conceptual-level model with parameter values that are consistent with observations of the processes being modeled.  
Soil moisture dynamics occur on the fast timescale, through infiltration, as well as the slow timescale, through evapotranspiration. A comprehensive set of appropriately resolved time series data on soil moisture for banded vegetation patterns would be useful to probe this modeling framework on its multiple timescales.

We do not choose parameters to fit predicted pattern characteristics to observations of banded vegetation patterns, as has been done for models that were developed for the slow timescale of the biomass dynamics alone (See, e.g. \cite{barbier2008spatial}). Even still, the fast-slow switching model is able to accurately capture certain  details of the phenomenon with hydrologically-informed parameter values.  Numerical simulation of the model reveals that the spacing of the vegetation bands, and their up-slope colonization speeds are of the right order of magnitude. Another focus of our investigation relates to the distribution of the soil moisture relative to the biomass. Our expectation, based on limited ground-based studies of this, is that the soil moisture should be trapped for much of the year where the biomass is located.
Our simulation results are consistent with this.  
However, we make a number of simplifying modeling assumptions, including to neglect subsurface water transport and the influence of lateral spreading of roots.
More careful thought must also be given to the biomass model in future work to ensure that their key processes are being captured.   

Many model studies suggest that bare-soil areas should increase in size to compensate for a decrease in precipitation level~\cite{meron2019vegetation}. The fast-slow switching model further predicts that whether this compensation occurs by adjusting the band spacing or the band width depends on whether the rainfall is increasing or decreasing. 
Our most extensive model simulation results focus on a situation where there are two heavy rain events of equal strength per year, which drive the pattern formation process. From those simulations, there is strong evidence for multi-stability, history-dependence, and hysteresis. In some of the longest simulations we gradually decrease and then increase the annual precipitation levels.  As precipitation levels decrease, a gradual decrease in band width is observed. This narrowing is accompanied by occasional losses of bands leading to increases in band spacing.  For low precipitation levels, the vegetation bands are narrow, with large swaths of bare soil in between them. As the precipitation level increases, the widths of the bands increase without any change to the wavelength of the pattern.  
These results are consistent with predictions suggested to Hemming by his detailed observations of banded patterns in Northern Somalia that he reported in 1965~\cite{hemming1965vegetation}.  In particular, he predicted that increasing rainfall might widen individual arcs while decreasing rainfall might reduce the number of arcs. 

The fast-slow modeling framework is well-suited for exploring the impact that the temporal rainfall pattern has on spatial vegetation pattern formation in these ecosystems, and this will be a major focus of future efforts.  Preliminary simulation results, which are described further in the online supplement,  indicate that the temporal distribution of rain events influences whether patterns will occur at a given level of mean annual precipitation, but more work is needed to understand this in detail.  With fewer rain events (at the same annual precipitation level), patterns occur within a smaller $MAP$ range.  For one twelve-hour rain event each year, at our standard parameters (see \cref{tab:nondim}), patterns occur at a range of $89$ to $147$ $mm/year$. With the same rain amount spread between two six hour events each year, the range grows to $34$ to $201$ $mm/year$ (\cref{fig:pscankn}).  A previous study by Guttal et al.~\cite{guttal2007self} on the effect of seasonal rain input within the three-field model by Rietkerk et al.~\cite{rietkerk2002self} found the opposite trend.  However they did not attempt to incorporate timescales associated with individual rain events and instead focused on capturing the impact of drought on the biomass dynamics.  

It may be tempting to interpret the presence of patterns as a sign of stress in the ecosystem, with their existence at  higher $MAP$ levels indicating reduced resilience. However, we caution against this. In the fast-slow model presented here, the patterned states have a higher total biomass than the corresponding uniform vegetation state for the same precipitation level. In our ramped precipitation simulations summarized by \cref{fig:pscankn}, we find that the pattern state only disappears (around 201~$mm/yr$) when it's no longer ``an advantage" to the system over the uniform vegetation case.

While the fast-slow switching model is motivated by a small parameter in a three-field coupled timescale model, \cref{eq:nondim}, we do not consider the fast-slow model as an approximation of the coupled model in the sense of singular perturbation theory. However we do show that linear predictions of the coupled model are consistent with simulations of the fast-slow model when the same temporal rain profile is used.  
The mathematical relationship between these two models will be explored at a more formal level in future work.
Our goal is to be able to start from a more detailed model that captures fast hydrological processes and formally make the reduction to a two-field model that operates on the slow timescale.  While this has been done in the high infiltration limit where there is no overland water flow~\cite{zelnik2015gradual}, capturing the influence of surface water transport on soil moisture through an effective transport could help make a direct link to models like those of Klausmeier.   Moreover, it may provide insight into how well rainfall is approximated by the inclusion of an additive constant to the soil moisture rate equation, as is commonly done, or suggest alternate approaches that may be more appropriate.

 In future work, we are particularly interested in  considering the influence of stochastic rainfall within the fast-slow switching framework. One approach for doing so, conceptually similar to the work of Siteur et al.~\cite{siteur2014will}, is to develop the interpretation that the fast system  provides a kick to the slow evolution of soil moisture and biomass.  The kicks from the fast system, under certain simplifying approximations, can be solved in closed form via the method of characteristics. 
 This may lead to progress towards analytic insights as well as an increased speed-up of numerical simulations.  Alternatively, numerical speed-up could be achieved by machine learning approaches~\cite{crompton2019emulation}, or by more parsimonious modeling of the fast system.
 
  The fast-slow framework  introduced in this paper is a conceptual model that captures the natural timescales of important processes. In particular, it resolves the fast hydrological processes that contribute important positive feedbacks between the biomass and the water resource.
 It produces physically reasonable predictions in the absence of parameter tuning and provides a testbed for exploring the impact of different precipitation regimes on patterns.  At a practical level, this framework's advantage is two-fold: (1) capturing both the fast and slow timescales in the model allows for the parameters to be chosen based on information about the particular processes being modeled, and (2) separating fast and slow processes in a switching model provides a computational speed-up over a coupled timescale model.  We also believe that there is a conceptual advantage in that the fast-slow switching framework captures the qualitatively different behaviors of the system when surface water is and is not present in a very transparent way.      
 \vspace{2mm}\\

\noindent \textbf{Acknowledgements.} The work by PG was supported in part by the National Science Foundation grant DMS-1440386 to the Mathematical Biosciences Institute. The research of MS  was supported by NSF-DMS-1517416. PG and MS also benefited from participating in the wonderful birthday conference, \textit{Advances in Pattern Formation: New Questions Motivated by Applications}, honoring Ehud Meron's pioneering work in the field of pattern formation. 

\bibliographystyle{unsrt}
\bibliography{vegetation}

\begin{thebibliography}{10}

\bibitem{noy1973desert}
I.~Noy-Meir.
\newblock Desert ecosystems: environment and producers.
\newblock {\em Annu. Rev. Ecol. Syst.}, 4(1):25--51, 1973.

\bibitem{Schwinning2004}
S.~Schwinning, O.~E. Sala, M.~E. Loik, and J.~R. Ehleringer.
\newblock {Thresholds, memory, and seasonality: understanding pulse dynamics in
  arid/semi-arid ecosystems}.
\newblock {\em Oecologia}, 141(2):191--193, 2004.

\bibitem{collins2014multiscale}
S.~L. Collins, J.~Belnap, N.~B. Grimm, J.~A. Rudgers, C.~N. Dahm, P.~D'odorico,
  M.~Litvak, D.~O. Natvig, D.~C. Peters, W.~T. Pockman, R.~L. Sinsabaugh, and
  B.~O. Wolf.
\newblock A multiscale, hierarchical model of pulse dynamics in arid-land
  ecosystems.
\newblock {\em Annu. Rev. Ecol. Evol. Syst.}, 45:397--419, 2014.

\bibitem{deblauwe2008global}
V.~Deblauwe, N.~Barbier, P.~Couteron, O.~Lejeune, and J.~Bogaert.
\newblock The global biogeography of semi-arid periodic vegetation patterns.
\newblock {\em Global Ecol. and Biogeogr.}, 17(6):715--723, 2008.

\bibitem{deblauwe2012determinants}
V.~Deblauwe, P.~Couteron, J.~Bogaert, and N.~Barbier.
\newblock Determinants and dynamics of banded vegetation pattern migration in
  arid climates.
\newblock {\em Ecol. Monogr.}, 82(1):3--21, 2012.

\bibitem{DRUSCH201225}
M.~Drusch, U.~Del Bello, S.~Carlier, O.~Colin, V.~Fernandez, F.~Gascon,
  B.~Hoersch, C.~Isola, P.~Laberinti, P.~Martimort, A.~Meygret, F.~Spoto,
  O.~Sy, F.~Marchese, and P.~Bargellini.
\newblock Sentinel-2: {ESA}'s optical high-resolution mission for {GMES}
  operational services.
\newblock {\em Remote Sensing of Environment}, 120(Supplement C):25--36, 2012.
\newblock The Sentinel Missions - New Opportunities for Science.

\bibitem{gelaro2017modern}
R.~Gelaro, W.~McCarty, M.~J. Su{\'a}rez, R.~Todling, A.~Molod, L.~Takacs, C.~A.
  Randles, A.~Darmenov, M.~G. Bosilovich, R.~Reichle, et~al.
\newblock The modern-era retrospective analysis for research and applications,
  version 2 ({MERRA}-2).
\newblock {\em J. Climate}, 30(14):5419--5454, 2017.

\bibitem{macfadyen1950vegetation}
W.~A. Macfadyen.
\newblock Vegetation patterns in the semi-desert plains of {B}ritish
  {S}omaliland.
\newblock {\em Geograph. J.}, 116(4/6):199--211, 1950.

\bibitem{greenwood1957development}
J.~E. G.~W. Greenwood.
\newblock The development of vegetation patterns in {S}omaliland
  {P}rotectorate.
\newblock {\em Geogr. J.}, pages 465--473, 1957.

\bibitem{hemming1965vegetation}
C.~F. Hemming.
\newblock Vegetation arcs in {S}omaliland.
\newblock {\em J. Ecol.}, pages 57--67, 1965.

\bibitem{boaler1964observations}
S.~B. Boaler and C.~A.~H. Hodge.
\newblock Observations on vegetation arcs in the northern region, {S}omali
  {R}epublic.
\newblock {\em J. Ecol.}, 52(3):511--544, 1964.

\bibitem{cornet1988dynamics}
A.~Cornet, J.-P. Delhoume, and C.~Montana.
\newblock Dynamics of striped vegetation patterns and water balance in the
  {C}hihuahuan {D}esert.
\newblock {\em Diversity and pattern in plant communities}, pages 221--231,
  1988.

\bibitem{dunkerley2002infiltration}
D.~L. Dunkerley.
\newblock Infiltration rates and soil moisture in a groved mulga community near
  {A}lice {S}prings, arid central {A}ustralia: evidence for complex internal
  rainwater redistribution in a runoff--runon landscape.
\newblock {\em J. Arid Environ.}, 51(2):199--219, 2002.

\bibitem{casenave1992runoff}
A.~Casenave and C.~Valentin.
\newblock A runoff capability classification system based on surface features
  criteria in semi-arid areas of {W}est {A}frica.
\newblock {\em J. Hydrol.}, 130(1-4):231--249, 1992.

\bibitem{worrall1959butana}
G.~A. Worrall.
\newblock The {B}utana grass patterns.
\newblock {\em J. Soil Sci.}, 10(1):34--53, 1959.

\bibitem{meron2015nonlinear}
E.~Meron.
\newblock {\em Nonlinear physics of ecosystems}.
\newblock CRC Press, 2015.

\bibitem{meron2016pattern}
E.~Meron.
\newblock Pattern formation--a missing link in the study of ecosystem response
  to environmental changes.
\newblock {\em Math. Biosci.}, 271:1--18, 2016.

\bibitem{meron2018patterns}
E.~Meron.
\newblock From patterns to function in living systems: {D}ryland ecosystems as
  a case study.
\newblock {\em Annu. Rev. Condens. Matter Phys.}, 9:79--103, 2018.

\bibitem{bastiaansen2019stable}
Robbin Bastiaansen, Paul Carter, and Arjen Doelman.
\newblock Stable planar vegetation stripe patterns on sloped terrain in dryland
  ecosystems.
\newblock {\em Nonlinearity}, 32(8):2759, 2019.

\bibitem{bastiaansen2019dynamics}
R.~Bastiaansen and A.~Doelman.
\newblock The dynamics of disappearing pulses in a singularly perturbed
  reaction--diffusion system with parameters that vary in time and space.
\newblock {\em Physica D}, 388:45--72, 2019.

\bibitem{bastiaansen2020effect}
R.~Bastiaansen, A.~Doelman, M.~B. Eppinga, and M.~Rietkerk.
\newblock The effect of climate change on the resilience of ecosystems with
  adaptive spatial pattern formation.
\newblock {\em Ecol. Lett.}, 2020.

\bibitem{gilad2004ecosystem}
E.~Gilad, J.~von Hardenberg, A.~Provenzale, M.~Shachak, and E.~Meron.
\newblock Ecosystem engineers: from pattern formation to habitat creation.
\newblock {\em Phys. Rev. Lett.}, 93(9):098105, 2004.

\bibitem{rietkerk2002self}
M.~Rietkerk, M.~C. Boerlijst, F.~van Langevelde, R.~HilleRisLambers, J.~van~de
  Koppel, L.~Kumar, H.~H.~T. Prins, and A.~M. de~Roos.
\newblock Self-organization of vegetation in arid ecosystems.
\newblock {\em Amer. Nat.}, 160(4):524--530, 2002.

\bibitem{istanbulluoglu2005vegetation}
E.~Istanbulluoglu and R.~L. Bras.
\newblock Vegetation-modulated landscape evolution: Effects of vegetation on
  landscape processes, drainage density, and topography.
\newblock {\em J. Geophys. Res.}, 110(F2), 2005.

\bibitem{thompson2011vegetation}
S.~E. Thompson, C.~J. Harman, P.~Heine, and G.~G. Katul.
\newblock Vegetation-infiltration relationships across climatic and soil type
  gradients.
\newblock {\em J. Geophys. Res. Biogeosci.}, 115(G2), 2011.

\bibitem{fatichi2012mechanistic}
S.~Fatichi, V.~Y. Ivanov, and E.~Caporali.
\newblock A mechanistic ecohydrological model to investigate complex
  interactions in cold and warm water-controlled environments: 1. theoretical
  framework and plot-scale analysis.
\newblock {\em J. Adv. Model. Earth Sys.}, 4(2), 2012.

\bibitem{paschalis2016matching}
A.~Paschalis, G.~G. Katul, S.~Fatichi, G.~Manoli, and P.~Molnar.
\newblock Matching ecohydrological processes and scales of banded vegetation
  patterns in semiarid catchments.
\newblock {\em Water Resour. Res.}, 52(3):2259--2278, 2016.

\bibitem{Klausmeier1999}
C.~A. Klausmeier.
\newblock Regular and irregular patterns in semiarid vegetation.
\newblock {\em Science}, 284(5421):1826--1828, 1999.

\bibitem{ursino2005influence}
N.~Ursino.
\newblock The influence of soil properties on the formation of unstable
  vegetation patterns on hillsides of semiarid catchments.
\newblock {\em Adv. Water Resour.}, 28(9):956--963, 2005.

\bibitem{guttal2007self}
V.~Guttal and C.~Jayaprakash.
\newblock Self-organization and productivity in semi-arid ecosystems:
  Implications of seasonality in rainfall.
\newblock {\em J. Theor. Biol.}, 248(3):490--500, 2007.

\bibitem{ursino2006stability}
N.~Ursino and S.~Contarini.
\newblock Stability of banded vegetation patterns under seasonal rainfall and
  limited soil moisture storage capacity.
\newblock {\em Adv. Water Resour.}, 29(10):1556--1564, 2006.

\bibitem{kletter2009patterned}
A.~Y. Kletter, J.~Von~Hardenberg, E.~Meron, and A.~Provenzale.
\newblock Patterned vegetation and rainfall intermittency.
\newblock {\em J. Theor. Biol.}, 256(4):574--583, 2009.

\bibitem{baudena2013vegetation}
M.~Baudena, J.~von Hardenberg, and A.~Provenzale.
\newblock Vegetation patterns and soil-atmosphere water fluxes in drylands.
\newblock {\em Adv. Water Resour.}, 53:131--138, 2013.

\bibitem{Dodorico2006}
P.~D'Odorico, F.~Laio, and L.~Ridolfi.
\newblock Vegetation patterns induced by random climate fluctuations.
\newblock {\em Geophys. Res. Lett.}, 33(19), 2006.

\bibitem{siteur2014will}
K.~Siteur, M.~B. Eppinga, D.~Karssenberg, M.~Baudena, M.~F.~P. Bierkens, and
  M.~Rietkerk.
\newblock How will increases in rainfall intensity affect semiarid ecosystems?
\newblock {\em Water Resour. Res.}, 50(7):5980--6001, 2014.

\bibitem{seghieri1997relationships}
J.~Seghieri, S.~Galle, J.-L. Rajot, and M.~Ehrmann.
\newblock Relationships between soil moisture and growth of herbaceous plants
  in a natural vegetation mosaic in {N}iger.
\newblock {\em J. Arid Environ.}, 36(1):87--102, 1997.

\bibitem{barbier2008spatial}
N.~Barbier, P.~Couteron, R.~Lefever, V.~Deblauwe, and O.~Lejeune.
\newblock Spatial decoupling of facilitation and competition at the origin of
  gapped vegetation patterns.
\newblock {\em Ecology}, 89(6):1521--1531, 2008.

\bibitem{kinast2014interplay}
S.~Kinast, Y.~R. Zelnik, G.~Bel, and E.~Meron.
\newblock Interplay between turing mechanisms can increase pattern diversity.
\newblock {\em Phys. Rev. Lett.}, 112(7):078701, 2014.

\bibitem{ursino2007modeling}
N~Ursino.
\newblock Modeling banded vegetation patterns in semiarid regions:
  Interdependence between biomass growth rate and relevant hydrological
  processes.
\newblock {\em Water Resour. Res.}, 43(4), 2007.

\bibitem{vereecken2019infiltration}
H.~Vereecken, L.~Weiherm{\"u}ller, S.~Assouline, J.~{\v{S}}im{\u u}nek,
  A.~Verhoef, M.~Herbst, N.~Archer, B.~Mohanty, C.~Montzka, J.~Vanderborght,
  et~al.
\newblock Infiltration from the pedon to global grid scales: An overview and
  outlook for land surface modeling.
\newblock {\em Vadose Zone J.}, 18(1), 2019.

\bibitem{rigby2006simplified}
J.~R. Rigby and A.~Porporato.
\newblock Simplified stochastic soil-moisture models: a look at infiltration.
\newblock {\em Hydrol. Earth Syst. Sci.}, (10):861--871, 2006.

\bibitem{gilad2007mathematical}
E.~Gilad, J.~von Hardenberg, A.~Provenzale, M.~Shachak, and E.~Meron.
\newblock A mathematical model of plants as ecosystem engineers.
\newblock {\em J. Theor. Biol.}, 244(4):680--691, 2007.

\bibitem{chow1959open}
V.~T. Chow.
\newblock Open-channel hydraulics.
\newblock In {\em Open-channel hydraulics}. McGraw-Hill, 1959.

\bibitem{bonetti2017manning}
Sara Bonetti, Gabriele Manoli, Costantino Manes, Amilcare Porporato, and
  Gabriel~G Katul.
\newblock Manning’s formula and strickler’s scaling explained by a
  co-spectral budget model.
\newblock {\em Journal of Fluid Mechanics}, 812:1189--1212, 2017.

\bibitem{rodriguez2007ecohydrology}
I.~Rodr{\'\i}guez-Iturbe and A.~Porporato.
\newblock {\em Ecohydrology of water-controlled ecosystems: soil moisture and
  plant dynamics}.
\newblock Cambridge University Press, 2007.

\bibitem{hillerislambers2001vegetation}
R.~HilleRisLambers, M.~Rietkerk, F.~van~den Bosch, H.~H.~T. Prins, and
  H.~de~Kroon.
\newblock Vegetation pattern formation in semi-arid grazing systems.
\newblock {\em Ecology}, 82(1):50--61, 2001.

\bibitem{mauchamp1994simulating}
A.~Mauchamp, S.~Rambal, and J.~Lepart.
\newblock Simulating the dynamics of a vegetation mosaic: a spatialized
  functional model.
\newblock {\em Ecol. Model.}, 71(1-3):107--130, 1994.

\bibitem{gandhi2019vegetation}
P.~Gandhi, S.~Iams, S.~Bonetti, and M.~Silber.
\newblock Vegetation pattern formation in drylands.
\newblock {\em arXiv preprint arXiv:1905.05115}, 2019.

\bibitem{meiss2017}
J.D. Meiss.
\newblock {\em Differentiable Dynamical Systesm}.
\newblock SIAM, 2017.

\bibitem{busse1978non}
FH~Busse.
\newblock Nonlinear properties of thermal convection.
\newblock {\em Rep.Prog. Phys.}, 41(12):1929, 1978.

\bibitem{shampine1997matlab}
L.~F. Shampine and M.~W. Reichelt.
\newblock The matlab ode suite.
\newblock {\em SIAM J, Sci. Comput.}, 18(1):1--22, 1997.

\bibitem{gowda2018signatures}
K.~Gowda, S.~Iams, and M.~Silber.
\newblock Signatures of human impact on self-organized vegetation in the horn
  of africa.
\newblock {\em Sci. Rep.}, 8(1):3622, 2018.

\bibitem{bastiaansen2018multistability}
R.~Bastiaansen, O.~Ja{\"\i}bi, V.~Deblauwe, M.~B. Eppinga, K.~Siteur, E.~Siero,
  S.~Mermoz, A.~Bouvet, A.~Doelman, and M.~Rietkerk.
\newblock Multistability of model and real dryland ecosystems through spatial
  self-organization.
\newblock {\em Proc. Natl. Acad. Sci.}, 115(44):11256--11261, 2018.

\bibitem{meron2019vegetation}
E.~Meron.
\newblock Vegetation pattern formation: The mechanisms behind the forms.
\newblock {\em Physics Today}, 72(11):30--37, 2019.

\bibitem{zelnik2015gradual}
Y.~R. Zelnik, E.~Meron, and G.~Bel.
\newblock Gradual regime shifts in fairy circles.
\newblock {\em Proc. Nat. Acad. Sci.}, 112(40):12327--12331, 2015.

\bibitem{crompton2019emulation}
O.~Crompton, A.~Sytsma, and S.~Thompson.
\newblock Emulation of the saint venant equations enables rapid and accurate
  predictions of infiltration and overland flow velocity on spatially
  heterogeneous surfaces.
\newblock {\em Water Resour. Res.}, 55(8):7108--7129, 2019.

\end{thebibliography}
 
\cleardoublepage

\appendix
\renewcommand{\thefigure}{A\arabic{figure}}
\setcounter{figure}{0}
\setcounter{page}{1}

{\Large \noindent A Fast-Slow Model of Banded Vegetation Pattern Formation in Drylands:}
 
	\hfill{\Large Supplemental Information }
\vspace{1cm}

{\hspace{5mm} \large Punit Gandhi\footnote{Department of Mathematics and Applied Mathematics, Virginia Commonwealth University, Richmond, VA 23284, USA}, 
	Sara Bonetti\footnote{Department of Environmental Systems Science, ETH Z\"urich, 8092 Z\"urich, Switzerland}, 
	Sarah Iams\footnote{John A. Paulson School of Engineering and Applied Sciences, Harvard University, Cambridge, MA 02138, USA}, 
	Amilcare Porporato\footnote{Department of Civil and Environmental Engineering and Princeton Environmental Studies, Princeton University, Princeton, NJ 08540, USA}, 
	Mary Silber\footnote{Department of Statistics and Committee on Computational and Applied Mathematics, University of Chicago, Chicago, IL 60637, USA}}

\vspace{2cm}

This supplement is divided into two parts: \ref{app:sim} provides additional simulation results in support of the main text, and \ref{app:linstab} provides details of the linear stability calculations presented in the main text.  In what follows, references labels that begin with ``A" and ``B" correspond to equations and figures contained within this supplement.  All other labels correspond to equations, figures, tables, etc. in the main text.

\section{Additional simulation results}
\label{app:sim}
We present additional simulation results to show more details of certain interesting aspects of the model, predictions and to support comments made in \cref{sec:sim,sec:discussion} of the main text.  

\subsection{Subsurface water transport}
We neglect subsurface water transport throughout the main text. In this section we show results from limited simulations  indicating that including linear diffusion of soil moisture does not qualitatively change the results. We fix the value of the diffusion coefficient associated with subsurface transport, which is not very well constrained, and further parameter exploration is required.  
     
In order to model subsurface transport, we consider the addition of a linear diffusion term of the form $\delta_s \frac{\partial^2 s}{ \partial x^2}$  in the soil moisture equation~\eqref{eq:slow:s} of the slow system in the fast-slow switching model, \cref{eq:fast,eq:slow}.  The value of the coefficient $\delta_s$, like the biomass diffusion constant $\delta_b$ in~\cref{eq:slow:b}, is largely unconstrained. We assume $\delta_s=10\delta_b$ based on previous modeling efforts that take the ratio of these coefficients to be between 1  and 100 [Rietkerk et al. (2002)  Amer. Nat. 160(4):524–530; Gilad et al. (2004) Phys. Rev. Lett. 93(9):098105].

\begin{figure*}
	\centering
	\includegraphics[width=\textwidth]{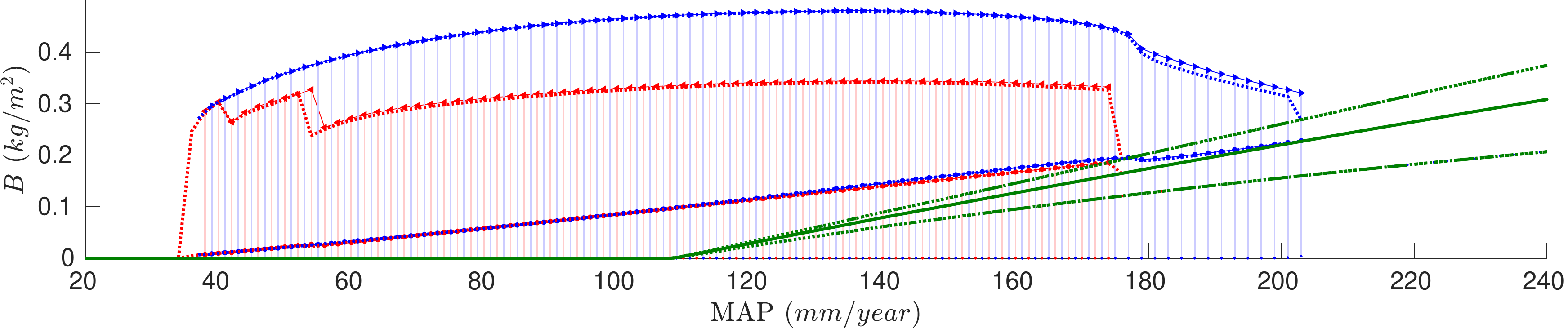}
	\caption{Including Soil Moisture diffusion.  The solid (dotted) green lines indicate the average (maximum and minimum) biomass values of the uniform vegetation state. The red left arrows indicate the maximum biomass values of two-band patterns obtained by decreasing precipitation with steps of 2 $mm/year$.   The blue right-arrows indicate the maximum biomass values for single-band patterns for increasing precipitation.  In both cases the dots indicate the (time and space) averaged biomass values, and the vertical lines indicate the range of biomass values within the pattern.  The corresponding results from \cref{fig:pscankn} are indicated by the blue and red dotted lines.  }
	\label{fig:pscan:soildiff}
\end{figure*}

We carry out the simulation shown in \cref{fig:pscankn:down,fig:pscankn} in which precipitation is slowly decreased and then increased with soil moisture diffusion, and the results are shown in \cref{fig:pscan:soildiff}.  For reference, the dotted lines in this figure show the maximum and average biomass values of the patterned states from the simulation with no soil moisture diffusion. There is no qualitative difference, and the simulations are only very slightly qualitatively different.    

\subsection{Random initial conditions and asymptotic behavior}

\begin{figure*}[h]
    \centering
    \setlength{\unitlength}{\linewidth}
    \begin{picture}(1,0.35)
    \put(0.01,0){\includegraphics[width=0.49\linewidth]{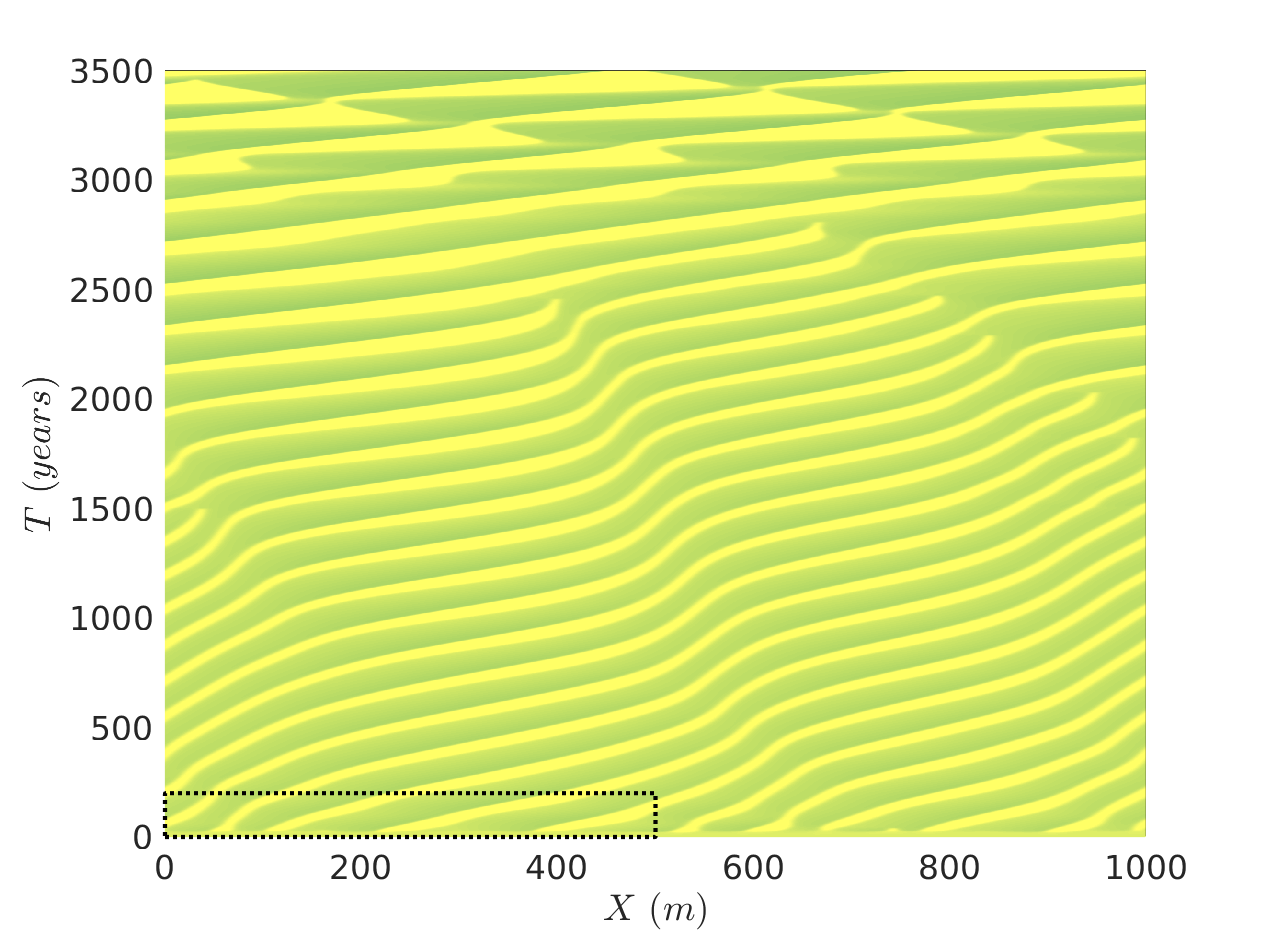}}
    \put(0.51,0.0){\includegraphics[width=0.49\linewidth]{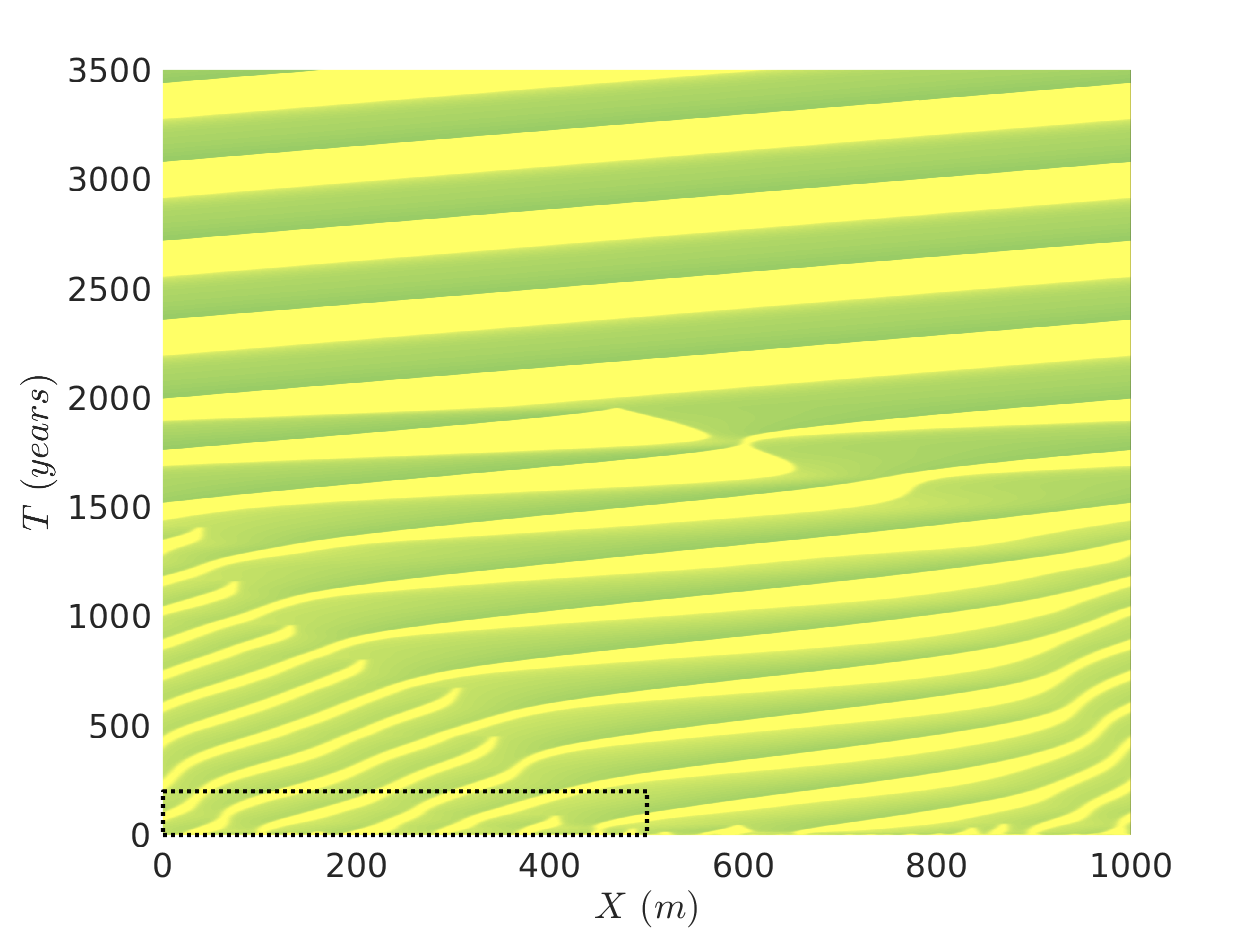}}
     \put(0,.33){(a)}
    \put(.5,.33){(b)}
    \end{picture}
    \caption{spacetime diagram of biomass from simulations initialized with 1\% random perturbation of (a) uniform vegetation state (b) bare soil state. For scale, the spacetime simulation window of \cref{fig:sim:spatialslice}(b) is indicated by a dotted box.
    }
    \label{fig:asymp:b}
\end{figure*}

\Cref{fig:asymp:b} illustrates the asymptotic behavior for two different random initial conditions on a one-kilometer spatial domain. The simulation shown in \cref{fig:asymp:b}(a) is initialized with a 1\% random noise on top of the predicted uniform vegetation state for constant precipitation of $P_0=160$ $mm/year$, given by \cref{eq:ss:uv}.  The simulation shown in \cref{fig:asymp:b}(b) is initialized with a 1\% random noise  on bare soil.
 
At approximately year 1500 the pattern formed in \cref{fig:asymp:b}(a) begins undergoing a series of band-merging events.  At around 2500 years, the bands begin to exhibit a periodic breathing dynamic in which the width of the bands oscillate as the trailing edge periodically moves downhill for a short time.  The  simulation in \cref{fig:asymp:b}(b) is initialized with the same perturbation of the bare soil state, given by \cref{eq:ss:bs}.  In this case the band merging events are initiated sooner, but the transient still persists for around two millennia before eventually reaching a pattern with a single band on the domain.  

\subsection{Nonlinear dependence of transport on surface water height}
\begin{figure*}
    \centering    
\setlength{\unitlength}{\linewidth}
\begin{picture}(1,0.65)
\put(0,0){\includegraphics[width=0.49\linewidth]{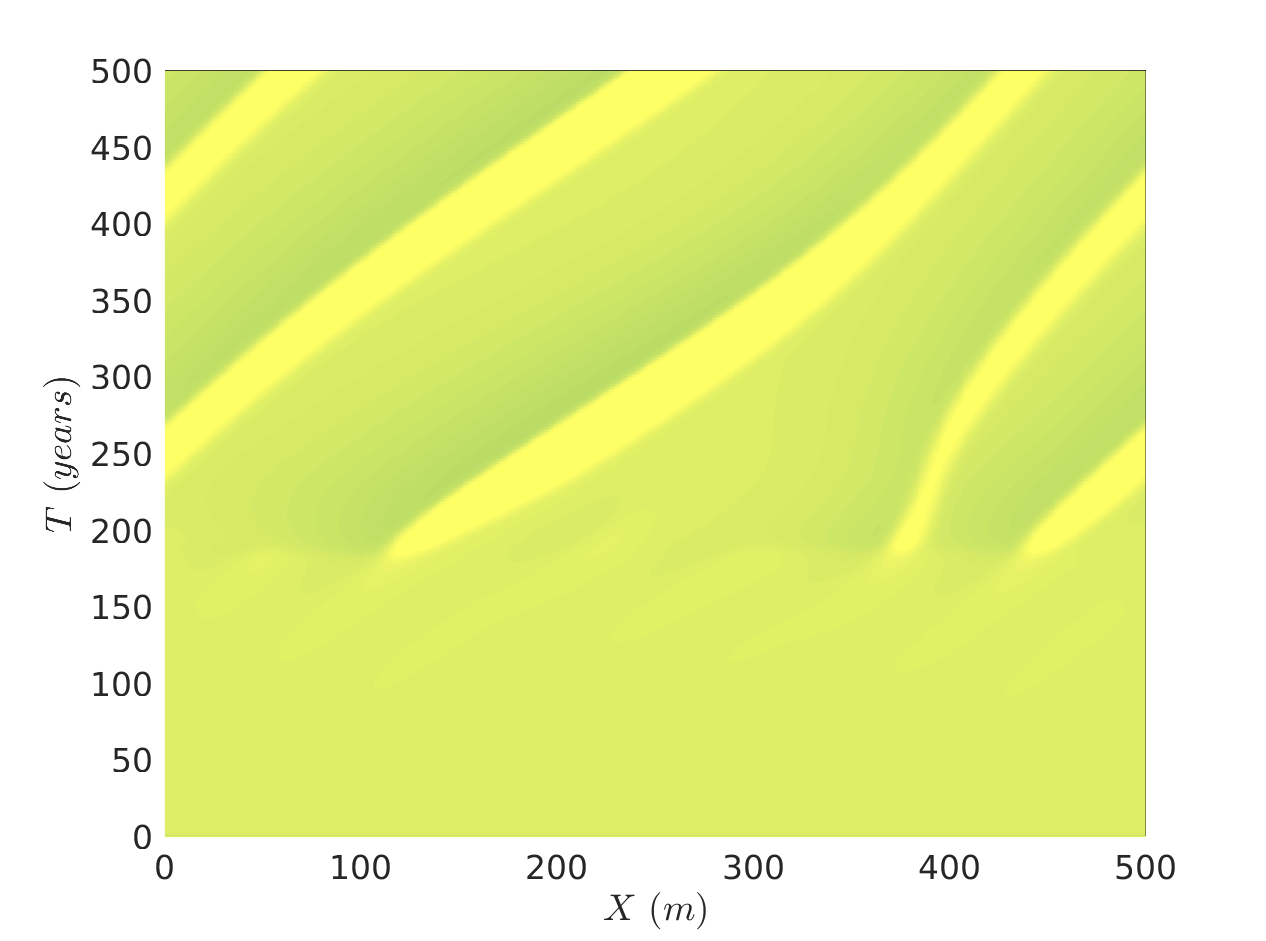}}
\put(0.5,0){\includegraphics[width=0.49\linewidth]{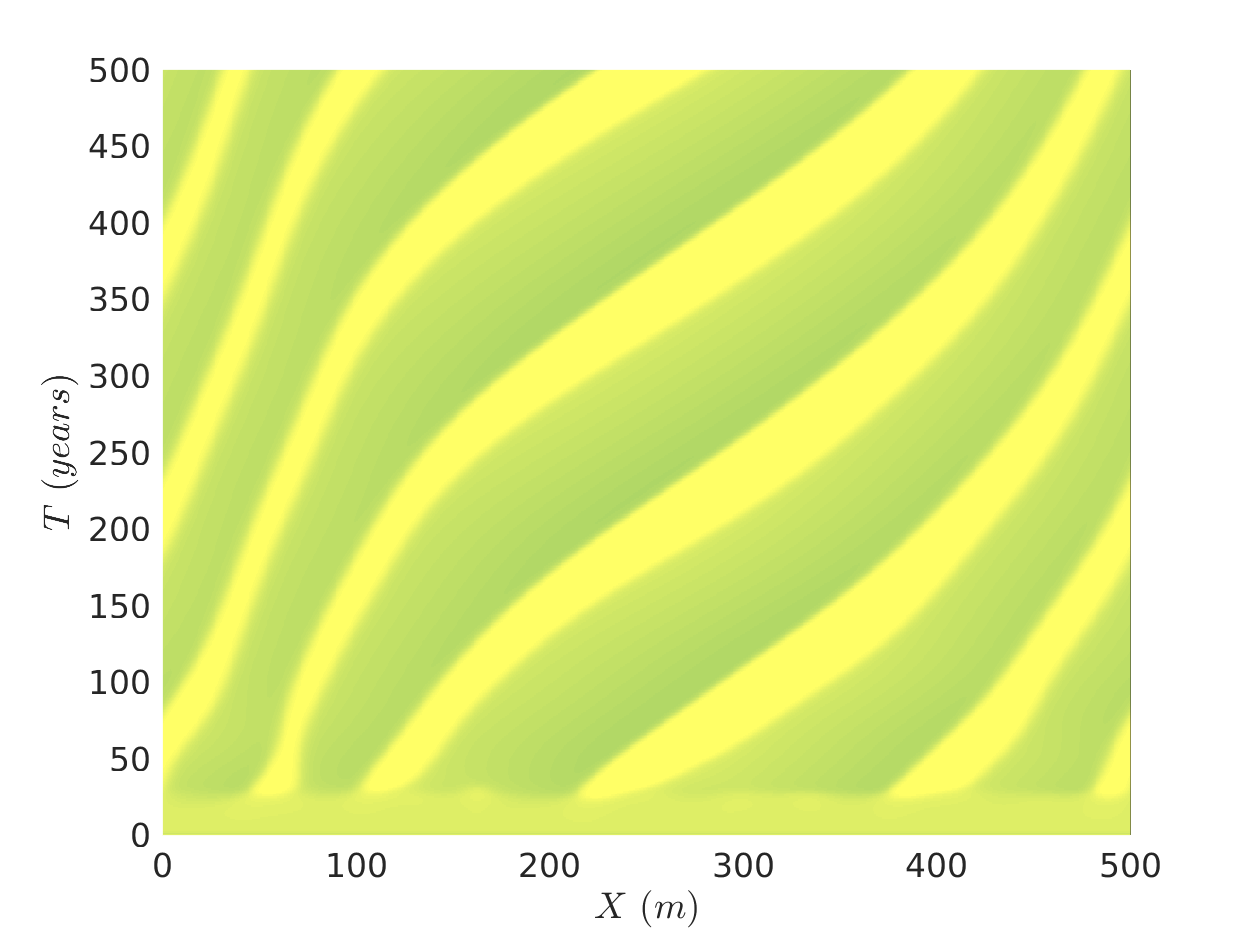}}
\put(0.02,0.35){\includegraphics[width=0.43\linewidth]{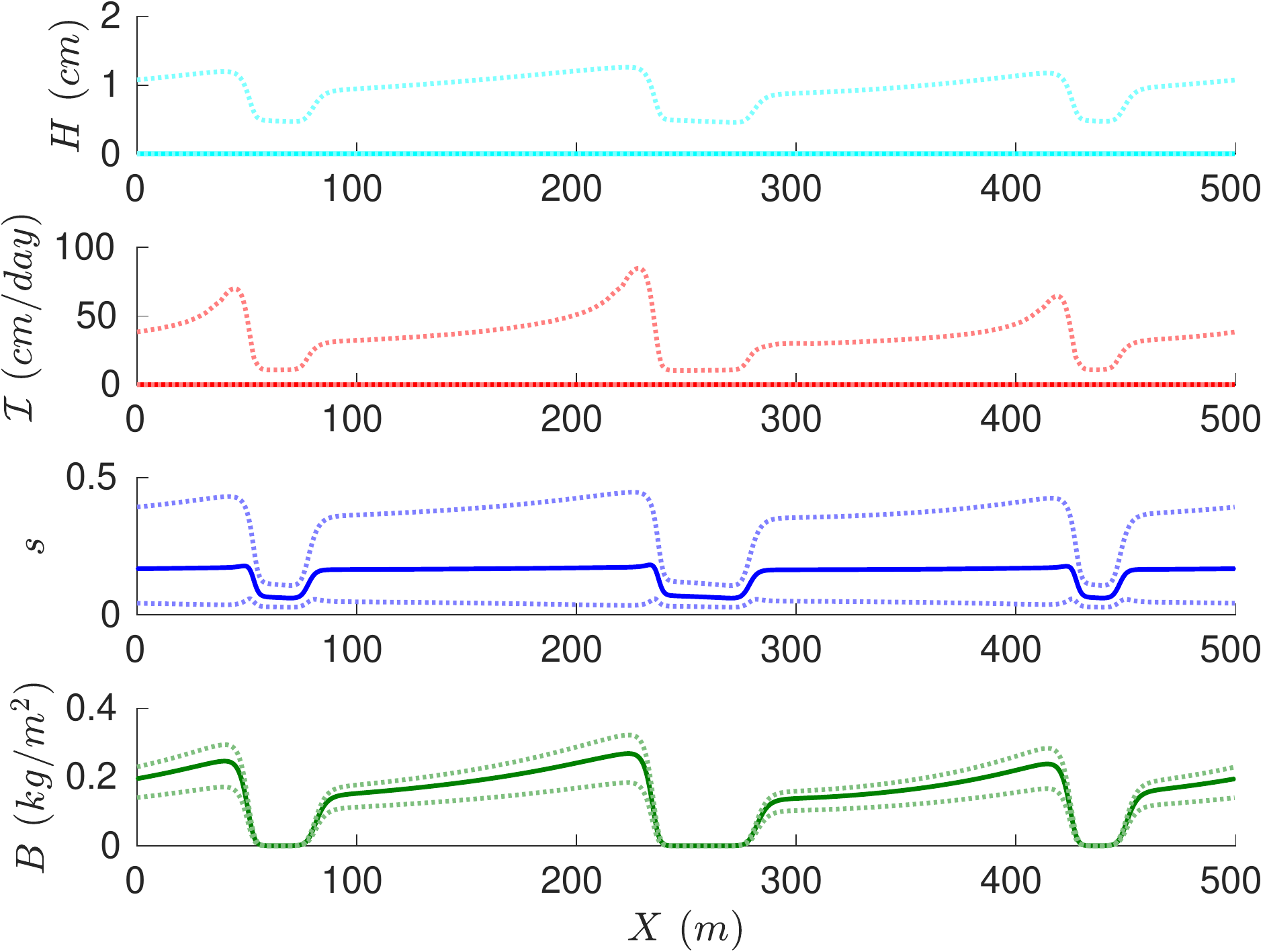}}
\put(0.52,0.35){\includegraphics[width=0.43\linewidth]{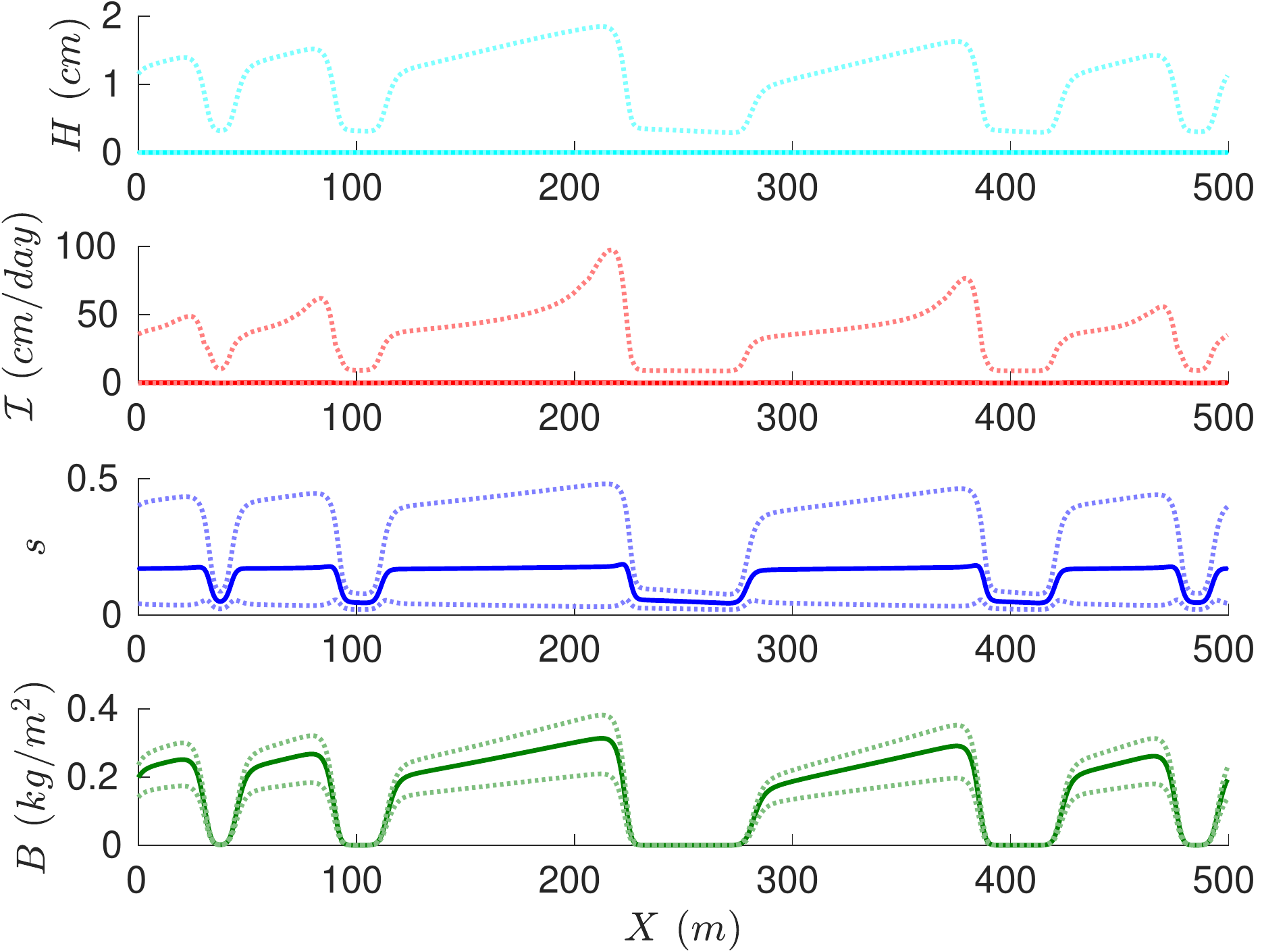}}
\put(0,.66){(a)}
\put(.5,.66){(b)}
\put(0,.33){(c)}
\put(.5,.33){(d)}
\end{picture}
    \caption{Nonlinearity in surface water advection, Left (a,c): $\delta=5/3$ and Right  (b,d): $\delta=1$. The upper plots (a,b) show spatial profile of surface water height $H$, Infiltration rate $\II$, soil moisture $s$, and biomass density $B$ at $t=500$ $years$.  The solid line is the annually averaged profile while the dotted lines show the pointwise minimum and maximum values over the course of the year.   The lower plots (c,d) show spacetime diagrams of annually averaged biomass $B$ in units of $kg/m^2$ over the course of the 500 $year$ simulation. 
    }
    \label{fig:nonlinh:spatialslice}
\end{figure*}

For a majority of the results presented in the main text, we take the exponent $\delta=1$  in the surface water transport speed term  $\nu(b,h)=h^{\delta-1}/(1+\eta b)$ of \cref{eq:fast:h}.  While $\delta=5/3$ would be more consistent with the theory for open channel flow, this increases computation time for our simulations. \Cref{fig:nonlinh:kn5} shows a comparison between solution profiles for a periodic pattern with wavenumber $k=2\pi/100\ m^{-1}$ for $\delta=1$ and $\delta=5/3$.  \Cref{fig:nonlinh:spatialslice} shows a comparison between $\delta=1$ and $\delta=5/3$ for simulations initialized with the same random initial perturbation to the uniform vegetation state given by \cref{eq:ss:uv}.  With $\delta=5/3$, the contrast in surface water height between vegetation bands and bare soil regions is less pronounced at peak values, which occur during rain storms. The $\delta=5/3$ state has wider bands with lower biomass values on average at $t=500$.  Despite these quantitative differences, the results are qualitatively similar.    

\subsection{Dependence on mean annual precipitation, rain intensity and timing}

\begin{figure}
    \centering
 \includegraphics[width=0.48\textwidth]{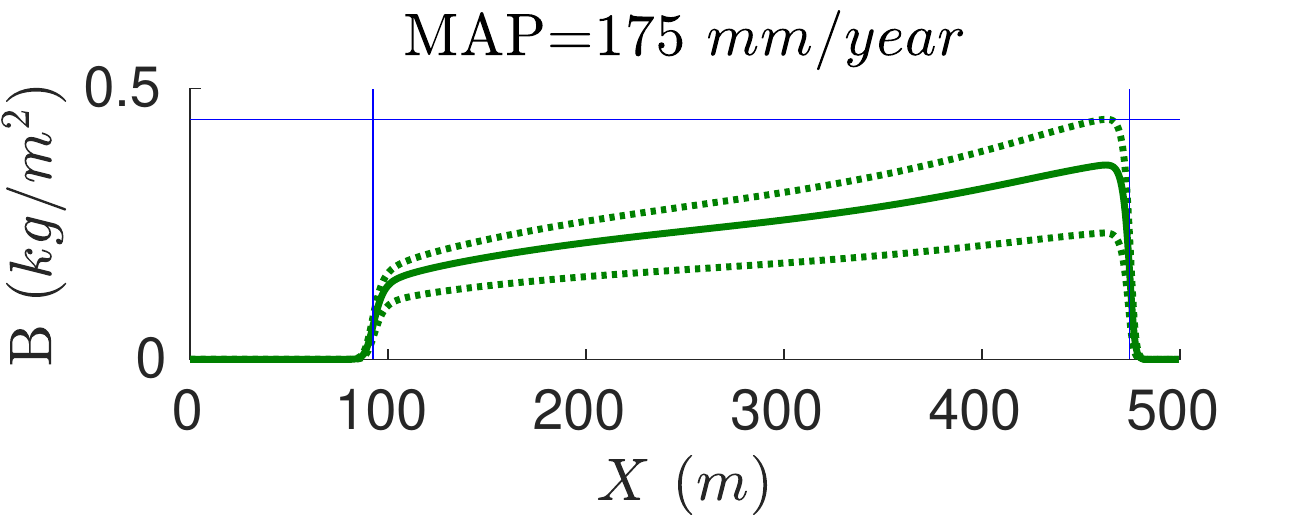}\\
 \includegraphics[width= 0.46\textwidth]{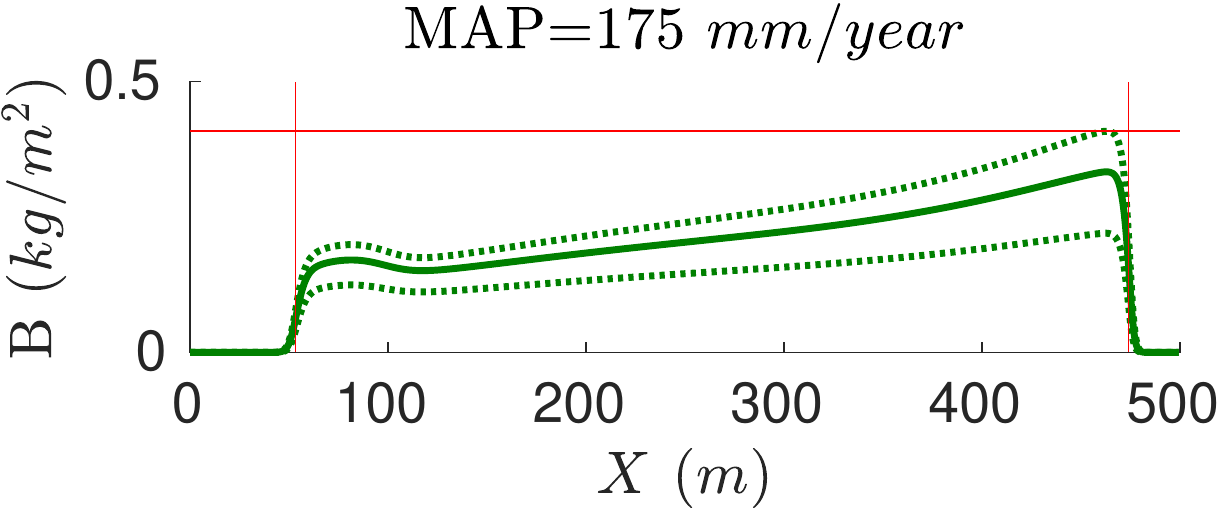}\\
    \includegraphics[width=0.48\textwidth]{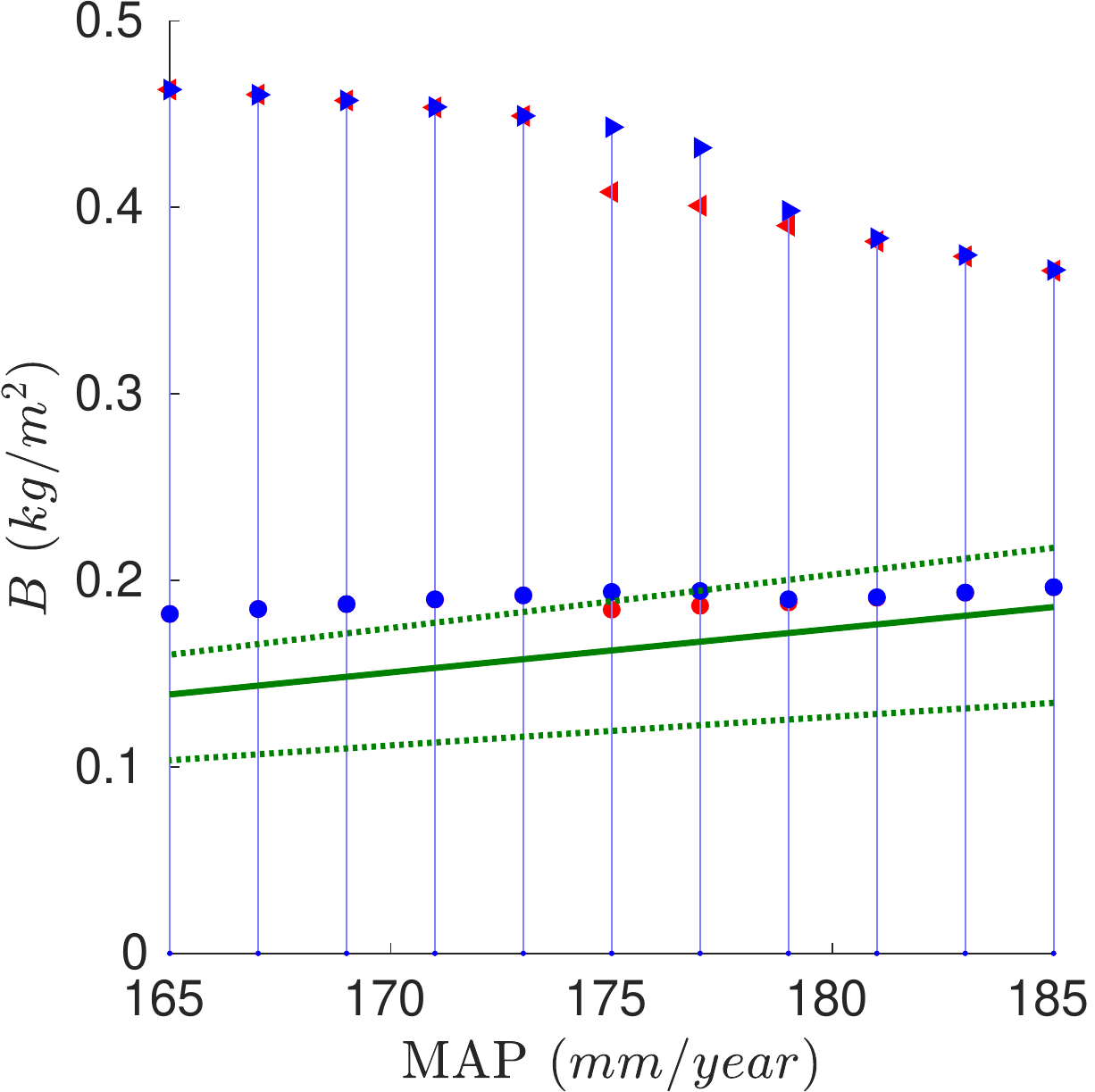}
    \caption{Bistability of one-band patterns.  The solid (dotted) green lines indicate the average (maximum and minimum) biomass values of the uniform vegetation state. The red left arrows indicate the maximum biomass values of one-band patterns obtained by decreasing precipitation with steps of 2 $mm/year$.   The blue right-arrows indicate the maximum biomass values for single-band patterns for increasing precipitation.  In both cases the dots indicate the (time and space) averaged biomass values, and the vertical lines indicate the range of biomass values within the pattern.  The solution profiles are obtained at 175~$mm/year$ during the precipitation scan up (boxed in blue) and the scan down (boxed in red).  }
    \label{fig:bistable1band}
\end{figure}

In \cref{fig:pscankn} of the main text, the single-band solutions undergo a jump in terms of maximum biomass value at around 177~$mm/year$ as precipitation is increased.  Scanning down in mean annual precipitation with a single-band solution reveals a hysteresis loop associated with this jump.  \Cref{fig:bistable1band} shows biomass profiles of the two distinct single-band solutions obtained at 175~$mm/year$.  The solution found by decreasing precipitation has a lower peak biomass value, a wider vegetation band, and has a bump in biomass at the trailing edge.

\begin{figure*}
    \centering
    \setlength{\unitlength}{\linewidth}
    \begin{picture}(1,0.65)
    \put(0,0){\includegraphics[width=0.49\linewidth]{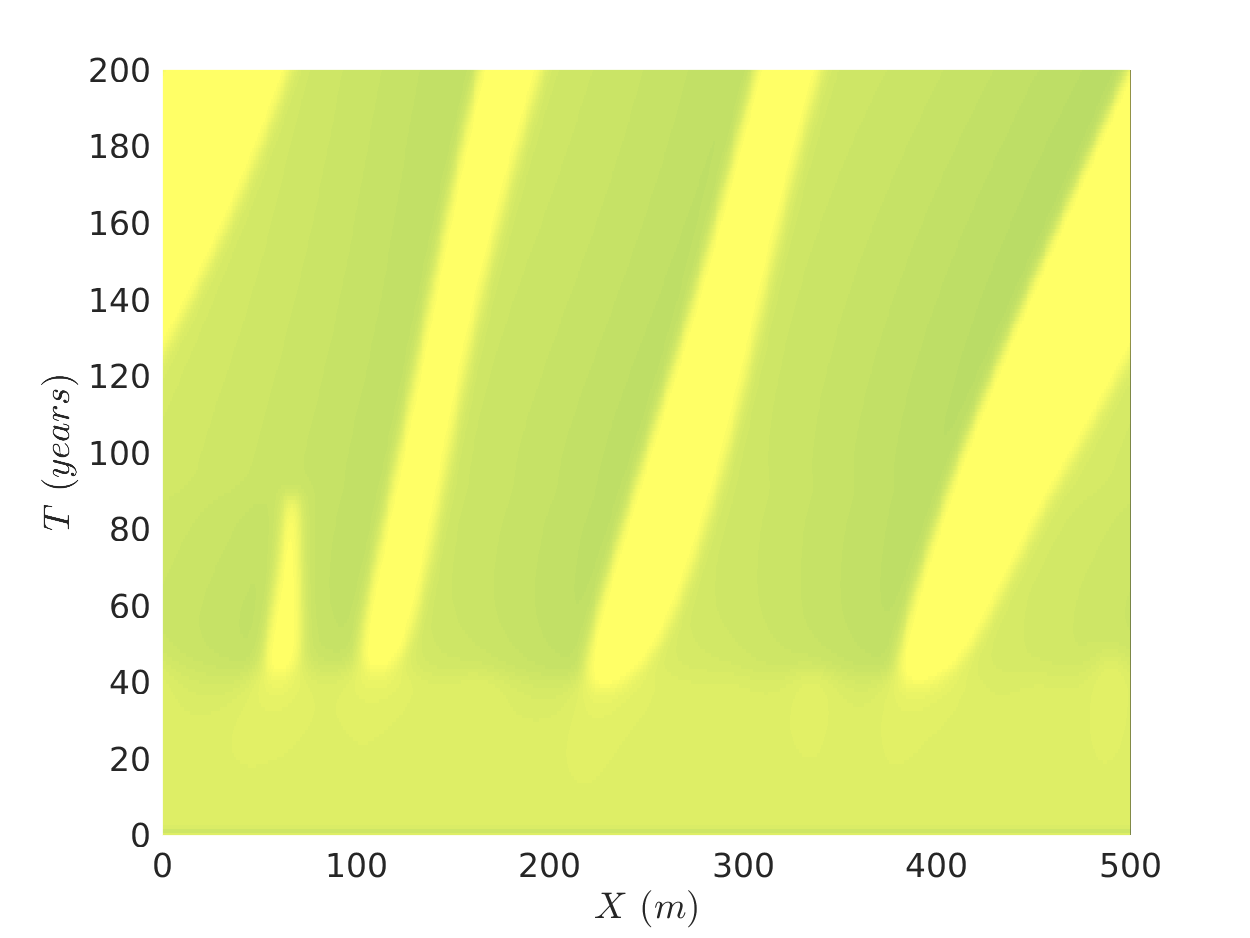}}
    \put(0.5,0){\includegraphics[width=0.49\linewidth]{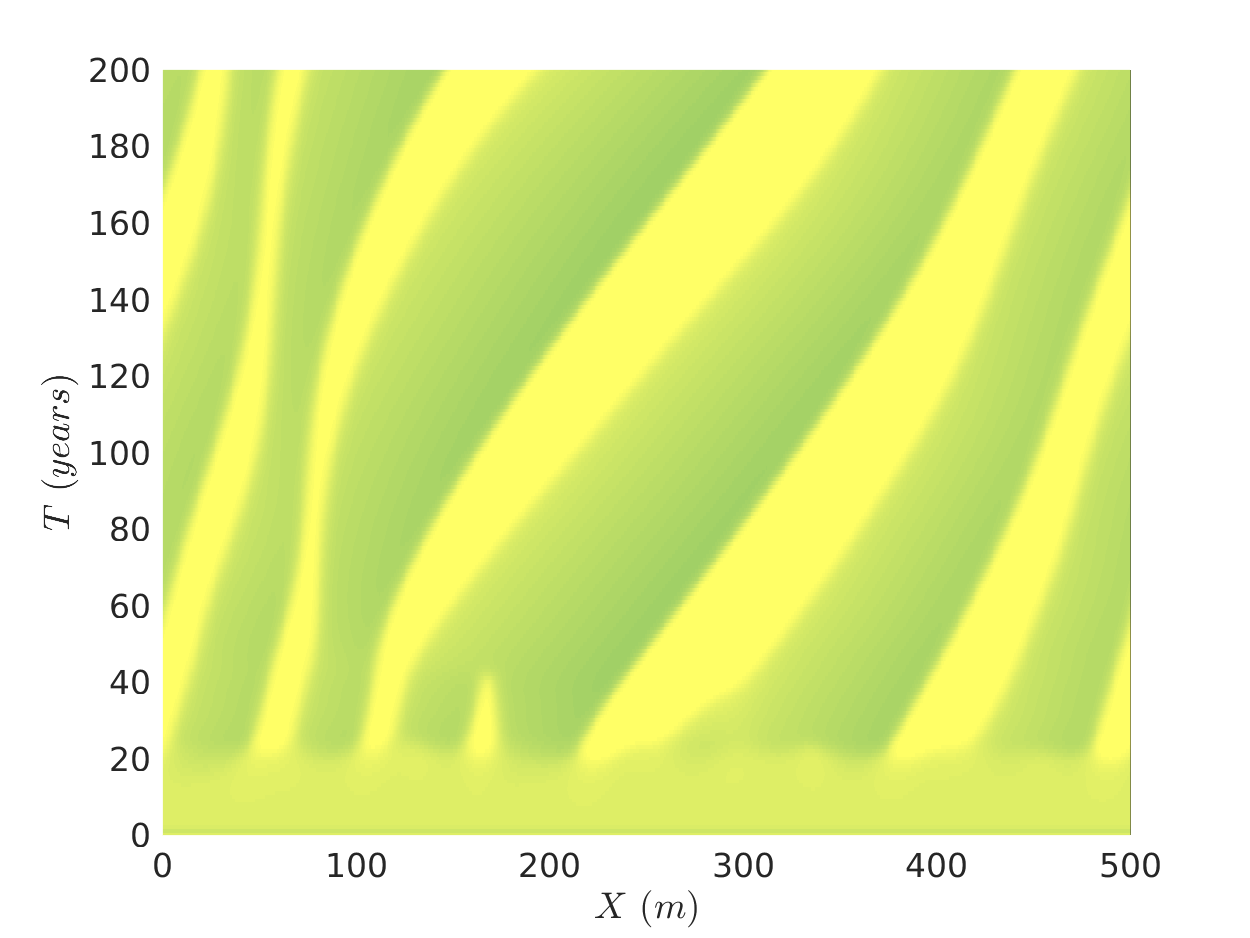}}
    \put(0.02,0.35){\includegraphics[width=0.43\linewidth]{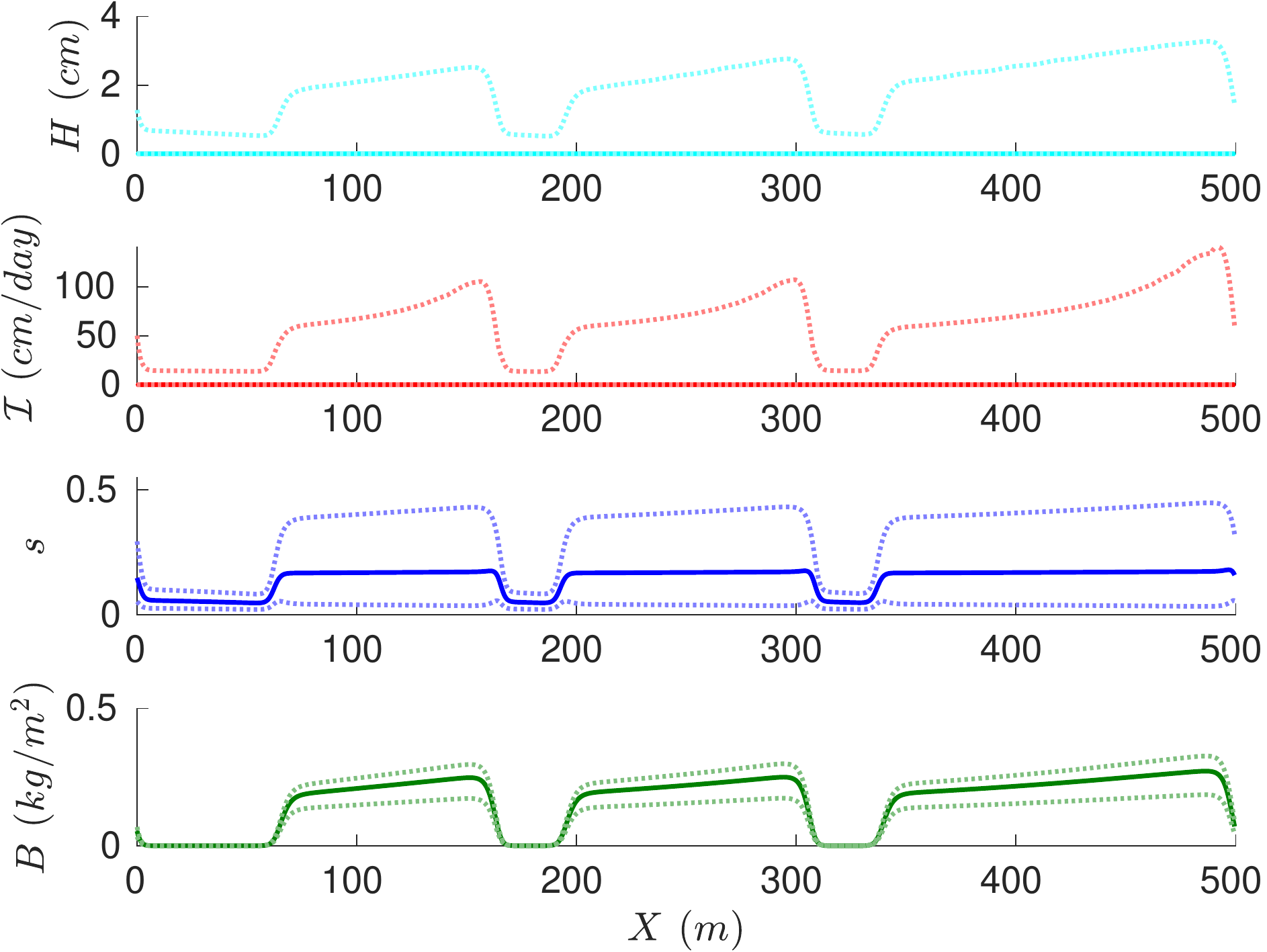}}
    \put(0.52,0.35){\includegraphics[width=0.43\linewidth]{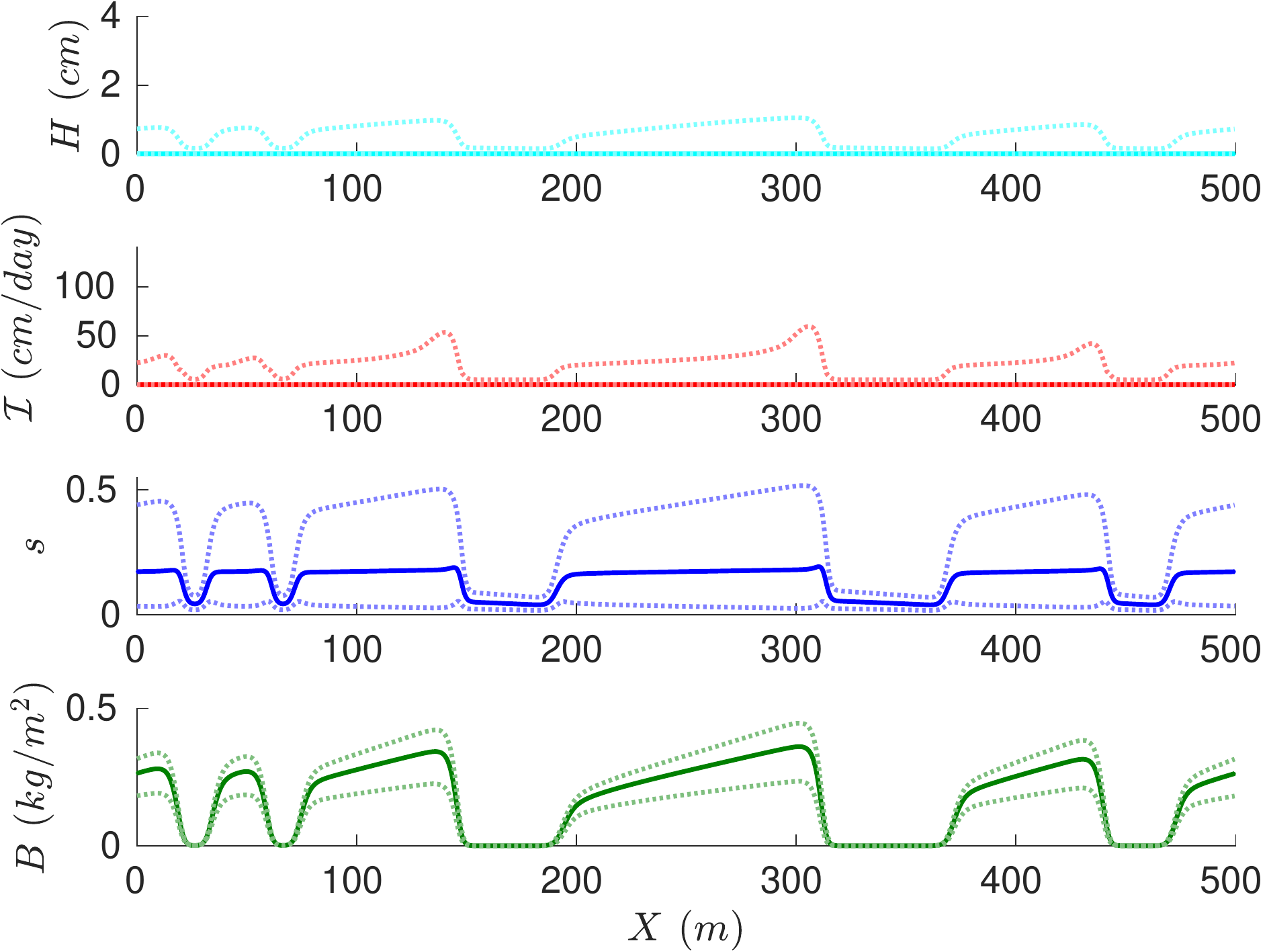}}
    \put(0,.66){(a)}
    \put(.5,.66){(b)}
    \put(0,.33){(c)}
    \put(.5,.33){(d)}
    \end{picture}
    \caption{Rain Intensity, Left (a,c) 3-hour rain events and Right (b,d): 12-hour rain events. (a,b)  Spatial profile of surface water height $H$, Infiltration rate $\II$, soil moisture $s$, and biomass density $B$ at $t=200$ $years$.  The solid line is the annually averaged profile while the dotted lines show the pointwise minimum and maximum values over the course of the year.   (c,d) Spacetime plot of annually averaged biomass $B$ in units of $kg/m^2$ over the course of the 200-$year$ simulation.     }
    \label{fig:intensity:spatialslice}
\end{figure*}    

The simulation results presented in the \cref{sec:sim} use a fixed rain input scheme of two six-hour events spaced 6 months apart, inspired by rainfall data shown in \cref{fig:rain}.  In order to explore the influence of the intensity of the rain events, we consider a fixed mean annual precipitation of 160~$mm/year$ and vary the length of the biannual rain events.  \Cref{fig:intensity:spatialslice} shows spacetime diagrams of biomass and spatial profiles for the case of 3~$hour$ and 12~$hour$ rainstorms.

\begin{figure*}
    \centering
    \includegraphics[width=\textwidth]{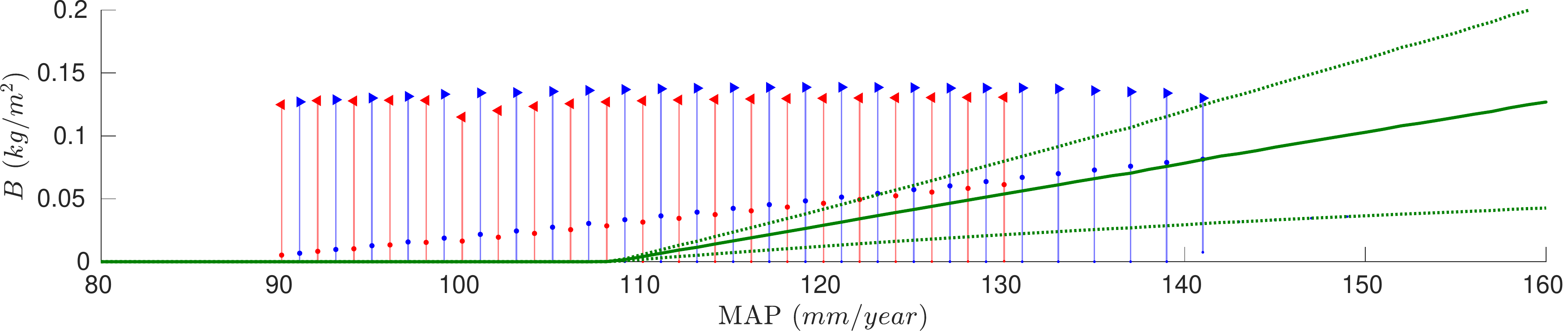}
    \caption{One rain event per year.  The solid (dotted) green lines indicate the average (maximum and minimum) biomass values of the uniform vegetation state. The red left arrows indicate the maximum biomass values of two- and one-band patterns obtained by decreasing precipitation with steps of 2 $mm/year$.   The blue right-arrows indicate the maximum biomass values for single-band patterns for increasing precipitation.  In both cases the dots indicate the (time and space) averaged biomass values, and the vertical lines indicate the range of biomass values within the pattern.  }
    \label{fig:pscan1rain}
\end{figure*}

Linear stability analysis for the coupled model, \cref{eq:3field}, shown in \cref{fig:uv:floquet}(b) indicates a drastic shrinking of the precipitation range over which the uniform state is unstable to patterns when a single 12-hour rain event per year is used instead of two equally spaced 6-hour rain events per year.  \Cref{fig:pscan1rain} shows that simulations of the fast-slow switching model, are consistent with these predictions.  With one rain event per year, patterns are found from 89 to 147 $mm/year$ with precipitation scans.  This is a significantly narrower range than the 34 to 201 $mm/year$ that is found with two rain events per year.

\begin{figure*}
	\centering
	\includegraphics[width=0.48\textwidth]{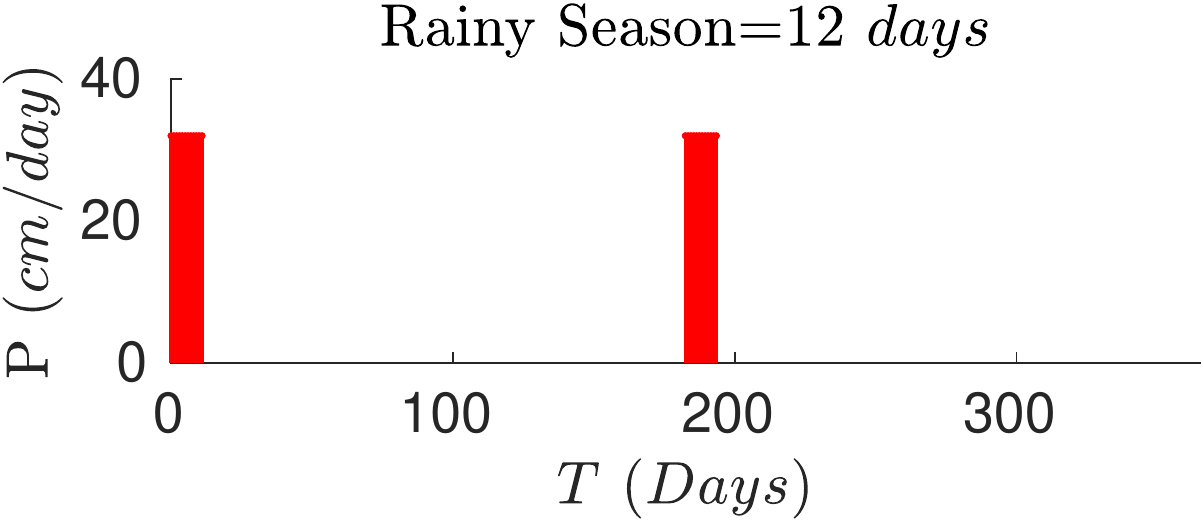}\hfill
	\includegraphics[width=0.48\textwidth]{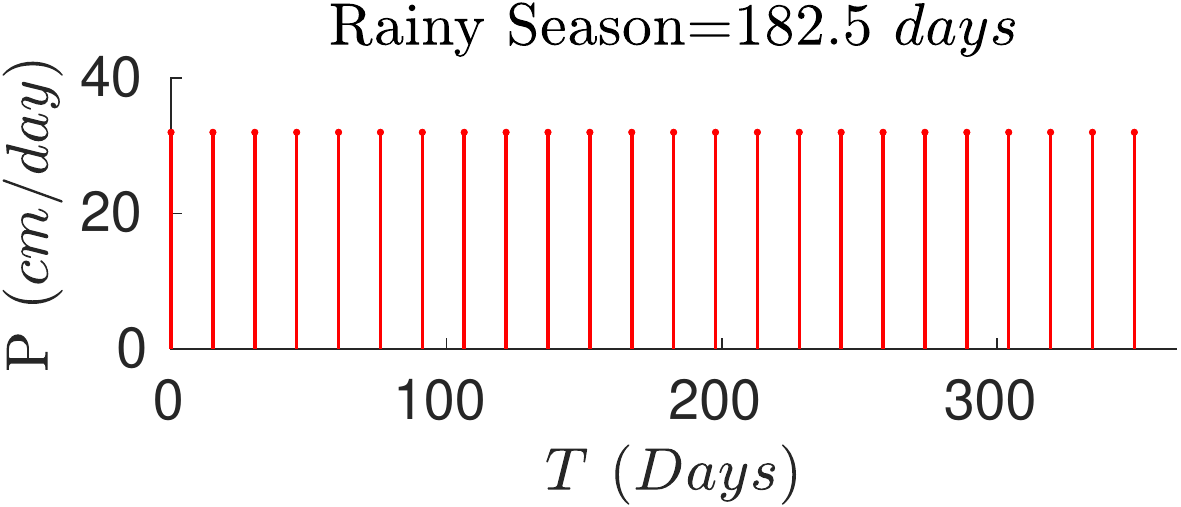}	\\ \vspace{5mm}
		\includegraphics[width=0.48\textwidth]{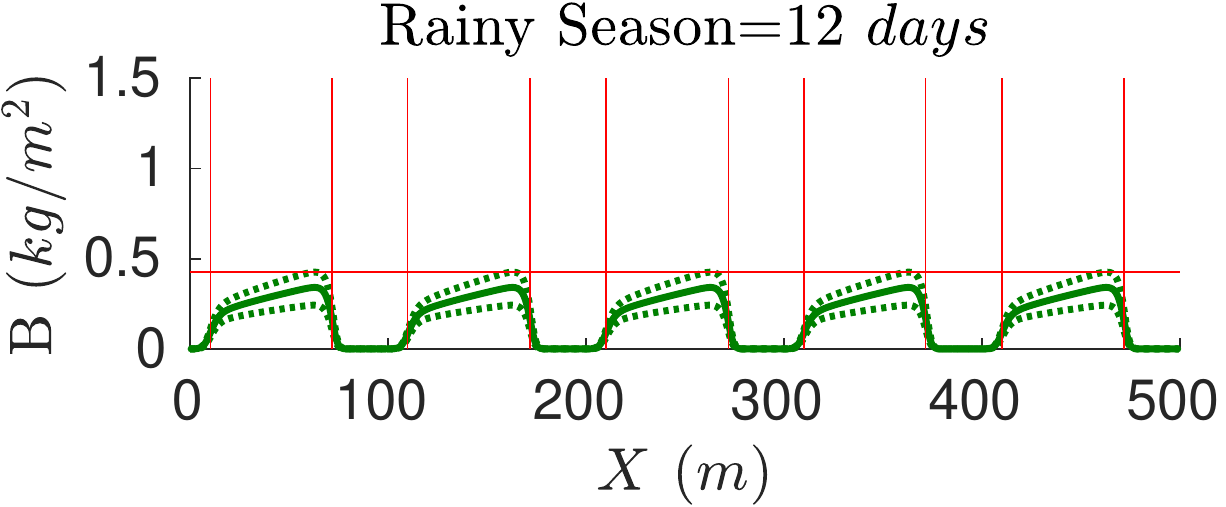}\hfill
	\includegraphics[width=0.48\textwidth]{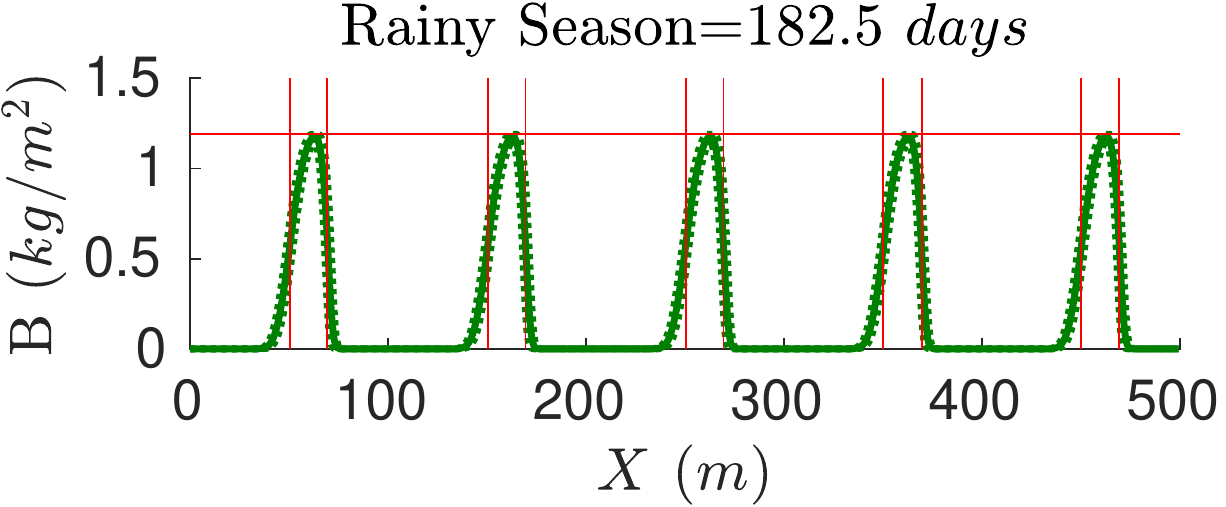}	\\ \vspace{5mm}
	\includegraphics[width=\textwidth]{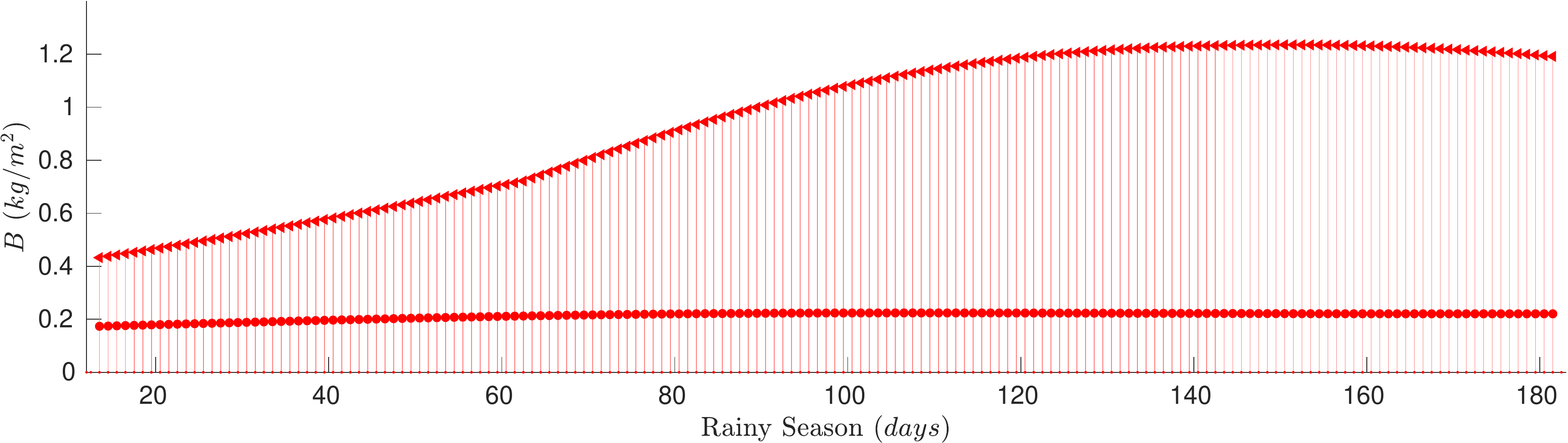}\\ 
		\includegraphics[width=\textwidth]{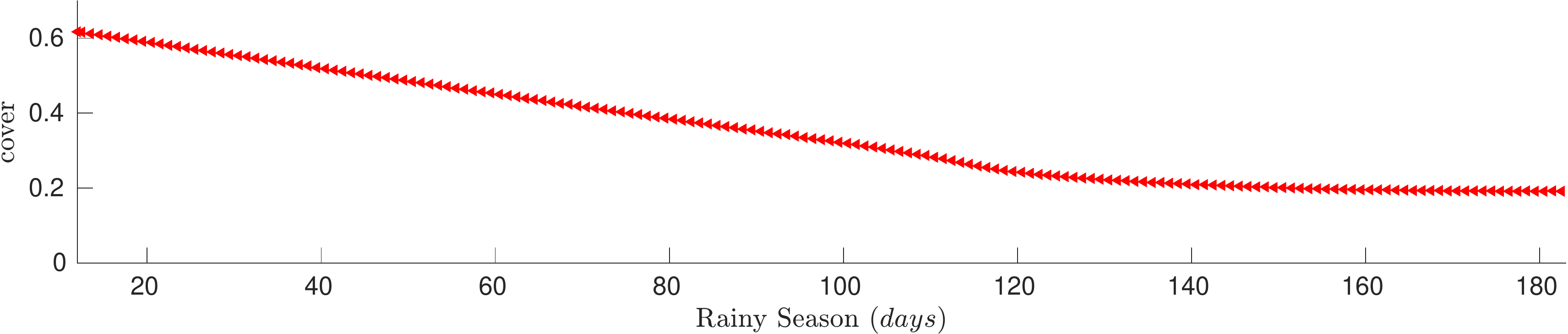}\\
	
	\caption{Gradual decrease in length of rainy season. (top row) Initial and final rain pattern, from right to left.  Initial pattern, on the right, consists of 24 equal and uniformly spaced rain events each lasting 30 minutes with an instantaneous rainfall rate of 32~$cm/day$.  The final rain pattern, on the left, consists of two 12-day long rainy seasons of 12 equal and uniformly spaced rain events also lasting 30 minutes with  instantaneous rainfall rate of 32~$cm/day$. Note that the rain events are too close together to be resolved individually in this left panel. (2nd row) Biomass profiles associated with the two rainfall patterns shown above.  The solid green lines indicate the time-averaged biomass while the dotted lines indicate the pointwise maximum and minimum values.  The red lines indicate the maximum biomass value indicated by the top of the vertical lines in the third row and the widths of the bands used to compute the fraction of the domain covered by biomass in the bottom row.  (3rd row) Biomass as a function of length of the rainy season.  The vertical bars indicate the range of biomass values obtained over the course of a year, and the circles within each bar indicate the time and space averaged value of biomass.  The peak biomass and average biomass values with a 12-day rainy season are approximately 0.36 and 0.79 times the corresponding values with a 6-month rainy season. (bottom row)  Fraction of domain covered by biomass as a function of the length of the rainy season.       }
	\label{fig:pspacing}
\end{figure*}

In order to explore the influence of moving towards seasonal rain input, we consider a case in which twelve rain events of 30 minutes each are evenly distributed over a rainy seasons that occurs every 6 months.  Initially each rainy season spans the entire 6-month period, but the length is gradually decreased in steps of 12 $hours$ every decade until the 12-event season occurs over a period of 12 days.  The first row of \cref{fig:pspacing} shows the initial and final patterns of instantaneous rainfall rates, from right to left.  We note that, for the 12 day rainy season, there are still 12 distinct rain events of 30 minutes even though they are not resolved in the figure. For the simulation, we take the parameters given in \cref{tab:nondim} and a MAP of 160~$mm/year$.  The second row shows the average biomass profiles (solid) along with the pointwise minimum and maximum (dotted) values. As the rainfall becomes more seasonal, the vegetation bands become wider with lower peak biomass value and larger fluctuations.  The last two rows of \cref{fig:pspacing} show the peak and averaged biomass, and the fraction of the domain covered by biomass as a function of the length of the rainy seasons.  While the more seasonal rain pattern leads to a larger fraction of the domain covered by biomass, the less seasonal rain leads to more biomass on the domain.  The peak biomass and average biomass values with a 12-day rainy season are approximately 0.36 and 0.79 times the corresponding values with a 6-month rainy season.  

\section{Linear stability of the uniformly vegetated state in the three-field coupled timescale model}
\label{app:linstab}

Here we summarize the equations used to compute the linear stability of the spatially uniform vegetation solution $(h_0(t),s_0(t),b_0(t))$ of the non-dimensionalized version of the three-field coupled timescale model~\eqref{eq:nondim}, in the case of temporally periodic precipitation $p(t)=p(t+\tau)$, where $\tau$ is the (minimal) period associated with the rain events. The uniform solution satisfies the following system of ordinary differential equations:
\begin{eqnarray}
\label{eq:periodicodes}
\dot{h}_0&=&p(t)-\iota(h_0,s_0,b_0)\nonumber\\
\dot{s_0}&=&\alpha\ \iota(b_0,s_0,h_0)-\epsilon(\sigma +\gamma b_0)s_0\\
\dot{b_0}&=&\epsilon(s_0(1-b_0)-\mu)b_0,\nonumber
\end{eqnarray}
where, recall,
\begin{equation}
\label{eq:iota}
    \iota(h,s,b)\equiv\Bigl(\frac{b+qf}{b+q}
    \Bigr)\Bigl(\frac{h}{h+1}\Bigr)(1-s)^\beta.
\end{equation}
The linear variational equations, associated with a perturbation of the form $e^{ikx}$ are obtained by inserting the ansatz
\begin{eqnarray*}
h(t,x)&=&h_0(t)+h_{1,k}(t)e^{ikx},\\
s(t,x)&=&s_0(t)+s_{1,k}(t)e^{ikx},\\
b(t,x)&=&b_0(t)+b_{1,k}(t)e^{ikx},\\
\end{eqnarray*}
into the coupled model~\eqref{eq:nondim} and linearizing in $(h_{1,k},s_{1,k},b_{1,k})$. The resulting linear ordinary differential equations, with time-periodic coefficients, are
\begin{eqnarray}
\label{eq:linFM}
\begin{pmatrix}
\dot{h}_{1,k}\\
\dot{s}_{1,k}\\
\dot{b}_{1,k}
\end{pmatrix}&=&
{\cal J}(t)\begin{pmatrix}
{h}_{1,k}\\
{s}_{1,k}\\
{b}_{1,k}
\end{pmatrix}\\ & & +
\begin{pmatrix}
ik\Bigl(\frac{\delta h_0^{\delta-1}}{1+\eta b_0}\Bigr)&\quad 0\quad &-ik\Bigl(\frac{\eta h_0^\delta}{(1+\eta b_0)^2}\Bigr)\\
0&0&0\\
0&0&-\delta_bk^2
\end{pmatrix}\begin{pmatrix}
{h}_{1,k}\\
{s}_{1,k}\\
{b}_{1,k}
\end{pmatrix}. \nonumber
\end{eqnarray}
Here the $\tau$-periodic Jacobian matrix ${\cal J}$, associated with~\eqref{eq:periodicodes}, is 
\begin{equation*}
    {\cal J}(t)=\begin{pmatrix}
-\partial_{h_0} \iota&-\partial_{s_0} \iota& -\partial_{b_0} \iota\\
\alpha\ \partial_{h_0} \iota &\alpha\ \partial_{s_0} \iota -\epsilon(\sigma+\gamma b_0)& \alpha\ \partial_{b_0} \iota -\epsilon\gamma s_0\\
\qquad 0\qquad&\epsilon(1-b_0)b_0&\epsilon ((1-2b_0)s_0-\mu)
\end{pmatrix},
\end{equation*}
where $\partial_{h_0}\iota$ is short-hand for $\frac{\partial \iota}{\partial h_0}$, etc.  The zero subscripts emphasize that the infiltration function $\iota(h_0,s_0,b_0)$, \cref{eq:iota},  is to be evaluated on the spatially uniform solution. 

The Floquet multipliers, for a given perturbation wavenumber $k$ and a given periodic precipitation function $p(t)$, are obtained as the eigenvalues of the Monodromy matrix ${\cal M}$.
In practice, to compute ${\cal M}$, we must first determine a good approximation of the initial condition for the $\tau$-periodic solution $(h_0(t),s_0(t),b_0(t))$. Since this solution is stable to perturbations with $k=0$
we can obtain an excellent approximation to the initial condition by forward-time integration of the ordinary differential equations~\eqref{eq:periodicodes}. Once we have the initial condition we simultaneously solve~\eqref{eq:periodicodes} with the linearized equations~\eqref{eq:linFM} over one period $\tau$
for three independent initial conditions $(h_{1,k},s_{1,k},b_{1,k})=(1,0,0),(0,1,0),(0,0,1)$. The resulting $(h_{1,k}(\tau),s_{1,k}(\tau), b_{1,k}(\tau))$ form the columns of the
$3\times 3$ Monodromy matrix ${\cal M}$. If all the eigenvalues of ${\cal M}$ are inside the unit circle then the uniform $\tau$-periodic solution is stable to perturbations of wavenumber $k$. (Our approach to computing the Floquet multipliers associated with Eq.~\eqref{eq:linFM} follows that described in the textbook [J.D. Meiss, \textit{Differentiable Dynamical Systems.} SIAM, 2017].)

\end{document}